%% file: arxiv.tex
\documentclass{article} 
\usepackage{arxiv,times}

\input{math_commands.tex}

\usepackage{hyperref}
\usepackage{url}
\usepackage{amsmath}
\usepackage[english]{babel}
\newtheorem{theorem}{Theorem}
\usepackage{wrapfig}
\usepackage{algpseudocode}
\usepackage{tikz}
\usepackage[ruled,linesnumbered]{algorithm2e}

\usepackage{caption}
\usepackage{graphicx}
\usepackage{multirow}
\usepackage{booktabs}
\usepackage{comment}
\usepackage{subfig} 

\usepackage[utf8]{inputenc} 
\usepackage[T1]{fontenc}    
\usepackage{booktabs}       
\usepackage{amsfonts}       
\usepackage{nicefrac}       
\usepackage{microtype}      
\usepackage{xcolor}         
\usepackage{colortbl}
\usepackage[symbol]{footmisc}

\DeclareCaptionFormat{nosublabel}{#3}
\DeclareCaptionLabelFormat{nosublabel}{}
\newtheorem{prop}{Proposition}
\newcommand{\sizecci}{0.32}
\newcommand{\sizeccj}{0.98}
\newtheorem{definition}[theorem]{Definition}

\title{\includegraphics[scale=0.05]{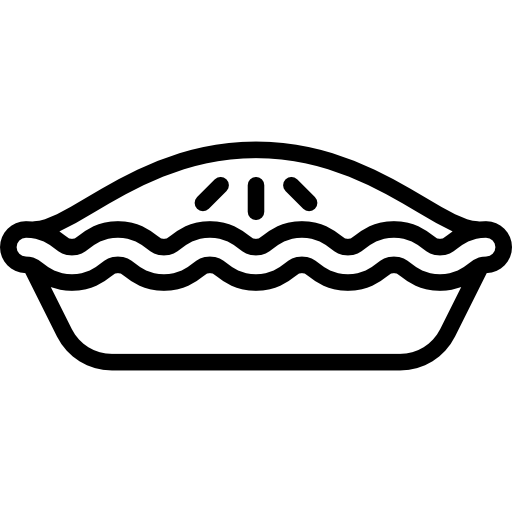} \color{violet}P\color{cyan}I\color{red}E\color{black}: Simulating Disease Progression via \color{violet}P\color{black}rogressive \color{cyan}I\color{black}mage \color{red}E\color{black}diting}

\iclrfinalcopy
\author{%
  \textbf{Kaizhao Liang$^{1*}$}, \textbf{Xu Cao$^{2,6*}$}, \textbf{Kuei-Da Liao$^{3}$}, \textbf{Tianren Gao$^{4}$}, \textbf{Wenqian Ye$^{5,6}$}, \textbf{Zhengyu Chen$^{7}$}, \\ \textbf{Jianguo Cao$^{6}$}, \textbf{Tejas Nama$^{8}$}, \textbf{Jimeng Sun$^{2}$} \\
  $^{1}$The University of Texas at Austin \space $^{2}$University of Illinois Urbana-Champaign \\
  $^{3}$University of California San Diego \space
  $^{4}$University of California, Berkeley \space $^{5}$University of Virginia \\ 
  $^{6}$PediaMed AI \space $^{7}$ Northwestern University \space $^{8}$Carnegie Mellon University 
  \\
  \texttt{kaizhaol@utexas.edu, xucao2@illinois.edu, wenqian@virginia.edu}
  \\
  \texttt{jimeng@illinois.edu}
}

\begin{document}

\maketitle

\begin{figure}[!htbp]
    \centering
    \includegraphics[width=0.95\linewidth]{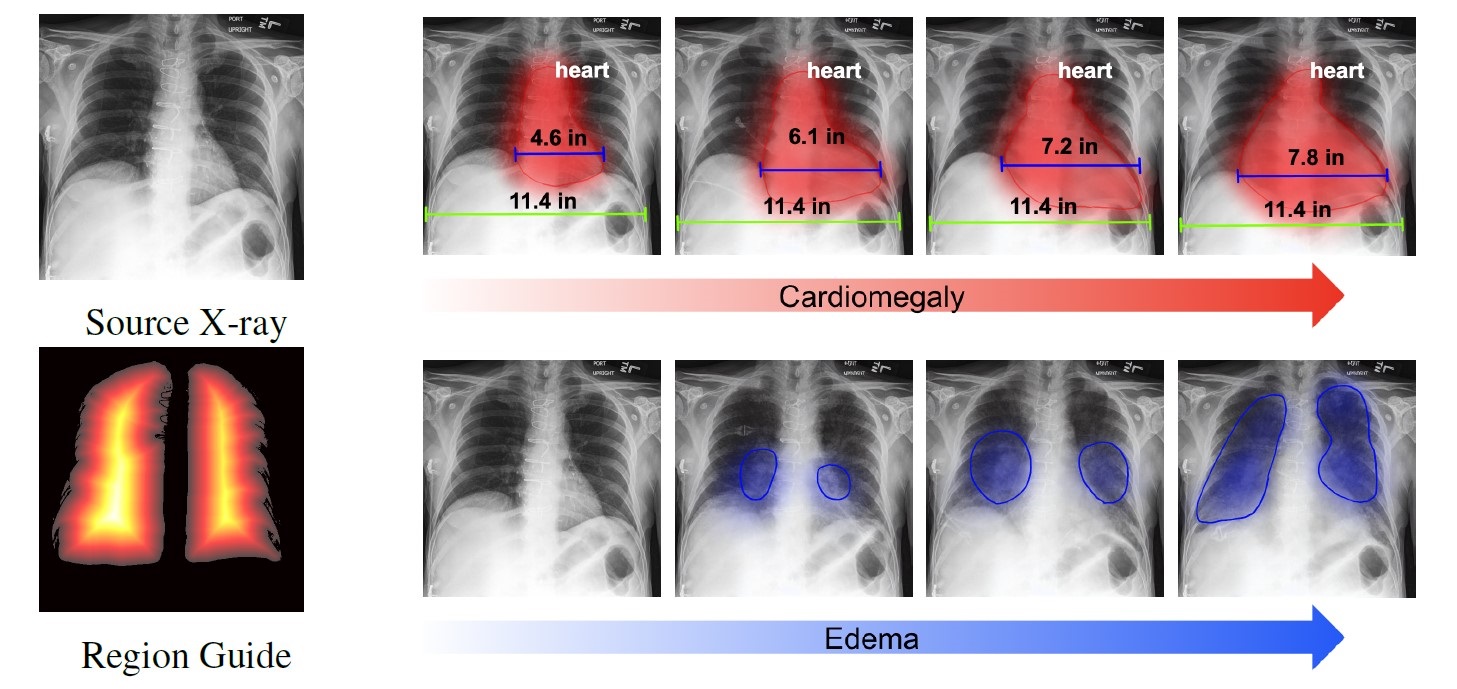}
    \caption{Illustrative examples of disease progression simulation using PIE. The top progression sequence depicts a patient's heart increasing in size (\textcolor{red}{red}), indicating Cardiomegaly. The bottom sequence demonstrates the expanding mass areas (\textcolor{blue}{blue}) in a patient's lung, indicating Edema.}
    \label{fig:repaint}
\end{figure}

\begin{abstract}
The trajectories of disease progression could greatly affect the quality and efficacy of clinical diagnosis, prognosis, and treatment. However, one major challenge is the lack of longitudinal medical imaging monitoring of individual patients over time. To address this issue, we develop a novel framework termed Progressive Image Editing (PIE) that enables controlled manipulation of disease-related image features, facilitating precise and realistic disease progression simulation in imaging space. Specifically, we leverage recent advancements in text-to-image generative models to simulate disease progression accurately and personalize it for each patient. We also theoretically analyze the iterative refining process in our framework as a gradient descent with an exponentially decayed learning rate. To validate our framework, we conduct experiments in three medical imaging domains. Our results demonstrate the superiority of PIE over existing methods such as Stable Diffusion Video and Style-Based Manifold Extrapolation based on CLIP score (Realism) and Disease Classification Confidence (Alignment). Our user study collected feedback from 35 veteran physicians to assess the generated progressions. Remarkably, $76.2\%$ of the feedback agrees with the fidelity of the generated progressions. PIE can allow healthcare providers to model disease imaging trajectories over time, predict future treatment responses, fill in missing imaging data in clinical records, and improve medical education. \footnote{Equal Contribution. Code and checkpoints for replicating our results can be found at \href{https://github.com/IrohXu/PIE}{github.com/IrohXu/PIE} and \href{https://huggingface.co/IrohXu/stable-diffusion-mimic-cxr-v0.1}{huggingface.co/IrohXu/stable-diffusion-mimic-cxr-v0.1}.}
\end{abstract}

\input{tex_files/introduction}

\input{tex_files/related_works}

\input{tex_files/problem_statement}


\input{tex_files/method}

\input{tex_files/experiment}

\input{tex_files/conclusion}




\bibliography{arxiv}
\bibliographystyle{iclr2024_conference}

\appendix
\input{tex_files/supplementary}


\end{document}

%% file: math_commands.tex
\usepackage{amsmath,amsfonts,bm}

\def\eqref#1{equation~\ref{#1}}

\def\1{\bm{1}}

\DeclareMathAlphabet{\mathsfit}{\encodingdefault}{\sfdefault}{m}{sl}
\SetMathAlphabet{\mathsfit}{bold}{\encodingdefault}{\sfdefault}{bx}{n}



%% file: tex_files/introduction.tex
\section{Introduction}
Disease progression refers to how an illness develops over time in an individual. By studying the progression of diseases, healthcare professionals can create effective treatment strategies and interventions. It allows them to predict the disease's course, identify possible complications, and adjust treatment plans accordingly.  Furthermore, monitoring disease progression allows healthcare providers to assess the efficacy of treatments, measure the impact of interventions, and make informed decisions about patient care. A comprehensive understanding of disease progression is essential for improving patient outcomes, advancing medical knowledge, and finding innovative approaches to prevent and treat diseases.

However,  disease progression modeling in the imaging space poses a formidable challenge primarily due to the lack of continuous monitoring of individual patients over time and the high cost to collect such longitudinal data~\citep{sukkar2012disease,wang2014unsupervised,liu2015efficient,cook2016disease,severson2020personalized}. 
The intricate and multifaceted dynamics of disease progression, combined with the lack of comprehensive and continuous image data of individual patients, result in the absence of established methodologies~\citep{hinrichs2011predictive,ray2011multimodality,lee2019predicting}. Moreover, disease progression exhibits significant variability and heterogeneity across patients and disease sub-types, rendering a uniform approach impracticable. 

Past disease progression simulation research has limitations in terms of its ability to incorporate clinical textual information, generate individualized predictions based on individualized conditions, and utilize non-longitudinal data. This highlights the need for more advanced and flexible simulation frameworks to accurately capture the complex and dynamic nature of disease progression in imaging data. To incorporate the generation model into a conditioned simulation of disease progression, we propose a progressive framework PIE, for disease progression simulation that combines text and image modalities. Specifically, we aim to progressively add and subtract disease-related features, controlled by a text encoder, to conditionally progress the disease without significantly altering the original base image features (see Figure~\ref{fig:repaint}). Our framework is built based on the invertibility of denoising diffusion probabilistic models~\citep{ho2020denoising,song2020denoising}. Our theoretical analysis shows PIE can be viewed as a gradient descent toward the objective maximum log-likelihood of given text conditioning. The learning rate in this iterative process is decaying exponentially with each iteration forward, which means that the algorithm is effectively exploring the solution space while maintaining a balance between convergence speed and stability. This theoretical analysis guarantees that our framework is moving the instance toward the targeted manifold and ensures modification is bounded.

We evaluate PIE on three distinct medical imaging datasets with non-longitudinal disease progression data, including Chexpert~\citep{irvin2019chexpert}, Diabetic Retinopathy Detection~\citep{CHF2015retino} and ISIC 2018~\citep{codella2019skin}. We demonstrate that our framework leads to more accurate and individualized disease progression predictions on these datasets, which can improve clinical diagnosis, treatment planning, and enhance patient records by filling in missing imaging data and potentially helping medical education.
 We also conducted a user study with physicians to evaluate the effectiveness of PIE for disease progression simulation. The study presented physicians with a set of simulated disease images and progressions, and then asked them to assess the accuracy and quality of each generated image and progression.  

\begin{itemize}
    \item We propose a temporal medical imaging simulation framework PIE, which allows for more precise and controllable manipulation of disease-related image features and leads to more accurate and individualized longitudinal disease progression simulation.
    \item We provide theoretical evidence that our iterative refinement process is equivalent to gradient descent with an exponentially decaying learning rate, which helps to establish a deeper understanding of the underlying mechanism and provides a basis for further improvement.
    \item We demonstrate the superior performance of PIE over baselines in disease progression prediction with three medical domains. The results show that PIE produces more accurate and high-quality disease progression prediction.
    \item We also conducted a user study with physicians to evaluate the effectiveness of our proposed framework for disease progression simulation. The physicians agree that simulated disease progressions generated by PIE closely matched physicians' expectations $76.2\%$ of the time, indicating high accuracy and quality.
\end{itemize}

%% file: tex_files/related_works.tex
\section{Related Works}

\textbf{Disease Progression Simulation} Longitudinal disease progression data derived from individual electronic health records offer an exciting avenue to investigate the nuanced differences in the progression of diseases over time~\citep{schulam2016disease,stankeviciute2021conformal,chen2022clustering,mikhael2023sybil,koval2021ad}. Most of the previous works are based on HMM~\citep{wang2014unsupervised,liu2015efficient,alaa2017learning} and deep probabilistic models~\citep{alaa2019attentive}. Some recent works start to resolve disease progression simulation by using deep generation models. \citep{ravi2022degenerative} utilized GAN-based model and linear regressor with individual's sequential monitoring data for Alzheimer's disease progression simulation in MRI imaging space. However, all these methods have to use full sequential images and fail to address personalized healthcare in the imaging space. The lack of such time-series data, in reality, poses a significant challenge for disease progression simulation~\citep{xue2020learning,chen2022machine,berrevoets2023impute}.

\textbf{Generative Models}
Generative models like Variational Autoencoders (VAEs)~\citep{kingma2013auto} and Generative Adversarial Networks (GANs)~\citep{goodfellow2020generative} have been widely employed in medical imaging applications~\citep{nie2017medical,isola2017image,cao2020auto}. Recent GAN models ~\citep{kang2023scaling, Patashnik_2021_ICCV} have harnessed the power of CLIP~\citep{radford2021learning} embedding to guide image editing based on contextual prompts. However, GAN-based models are unstable and difficult to optimize in general. Denoising Diffusion Models~\citep{sohl2015deep, ho2020denoising,song2020denoising,rombach2022high,karras2022elucidating} have become increasingly popular in recent years due to their ability to create photo-realistic images from textual descriptions. One major advantage of these models is their ability to learn from large-scale datasets. Among the various text-to-image models, Stable Diffusion ~\citep{rombach2022high} has received considerable attention because of its impressive performance in generating high-quality images and its relatively low cost to fine-tune. Its denoising process works similarly to the diffusion models but in a latent space, and this process results in a final image that is highly consistent with the input text, making it an excellent tool for text-guided image editing. Diffusion models can also be effortlessly incorporated into an image-to-image editing pipeline~\citep{brooks2022instructpix2pix,parmar2023zero,orgad2023editing}, thus providing users the ability to edit scenarios across multiple modalities and assess potential imaging progressive editing paths. However, existing image-to-image methods can only be used for single-step editing, which makes it difficult to simulate personalized time-series progression data in the medical domain.

%% file: tex_files/problem_statement.tex
\section{Problem Statement}

In the traditional disease progression simulation setting, assume having sequential time-series image-text data pairs $\{(\boldsymbol{x_0}, y_0), (\boldsymbol{x_1}, y_1), ..., (\boldsymbol{x_T}, y_T)\}$ from each patient. The clinical image-text data pair $(\boldsymbol{x}, y) \in \mathcal{X} \times \mathcal{Y}$ is sampled from a non-independent identically distribution, where $\mathcal{Y}=\mathbb{R}^n$ denote the medical report space and $\mathcal{X}=\mathbb{R}^m$ denote the medical imaging space. All the prior works either rely heavily on probability modeling: $f_\theta(y_{0:t-1}) \rightarrow y_{t}$~\citep{liu2015efficient,alaa2019attentive}, or rely on using longitude data to train regression models for imaging simulation: $f_\theta(x_{0:t-1}, y_{t-1}) \rightarrow y_{t}$~\citep{han2022image,ravi2022degenerative}. However, it is hard to obtain sequential longitude data as most patients may not go to the same hospital for follow-up treatment. And the hospitals also lack medical imaging and clinical reports in the early stages of the disease.

In this paper, we redefine disease progression simulation using a data-driven generative model without the need for sequential time-series data or clinical prior knowledge. Anyone with access to discrete imaging and medical report data could individually train the model to predict disease progressions without profound medical prior, significantly reducing the amount of work required for feature engineering and data collection. 

\begin{definition}[Simulate disease progression with non-sequential data]
\label{def:progress_discrete}
Assume $h_\phi$ is a generative model learned from the data space: $\Omega = \{ (\boldsymbol{x}, y) \in \chi \times \Gamma \}$, assuming it is independent identically distributed and each $(\boldsymbol{x}, y)$ is from different individuals. In training phase, $h_\phi$ models the mapping: $\Gamma \rightarrow \chi$. In the inference phase, given an initial test data sample $(x_{t}, y_{t})$ at progression stage $t$, $h_\phi$ converts input imaging $x_{t}$ and input clinical context $y_T$ to $x_{T}$, where $y_T$ is the language model inferred final step clinical report from $y_{0}$ and $x_{t}, x_{t+1}, ..., x_{T}$ is the simulated sequential imaging progression. 
\end{definition}

In the following sections, we picked DDIM as a base step of our proposed method, because of its reversible theoretical properties that allow smooth transitions and convergence based on Definition ~\ref{def:progress_discrete}. The proof is shown in the supplementary section.

%% file: tex_files/method.tex
\begin{figure}[!htbp]
    \centering
    \includegraphics[width=0.90\linewidth]{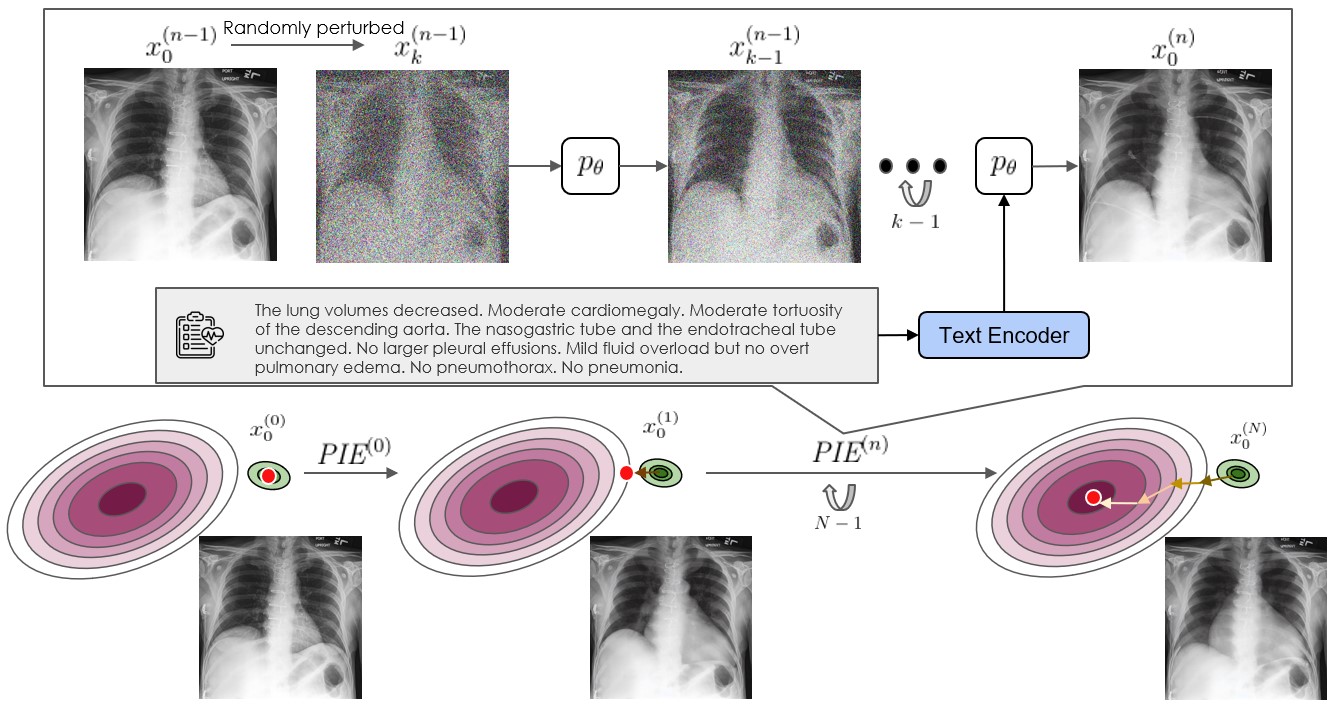}
    \caption{Overview of the PIE inference pipeline. PIE is illustrated using an example of disease progression editing X-ray from a healthy state to cardiomegaly. For any given step $n$ in PIE, we first utilize DDIM inversion to procure an inverted noise map. Subsequently, we denoise it using clinical reports imbued with progressive cardiomegaly information. The output of DDIM denoising serves as the input for step $n+1$, thus ensuring a gradual and controllable disease progression simulation. After simulating $N$ steps, the image is converged to the final state.}
    \label{fig:pipeline}
\end{figure}

\section{Progressive Image Editing (PIE)}

Progressive image editing (PIE) is a novel framework proposed to refine and enhance images in an iterative and discrete manner, allowing the use of additional prompts for small and precise adjustments to simulate semantic modification while keeping realism. Unlike traditional image editing techniques, PIE involves a multi-stage process where each step builds upon the previous one, with the aim of achieving a final result that is more refined and smooth than if all changes were made at once. The approach also enables precise control over specific semantic features of the image to be adjusted without significant impacts on other regions. The main purpose of PIE is to simulate disease progression from multi-modal input data.



\textbf{Procedure.} 
The inputs to PIE are a discrete medical image $x_0^{(0)}$ depicting any start or middle stage of a disease and a corresponding clinical report latent $y$ as the text conditioning~\citep{rombach2022high}. $y$ is generated from a pretrained text encoder from CLIP~\citep{radford2021learning} [clip-vit-large-patch14], where the raw text input could either be a real report or synthetic report, providing the potential hint of the patient's disease progression. The output generated is a sequence of images, $\{ x_0^{(0)}, x_0^{(1)}, ..., x_0^{(N)} \}$, illustrating the progression of the disease as per the input report. The iterative PIE procedure is defined as follows:

\begin{prop}
\label{prop:1}
Let $x_{0}^{(N)} \sim \chi$, where $\chi$ is distribution of photo-realistic images , $y$ be the text conditioning,  running $PIE^{(n)}(\cdot, \cdot)$ recursively is denoted as following, where $N \geq n \geq 1$,
\begin{equation}
    x_0^{(n)} = PIE^{(n)}(x_0^{(n - 1)}, y)
\end{equation}
Then, the resulting output $x_0^{(N)}$ maximizes the posterior probability $p(x_0^{(N)} |\ x^{(0)}_{0}, y)$.
\end{prop}

With each round of editing as shown in Figure ~\ref{fig:pipeline}, the image gets closer to the objective by moving in the direction of $-\nabla \log p(x|y)$. Due to the properties of DDIM, the step size would gradually decrease with a constant factor. Additional and more detailed proofs will be available in Supplementary \ref{appendix:proof:theorem1}.

\begin{prop}
    \label{prop:2}
    Assuming $\|x_0^{(0)}\|\leq C_1$ and $\|\epsilon_{\theta}(x, y)\|\leq C_2$, $(x, y)\in (\chi, \Gamma)$, for any $\delta > 0$, if
    \begin{equation}
        n > \frac{2}{\log(\alpha_0)}\cdot (log(\delta) - C)
    \end{equation}
    then,
    \begin{equation}
        \|x^{(n+1)}_{0} - x^{(n)}_{0}\| < \delta
    \end{equation}
    where, $\lambda =\frac{\sqrt{\alpha_0 - \alpha_0\alpha_1} - \sqrt{\alpha_1 - \alpha_0\alpha_1}}{\sqrt{\alpha_{1}}}$, $\chi$ is the image distribution, $\Gamma$ is the text condition distribution , and $ C = \log((\frac{1}{\sqrt{\alpha_0}} - 1)\cdot C_1 + \lambda \cdot C_2)$
\end{prop}

\begin{prop}
    \label{prop:3}
    For all $N > 1$, $\|x_0^{(N)} - x_0^{(0)}\| \leq [(\frac{1}{\sqrt{\alpha_0}} - 1)\cdot C_1 + \lambda \cdot C_2]$
\end{prop}

In addition, Proposition \ref{prop:2} and \ref{prop:3} show as $n$ grows bigger, the changes between steps would grow smaller. Eventually, the difference between steps will get arbitrarily small. Hence, the convergence of $PIE$ is guaranteed and modifications to any inputs are bounded by a constant.

%% file: tex_files/experiment.tex

\begin{algorithm}[H]
\SetAlgoLined
\KwIn{Original input image $x_{0}^{(0)}$ at the start point, input image $x_{0}^{(n-1)}$ at stage $n$, number of diffusion steps $T$, text conditional vector $y$, noise strength $\gamma$, stable diffusion parameterized denoiser $\epsilon_{\theta}$, a ROI mask $M_{ROI}$, $M_{ROI}^{i,j} \in [0,1]$}

\KwOut{Modified image $x'$ as $x_{0}^{n}$}
$x' \gets x_{0}^{(n-1)}$

$k \gets \gamma \cdot T$

$\epsilon \sim \mathcal{N}(0, \mathcal{I})$ 

$x' \gets \sqrt{\alpha_{k}}\cdot x' + \sqrt{1 - \alpha_{k}} \cdot \epsilon$

\For{$t=k$ \KwTo $1$}{
    $x' \gets \sqrt{\alpha_{t-1}}(\frac{x' - \sqrt{1 - \alpha_t}\epsilon_{\theta}^{(t)}(x', y)}{\sqrt{\alpha_t}}) + \sqrt{1 - \alpha_{t - 1}}\cdot \epsilon_{\theta}^{(t)}(x', y)$
}

$x' \gets ( \beta_1 \cdot (x' - x_{0}^{(0)}) + x_{0}^{(0)}) \cdot(1 - M_{ROI}) + ( \beta_2 \cdot (x' - x_{0}^{(0)}) + x_{0}^{(0)}) \cdot M_{ROI} $

\Return $x'$ \textbf{as} $x^{(n)}_{0}$
\caption{Progressive Image Editing $n$-th step ($\textit{PIE}^{(n)}$)}
\label{algo:1}
\end{algorithm}

\section{Experiments and Results}
In this section, we present experiments on various disease progression tasks. Experiments results demonstrate that PIE can simulate the disease-changing trajectory that is influenced by different medical conditions. 
Notably, PIE also preserves unrelated visual features from the original medical imaging report, even as it progressively edits the disease representation. Figure~\ref{fig:progression_visualization} showcases a set of disease progression simulation examples across three distinct types of medical imaging. Details for Stable Diffusion fine-tuning, pretraining model for confidence metrics settings are available in Supplementary~\ref{appendix:experiment}.

\begin{table}[!htbp]
\small
\centering
\caption{Comparisons with multi-step editing simulations. The backbone of PIE and baseline approaches are Stable Diffusion with the same pre-trained weight.}
\label{tab:performance}
\begin{tabular}{lcccccc}
\toprule
\multirow{2}{*}{\textbf{Method}} & \multicolumn{2}{c}{\textbf{Chest X-ray}} & \multicolumn{2}{c}{\textbf{Retinopathy}} & \multicolumn{2}{c}{\textbf{Skin Lesion Image}} \\ \cmidrule(lr){2-3} \cmidrule(lr){4-5} \cmidrule(lr){6-7}
& \textbf{Conf $(\uparrow)$} & \textbf{CLIP-I  $(\uparrow)$} & \textbf{Conf $(\uparrow)$} & \textbf{CLIP-I $(\uparrow)$} & \textbf{Conf $(\uparrow)$} & \textbf{CLIP-I  $(\uparrow)$} \\ 
\midrule
Stable Diffusion Video &  0.389 & 0.923 & 0.121 & 0.892 & 0.201 & 0.886 \\ 
Extrapolation & 0.0543 & \textbf{0.972} & 0.0742 & 0.991 & 0.226 & 0.951 \\ 
PIE & \textbf{0.690} &  0.968 & \textbf{0.807} &  \textbf{0.992} & \textbf{0.453} &  \textbf{0.958} \\ 
\bottomrule
\end{tabular}
\end{table}

\subsection{Experimental Setups}

\textbf{Implementation Details.} We present the details of single-step PIE in Algorithm~\ref{algo:1}. For $\textit{PIE}^{(n)}$, we define $\alpha_k$ according to the DDIM case. Line 8 in Algorithm~\ref{algo:1} ensures progressive and limited modifications between the original input image $x_0^{(0)}$, the single-step edited output $x'$, and the region guide selector $M_{ROI}$ through the utilization of interpolation average parameters $\beta_1$ and $\beta_2$. These parameters dictate the modification ratio between the ROI mask-guided space and the original input space. As $\beta_1$ increases, the multi-step editing process becomes smoother, though it may sacrifice some degree of realism.

\begin{figure}[!htbp]
    \centering
    \begin{tabular}{c}
    \begin{minipage}[t]{0.03\textwidth} 
        \rotatebox{90}{\centering \!\!\!\!\!\!\!\!\!\!\!\!\! Chest X-ray}
    \end{minipage}
    \begin{minipage}[t]{0.75\textwidth}
        \begin{minipage}{0.2\textwidth}
            \centering
            \scriptsize
            \includegraphics[width=\textwidth]{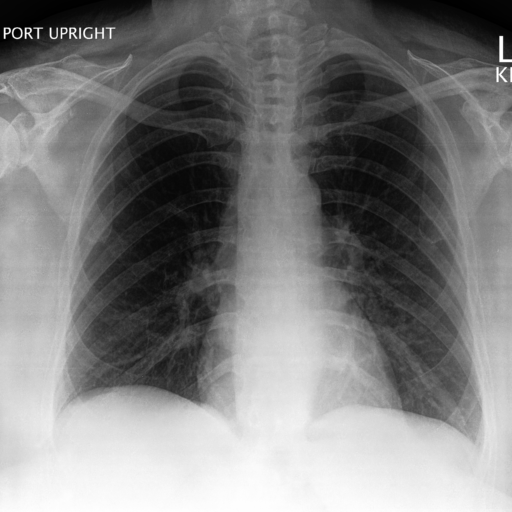}\\
        \end{minipage}
        \begin{minipage}{0.02\textwidth}
            \centering
            \scriptsize
            \includegraphics[width=1.0\textwidth]{figures/ar.png}\\
        \end{minipage}
        \begin{minipage}{0.2\textwidth}
            \centering
            \scriptsize
            \includegraphics[width=\textwidth]{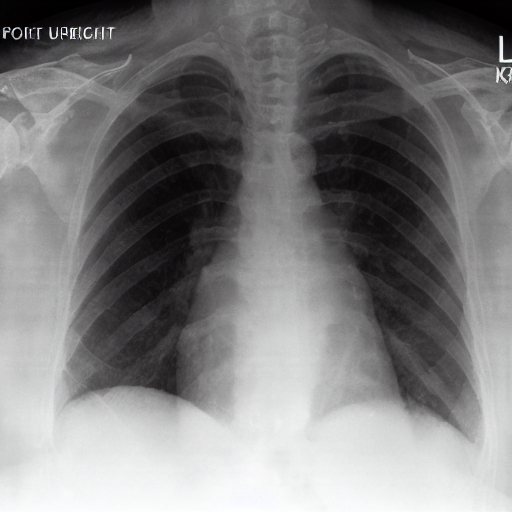}\\
        \end{minipage}
        \begin{minipage}{0.02\textwidth}
            \centering
            \scriptsize
            \includegraphics[width=1.0\textwidth]{figures/ar.png}\\
        \end{minipage}
        \begin{minipage}{0.2\textwidth}
            \centering
            \scriptsize
            \includegraphics[width=\textwidth]{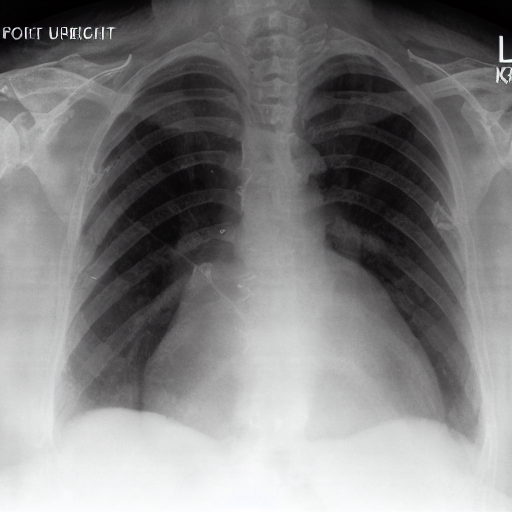}\\
        \end{minipage}
        \begin{minipage}{0.02\textwidth}
            \centering
            \scriptsize
            \includegraphics[width=1.0\textwidth]{figures/ar.png}\\
        \end{minipage}
        \begin{minipage}{0.2\textwidth}
            \centering
            \scriptsize
            \includegraphics[width=\textwidth]{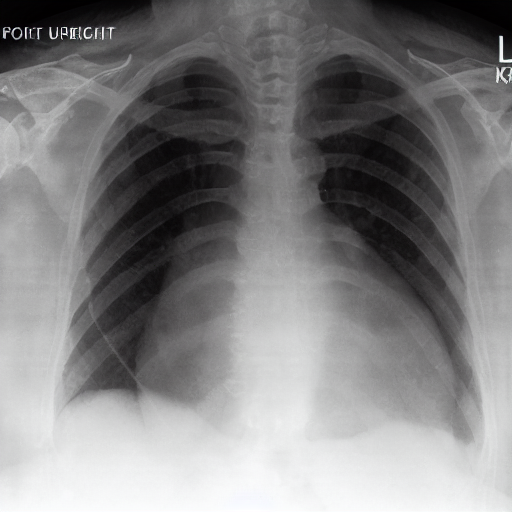}\\
        \end{minipage}
    \end{minipage}
    \\
    \begin{minipage}[t]{0.03\textwidth}
        \rotatebox{90}{\centering \!\!\!\!\!\!\!\!\!\!\!\!\!\! Retinopathy}
    \end{minipage}
    \begin{minipage}[t]{0.75\textwidth}
        \begin{minipage}{0.2\textwidth}
            \centering
            \scriptsize
            \includegraphics[width=\textwidth]{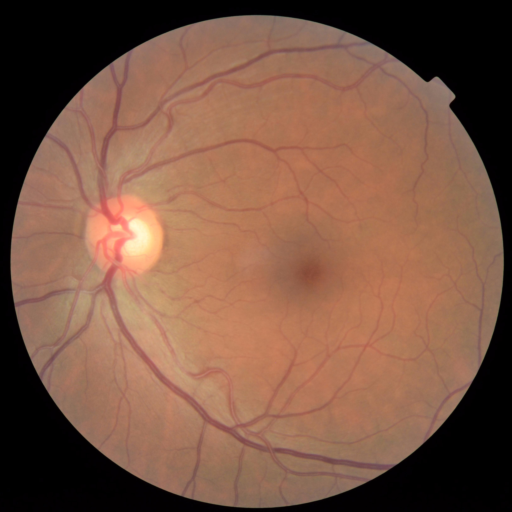}\\
        \end{minipage}
        \begin{minipage}{0.02\textwidth}
            \centering
            \scriptsize
            \includegraphics[width=1.0\textwidth]{figures/ar.png}\\
        \end{minipage}
        \begin{minipage}{0.2\textwidth}
            \centering
            \scriptsize
            \includegraphics[width=\textwidth]{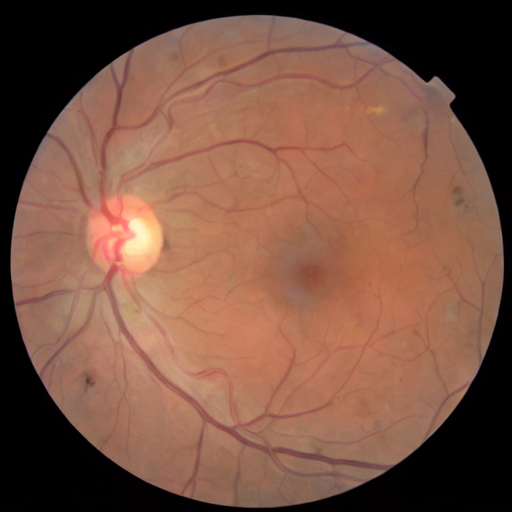}\\
        \end{minipage}
        \begin{minipage}{0.02\textwidth}
            \centering
            \scriptsize
            \includegraphics[width=1.0\textwidth]{figures/ar.png}\\
        \end{minipage}
        \begin{minipage}{0.2\textwidth}
            \centering
            \scriptsize
            \includegraphics[width=\textwidth]{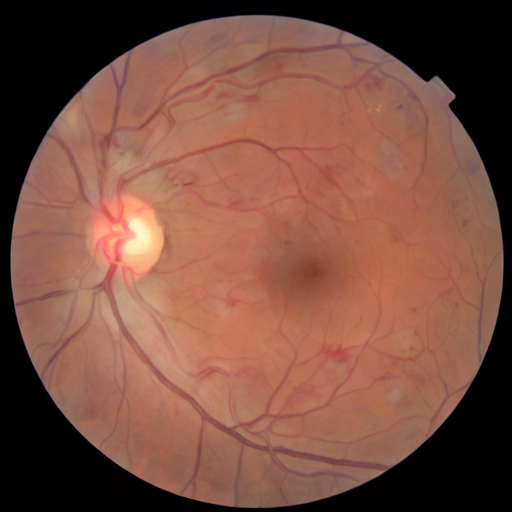}\\
        \end{minipage}
        \begin{minipage}{0.02\textwidth}
            \centering
            \scriptsize
            \includegraphics[width=1.0\textwidth]{figures/ar.png}\\
        \end{minipage}
        \begin{minipage}{0.2\textwidth}
            \centering
            \scriptsize
            \includegraphics[width=\textwidth]{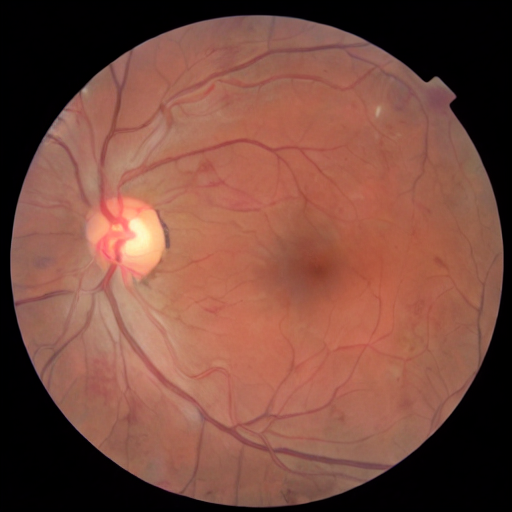}\\
        \end{minipage}
    \end{minipage}
    \\
    \begin{minipage}[t]{0.03\textwidth}
        \rotatebox{90}{\centering \!\!\! Skin}
    \end{minipage}
    \begin{minipage}[t]{0.75\textwidth}
        \begin{minipage}{0.2\textwidth}
            \centering
            \scriptsize
            \includegraphics[width=\textwidth]{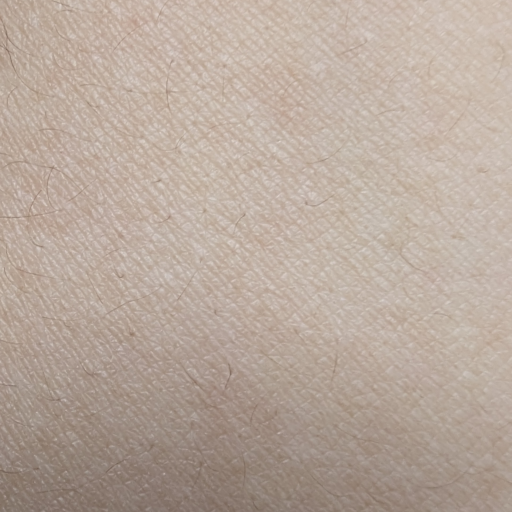}\\
            \textit{Input Image}
        \end{minipage}
        \begin{minipage}{0.02\textwidth}
            \centering
            \scriptsize
            \includegraphics[width=1.0\textwidth]{figures/ar.png}\\
        \end{minipage}
        \begin{minipage}{0.2\textwidth}
            \centering
            \scriptsize
            \includegraphics[width=\textwidth]{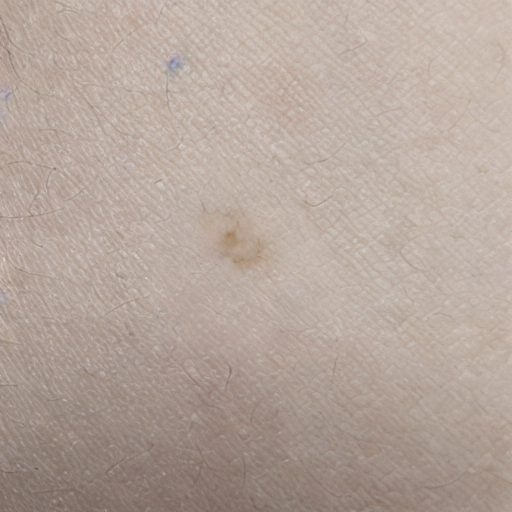}\\
            \textit{Step 1}
        \end{minipage}
        \begin{minipage}{0.02\textwidth}
            \centering
            \scriptsize
            \includegraphics[width=1.0\textwidth]{figures/ar.png}\\
        \end{minipage}
        \begin{minipage}{0.2\textwidth}
            \centering
            \scriptsize
            \includegraphics[width=\textwidth]{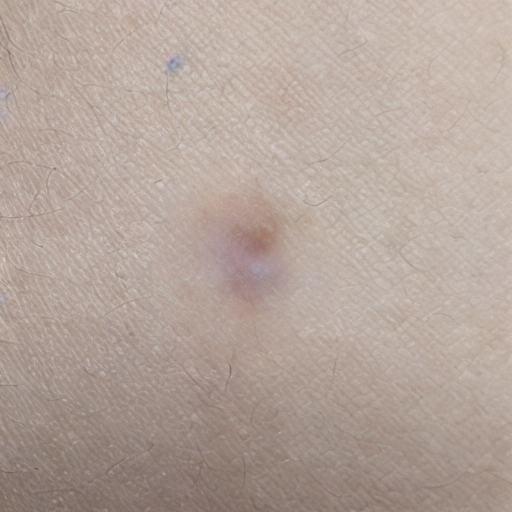}\\
            \textit{Step 4}
        \end{minipage}
        \begin{minipage}{0.02\textwidth}
            \centering
            \scriptsize
            \includegraphics[width=1.0\textwidth]{figures/ar.png}\\
        \end{minipage}
        \begin{minipage}{0.2\textwidth}
            \centering
            \scriptsize
            \includegraphics[width=\textwidth]{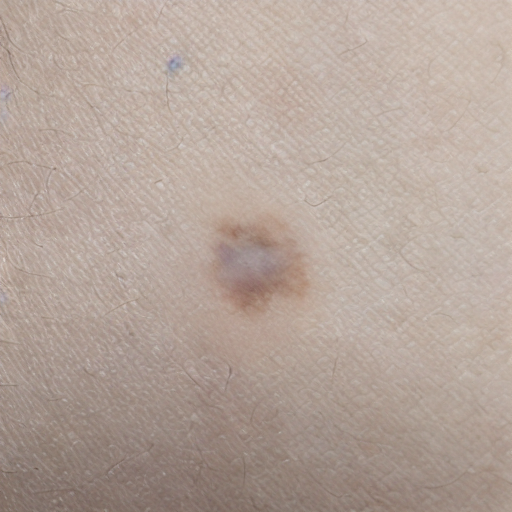}\\
            \textit{Step 10}
        \end{minipage}
    \end{minipage}
    \end{tabular}
    \caption{Disease Progression Simulation of PIE. The top progression is for Cardiomegarly. The middle progression is for Diabetic Retinopathy. The bottom progression is for Melanocytic Nevus.}
    \label{fig:progression_visualization}
\end{figure}

\textbf{Datasets for Disease Progression.}
We validate the disease progression analysis through end-to-end medical domain-specific image inference. Specifically, we evaluate the pretrained domain-specific stable diffusion model on three different types of disease datasets in classification tasks: CheXpert for chest X-ray classification ~\citep{irvin2019chexpert}, ISIC 2018 / HAM10000 ~\citep{codella2019skin,tschandl2018ham10000} for skin cancer prediction, and Kaggle Diabetic Retinopathy Detection Challenge ~\citep{CHF2015retino}. Each of these datasets presents unique challenges and differ in scale, making them suitable for testing the robustness and versatility of PIE. We also collected over 30 healthy data among the test set from these datasets. These data were used for disease progression simulation. Three groups of progression visualization results can be found in Figure~\ref{fig:progression_visualization}.

\begin{figure}[!htbp]
    \centering
    \begin{tabular}{c}
    \begin{minipage}[t]{0.85\textwidth}
        \centering
        \begin{minipage}{0.16\textwidth}
            \centering
            \scriptsize
            \includegraphics[width=\textwidth]{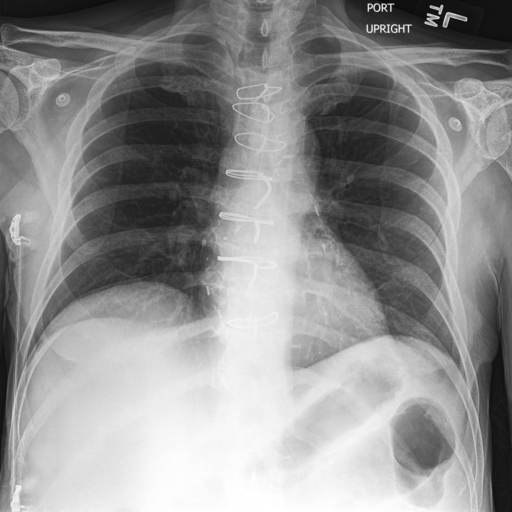}\\
            \textit{Input}
        \end{minipage}
        \fbox{
        \begin{minipage}{0.15\textwidth}
            \centering
            \scriptsize
            \includegraphics[width=\textwidth]{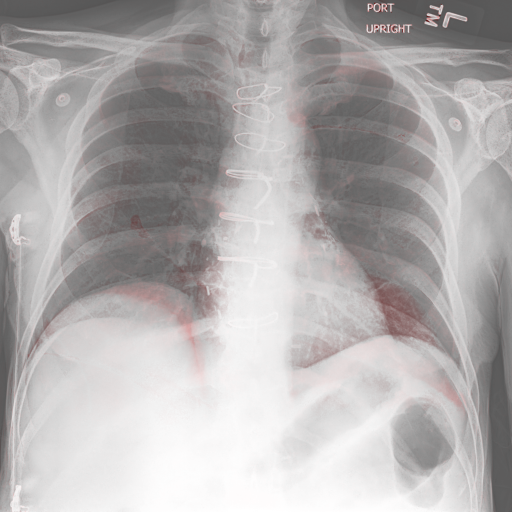}\\
            \textit{Absdiff step 1}
        \end{minipage}
        \begin{minipage}{0.15\textwidth}
            \centering
            \scriptsize
            \includegraphics[width=\textwidth]{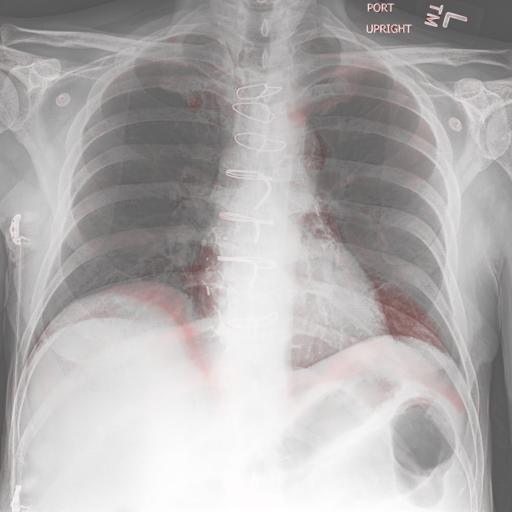}\\
            \textit{Absdiff step 2}
        \end{minipage}
        \begin{minipage}{0.15\textwidth}
            \centering
            \scriptsize
            \includegraphics[width=\textwidth]{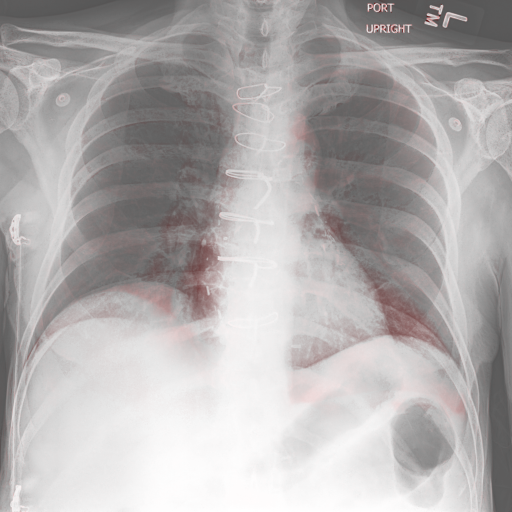}\\
            \textit{Absdiff step 4}
        \end{minipage}
        \begin{minipage}{0.15\textwidth}
            \centering
            \scriptsize
            \includegraphics[width=\textwidth]{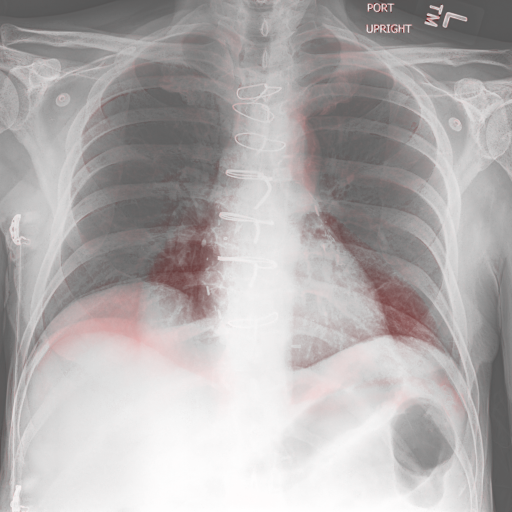}\\
            \textit{Absdiff step 10}
        \end{minipage}
        }
        \begin{minipage}{0.16\textwidth}
            \centering
            \scriptsize
            \includegraphics[width=\textwidth]{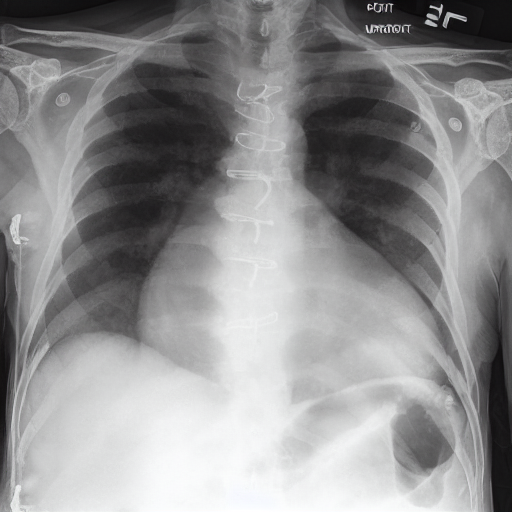}\\
            \textit{PIE Output}
        \end{minipage}
    \end{minipage}
    \end{tabular}
    \caption{Cardiomegarly disease progression absolute difference heatmap simulated by PIE. The highlighted \textcolor{red}{red} portion illustrates the progression of the pathology at each step.}
    \label{fig:progression_heatmap_visualization}
\end{figure}

\begin{figure}[htbp]
    \centering
    \begin{tabular}{c c}
    \begin{minipage}{0.15\textwidth}
        \centering
        \begin{minipage}{\textwidth}
            \centering
            \scriptsize
            \includegraphics[width=0.85\textwidth]{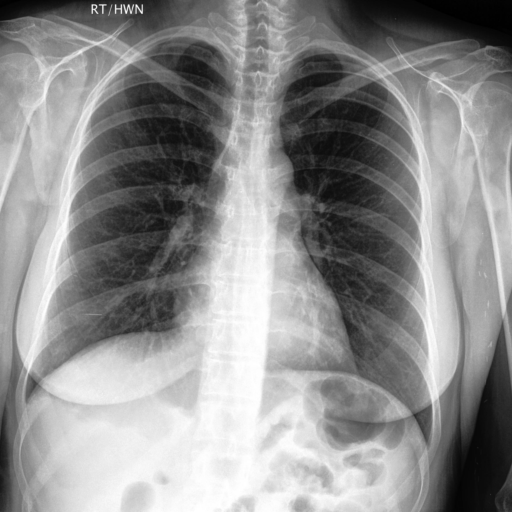}\\
            \textit{Input Image}
        \end{minipage}
        \\
        \centering
        \begin{minipage}{\textwidth}
            \centering
            \scriptsize
            \includegraphics[width=0.85\textwidth]{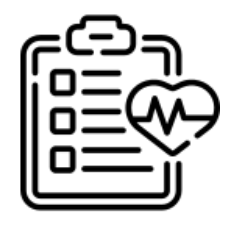}\\
            \textit{Input Prompt}
        \end{minipage}
    \end{minipage}
    
    \begin{minipage}{0.6\textwidth}
    \centering
    \begin{minipage}[t]{0.03\textwidth}
        \rotatebox{90}{\centering \!\!\!\! \tiny PIE}
    \end{minipage}
    \centering
    \begin{minipage}[t]{0.8\textwidth}
        \begin{minipage}{0.19\textwidth}
            \centering
            \scriptsize
            \includegraphics[width=\textwidth]{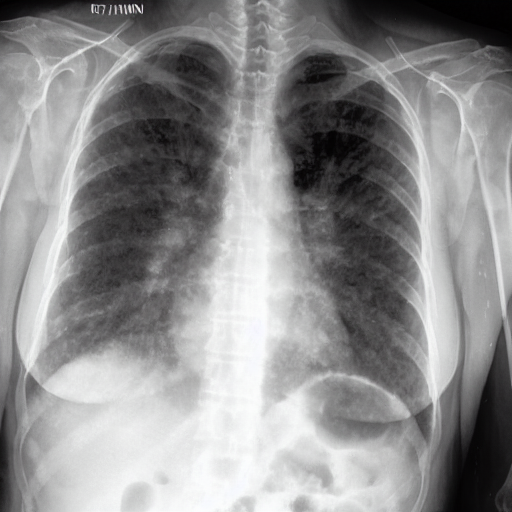}\\
        \end{minipage}
        \begin{minipage}{0.03\textwidth}
            \centering
            \scriptsize
            \includegraphics[width=1.0\textwidth]{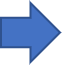}\\
        \end{minipage}
        \begin{minipage}{0.19\textwidth}
            \centering
            \scriptsize
            \includegraphics[width=\textwidth]{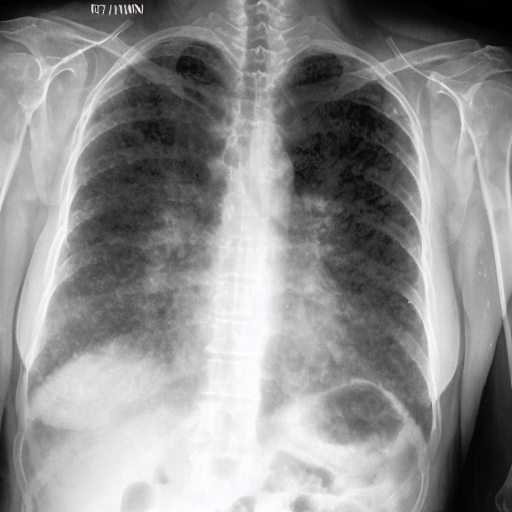}\\
        \end{minipage}
        \begin{minipage}{0.03\textwidth}
            \centering
            \scriptsize
            \includegraphics[width=1.0\textwidth]{figures/ar.png}\\
        \end{minipage}
        \begin{minipage}{0.19\textwidth}
            \centering
            \scriptsize
            \includegraphics[width=\textwidth]{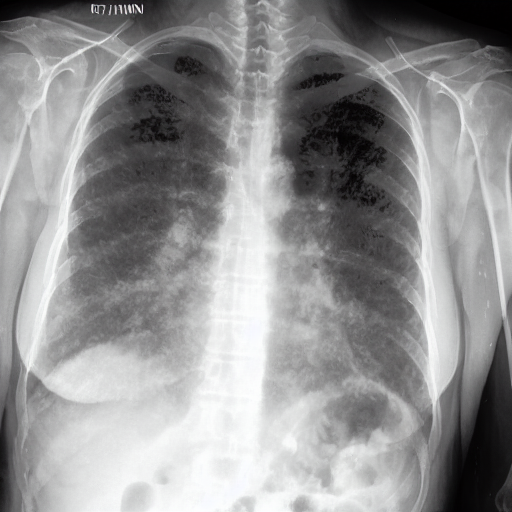}\\
        \end{minipage}
        \begin{minipage}{0.03\textwidth}
            \centering
            \scriptsize
            \includegraphics[width=1.0\textwidth]{figures/ar.png}\\
        \end{minipage}
        \begin{minipage}{0.19\textwidth}
            \centering
            \scriptsize
            \includegraphics[width=\textwidth]{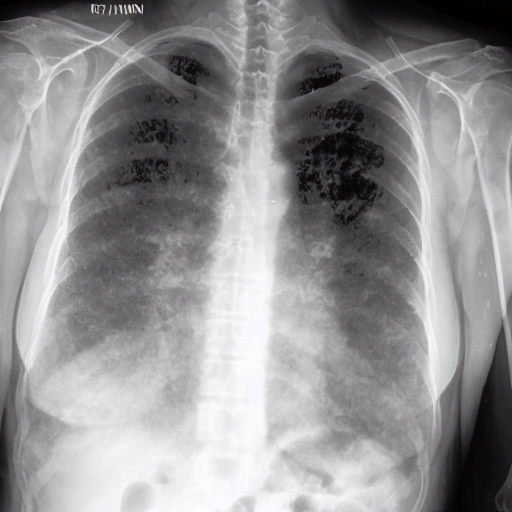}\\
        \end{minipage}
    \end{minipage}
    \\
    \centering
    \begin{minipage}[t]{0.03\textwidth}
        \rotatebox{90}{\centering  \!\!\!\!\!\!\!\! \tiny SD Video}
    \end{minipage}
    \centering
    \begin{minipage}[t]{0.8\textwidth}
        \begin{minipage}{0.19\textwidth}
            \centering
            \scriptsize
            \includegraphics[width=\textwidth]{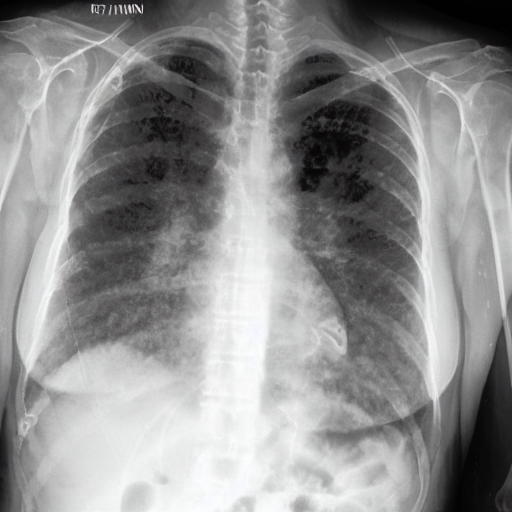}\\
        \end{minipage}
        \begin{minipage}{0.03\textwidth}
            \centering
            \scriptsize
            \includegraphics[width=1.0\textwidth]{figures/ar.png}\\
        \end{minipage}
        \begin{minipage}{0.19\textwidth}
            \centering
            \scriptsize
            \includegraphics[width=\textwidth]{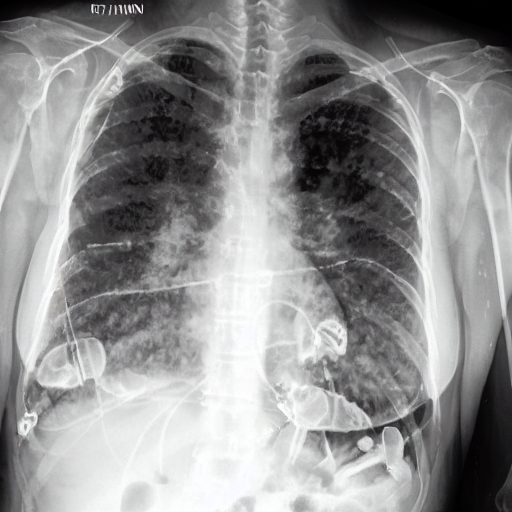}\\
        \end{minipage}
        \begin{minipage}{0.03\textwidth}
            \centering
            \scriptsize
            \includegraphics[width=1.0\textwidth]{figures/ar.png}\\
        \end{minipage}
        \begin{minipage}{0.19\textwidth}
            \centering
            \scriptsize
            \includegraphics[width=\textwidth]{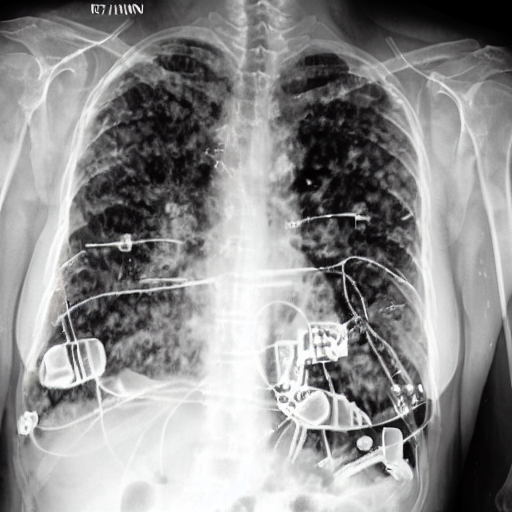}\\
        \end{minipage}
        \begin{minipage}{0.03\textwidth}
            \centering
            \scriptsize
            \includegraphics[width=1.0\textwidth]{figures/ar.png}\\
        \end{minipage}
        \begin{minipage}{0.19\textwidth}
            \centering
            \scriptsize
            \includegraphics[width=\textwidth]{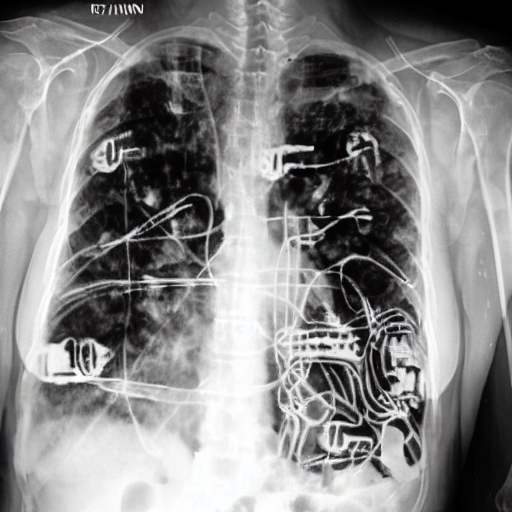}\\
        \end{minipage}
    \end{minipage}
    \\
    \centering
    \begin{minipage}[t]{0.03\textwidth}
        \rotatebox{90}{\centering \footnotesize \!\!\!\!\!\!\!\!\! \tiny Extrapolation}
    \end{minipage}
    \centering
    \begin{minipage}[t]{0.8\textwidth}
        \begin{minipage}{0.19\textwidth}
            \centering
            \scriptsize
            \includegraphics[width=\textwidth]{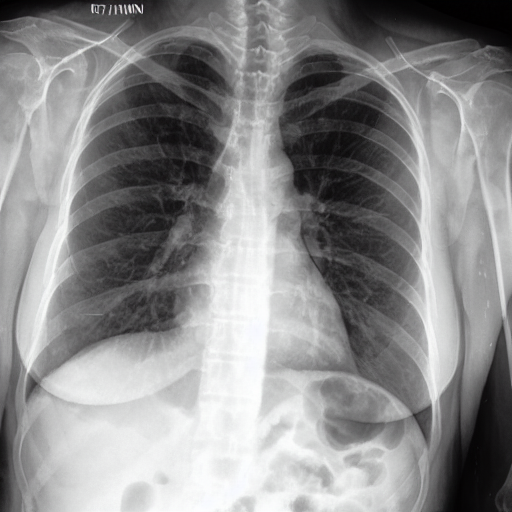}\\
            \textit{Step 1}
        \end{minipage}
        \begin{minipage}{0.03\textwidth}
            \centering
            \scriptsize
            \includegraphics[width=1.0\textwidth]{figures/ar.png}\\
        \end{minipage}
        \begin{minipage}{0.19\textwidth}
            \centering
            \scriptsize
            \includegraphics[width=\textwidth]{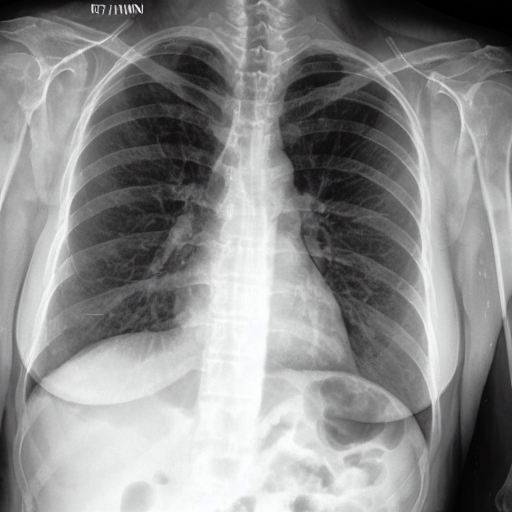}\\
            \textit{Step 2}
        \end{minipage}
        \begin{minipage}{0.03\textwidth}
            \centering
            \scriptsize
            \includegraphics[width=1.0\textwidth]{figures/ar.png}\\
        \end{minipage}
        \begin{minipage}{0.19\textwidth}
            \centering
            \scriptsize
            \includegraphics[width=\textwidth]{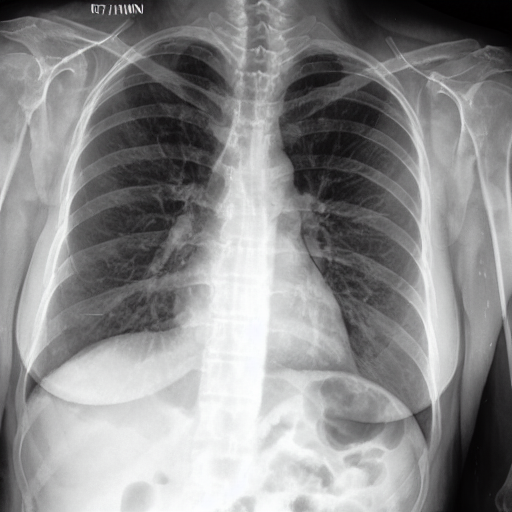}\\
            \textit{Step 4}
        \end{minipage}
        \begin{minipage}{0.03\textwidth}
            \centering
            \scriptsize
            \includegraphics[width=1.0\textwidth]{figures/ar.png}\\
        \end{minipage}
        \begin{minipage}{0.19\textwidth}
            \centering
            \scriptsize
            \includegraphics[width=\textwidth]{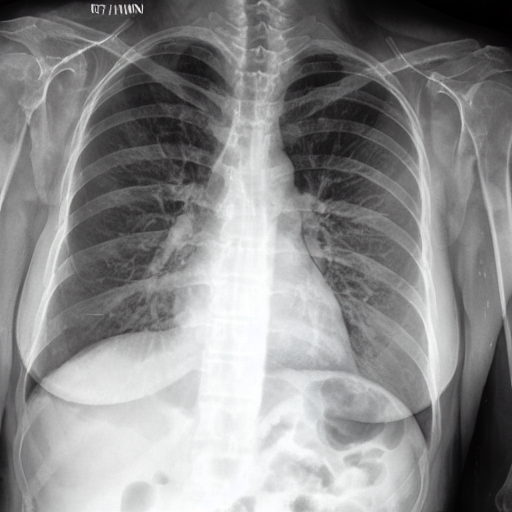}\\
            \textit{Step 10}
        \end{minipage}
    \end{minipage}
    \end{minipage}
    \end{tabular}
    \caption{Using PIE, SD Video, Extrapolation to simulate Edema progression with clinical reports as input prompt.}
\label{fig:progression_visualization_comp_edema}
\end{figure}

\textbf{Evaluation Metrics.}
The assessment of generated disease progression images relies on two crucial aspects: alignment to edited disease feature and subject fidelity. To measure these characteristics, we utilize two primary metrics: the CLIP-I score and the classification confidence score. The CLIP-I score (range from [-1, 1] in theory ) represents the average pairwise cosine similarity between the CLIP embeddings of generated and real images~\citep{radford2021learning,ruiz2022dreambooth}. The classification confidence score is determined using supervised train deep networks for binary classification between negative (healthy) and positive (disease) samples. It is denoted as $\textbf{Conf}=Sigmoid(f_\theta(x))$ and represent whether the simulation results are aligned to target disease. In our experiments, we train the DeepAUC maximization method~\citep{yuan2021large} (SOTA of Chexpert and ISIC 2018 task 3) using DenseNet121~\citep{huang2017densely} as the backbone to compute the classification confidence score. 

\textbf{Baselines}
To our knowledge, there are no existing image editing models specifically designed for simulating disease progression without sequential training data. To underscore the unique strengths of PIE, we compare it against two of the most promising state-of-the-art baseline methods. One of them is Stable Diffusion Video (SD Video)~\citep{nateraw2022} for short video generation. SD Video is the code implementation based on recent latent-based video generation methods~\citep{blattmann2023align,wu2022tune}. Another one is the Style-Based Manifold Extrapolation (Extrapolation)~\citep{han2022image} for generating progressive medical imaging, as it don't need diagnosis labelled data~\citep{ravi2019degenerative,han2022image}, which is similar to PIE's definition setting but need progression inference prior. During the comparison, all baseline methods are using the same Stable Diffusion finetuned weights and also applied $M_{ROI}$ for region guided. 

\subsection{Progression Simulation Comparison}

\begin{figure}[t]%
\centering
\begin{minipage}{\textwidth}
\centering
\begin{minipage}{\sizecci\textwidth}
    \centering
    \scriptsize
    \includegraphics[width=\sizeccj\linewidth]{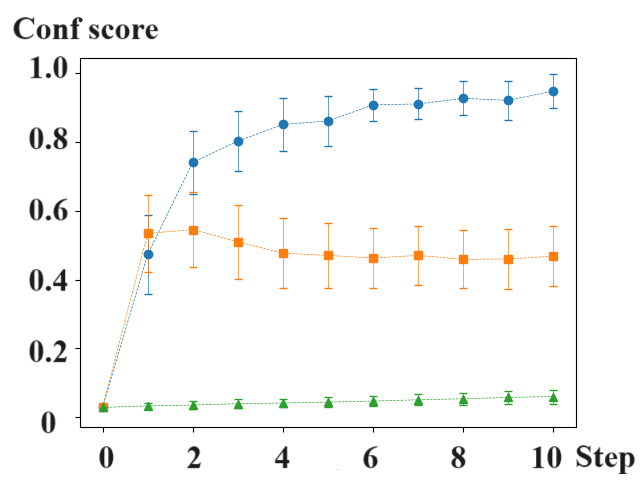}\\
    \textit{Cardiomegaly}
\end{minipage}
\begin{minipage}{\sizecci\textwidth}
    \centering
    \scriptsize
    \includegraphics[width=\sizeccj\linewidth]{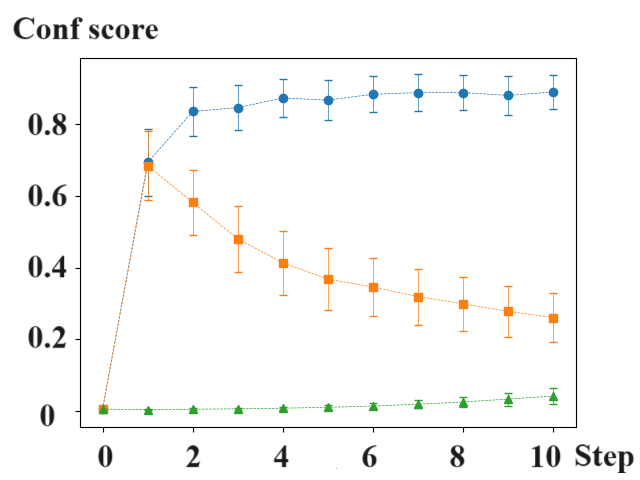}\\
    \textit{Edema}
\end{minipage}
\begin{minipage}{\sizecci\textwidth}
    \centering
    \scriptsize
    \includegraphics[width=\sizeccj\linewidth]{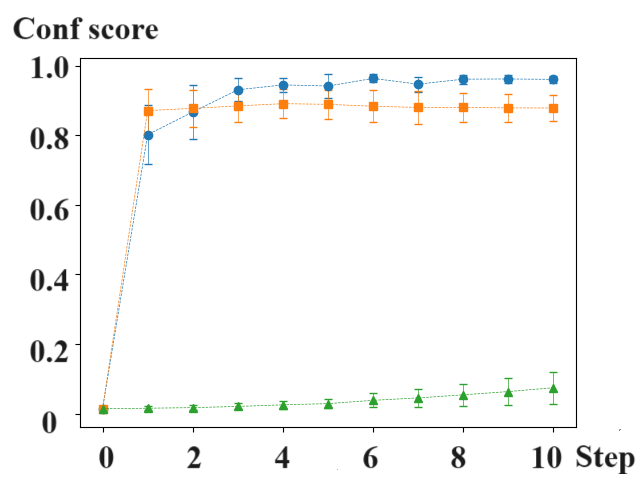}\\
    \textit{Pleural Effusion}
\end{minipage} 
\begin{minipage}{\sizecci\textwidth}
    \centering
    \scriptsize
    \includegraphics[width=\sizeccj\linewidth]{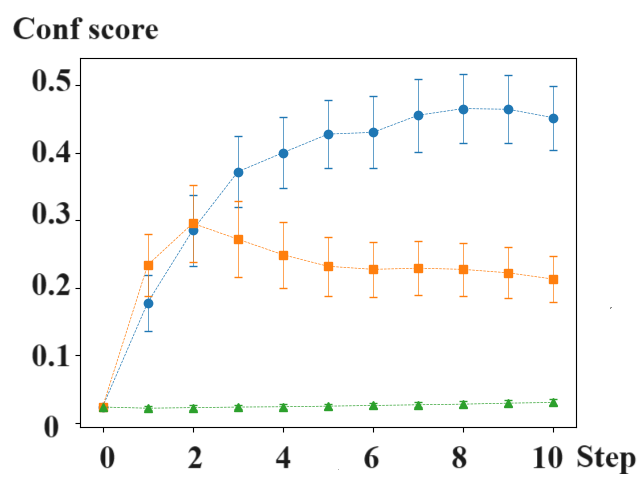}\\
    \textit{Consolidation}
\end{minipage}
\begin{minipage}{\sizecci\textwidth}
    \centering
    \scriptsize
    \includegraphics[width=\sizeccj\linewidth]{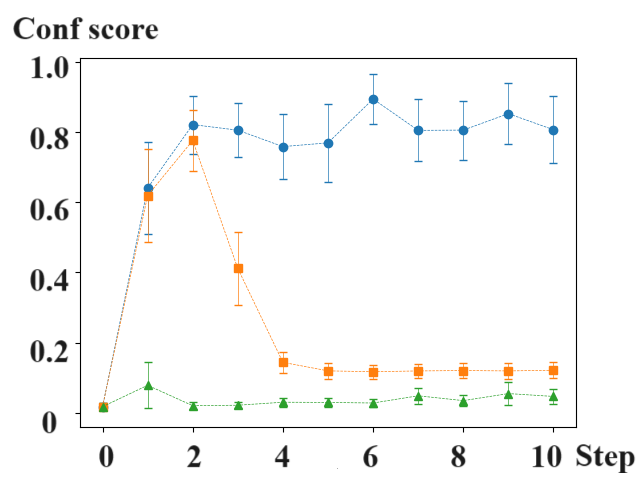}\\
    \textit{Diabetic Retinopathy}
\end{minipage}
\begin{minipage}{\sizecci\textwidth}
    \centering
    \scriptsize
    \includegraphics[width=\sizeccj\linewidth]{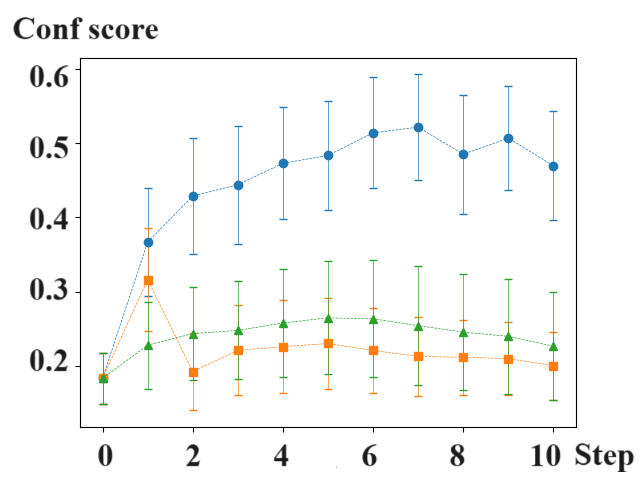}\\
    \textit{Melanocytic Nevi}
\end{minipage}
\begin{minipage}{\textwidth}
    \centering
    \scriptsize
    \includegraphics[width=0.70\linewidth]{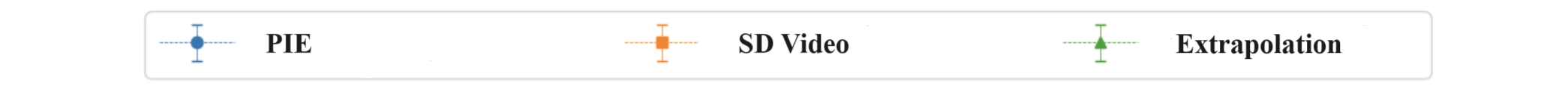}
\end{minipage}
\end{minipage} 
\caption{PIE excels in comparison to all the baseline methods across six different disease progression simulations. The inputs utilized are genuine healthy images from the test sets. For each image, we apply five random seeds to simulate disease progression over ten steps. The confidence score, a value that ranges from 0 to 1, signifies the classification confidence for a specific disease.}
\label{fig:comp_plots}
\end{figure}

In order to demonstrate the superior performance of PIE in disease progression simulation over other single-step editing methods, we perform experiments on three datasets previously mentioned. For each disease in these datasets, we used 10 healthy samples in the test set as simulation start point and run PIE, SD Video, Extrapolation with 5 random seeds. We obtain at least 50 disease imaging trajectories for each patient. Table~\ref{tab:performance} showcases that PIE consistently surpasses both SD Video and Extrapolation in terms of disease confidence scores while maintaining high CLIP-I scores. For Chexpert dataset, the 0.690 final confidence score is the average score among 5 classes. For Diabetic Retinopathy and ISIC 2018 datasets, we compare PIE with SD Video, Extrapolation for editing image to the most common seen class since these datasets are highly imbalanced. Figure~\ref{fig:comp_plots} illustrates the evolution of disease confidence scores during the progression simulation in each step. We observe that PIE is able to produce more faithful and realistic progressive editing compared to the other two baselines. Interestingly, while the CLIP-I score of Extrapolation is comparable to that of PIE, it fails to effectively edit the key disease features of the input images as its confidence scores are low throughout and at the end of the progression. We also visualize the absolute differences between initial stage and each progression stage of Cardiomegaly in Figure.~\ref{fig:progression_heatmap_visualization}.

Figure~\ref{fig:progression_visualization_comp_edema} showcases a group of progression simulation results for Edema in chest X-rays with CheXpert clinical report prompt. It is evident from our observations that while SD Video can significantly alter the input image in the initial step, it fails to identify the proper direction of progression in the manifold after a few steps and would easily create uncontrollable noise. Conversely, Extrapolation only brightens the Chest X-ray without making substantial modifications. PIE, on the other hand, not only convincingly simulates the disease trajectory but also manages to preserve unrelated visual features from the original medical imaging. Further visual comparisons among different datasets are presented in Supplementary~\ref{appendix:extra_ablation}.




\subsection{Ablation Study}

\textbf{Medical heuristic guidance.}
During the PIE simulation, the region guide masks play a big role as prior information. Unlike other randomly inpainting tasks~\citep{lugmayr2022repaint}, ROI mask for medical imaging can be extracted from real or synthetic clinical reports~\citep{boag2020baselines,lovelace2020learning} using domain-specific Segment Anything models~\citep{kirillov2023segment,ma2023segment}. It helps keep unrelated regions consistent through the progressive changes using PIE or baseline models. In order to generate sequential disease imaging data, PIE uses noise strength $\gamma$ to control the influence from the patient's clinically reported and expected treatment regimen at time $n$. $N$ is used to control the duration of the disease occurrence or treatment regimen. PIE allows the user to make such controls over the iterative process, and running $\textit{PIE}^{(n)}$ multiple times can improve the accuracy of disease imaging tracking and reduce the likelihood of missed or misinterpreted changes. Related ablation study results for $M_{ROI}$, $\gamma$, $N$, $\beta_1$, $\beta_2$ is available in Supplementary~\ref{appendix:extra_ablation}.

\textbf{Compare with real longitude medical imaging sequence.} Lack of longitudinal data is a common problem in current chest X-ray datasets. However, due to the spread of COVID, part of the latest released dataset contains limited longitudinal data. In order to validate the disease sequence modeling that PIE can match real disease trajectories, we conduct experiments on generating edema disease progression from 10 patients in BrixIA COVID-19 Dataset~\citep{signoroni2021bs}. The input image is the day 1 image, and we use PIE to generate future disease progression based on real clinical reports for edema.

\begin{figure}[htbp]
    \centering
    \begin{minipage}[t]{0.95\textwidth}
    \centering
    \begin{minipage}{0.18\textwidth}
        \centering
        \scriptsize
        \includegraphics[width=\textwidth]{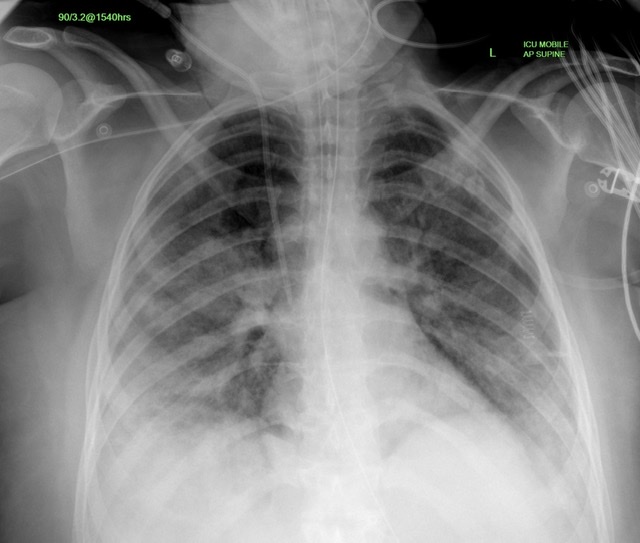}
        \textit{Input image}
    \end{minipage}
    \begin{minipage}{0.02\textwidth}
        \centering
        \scriptsize
        \includegraphics[width=1.0\textwidth]{figures/ar.png}
    \end{minipage}
    \fbox{
        \begin{minipage}{0.16\textwidth}
            \centering
            \scriptsize
            \includegraphics[width=\textwidth]{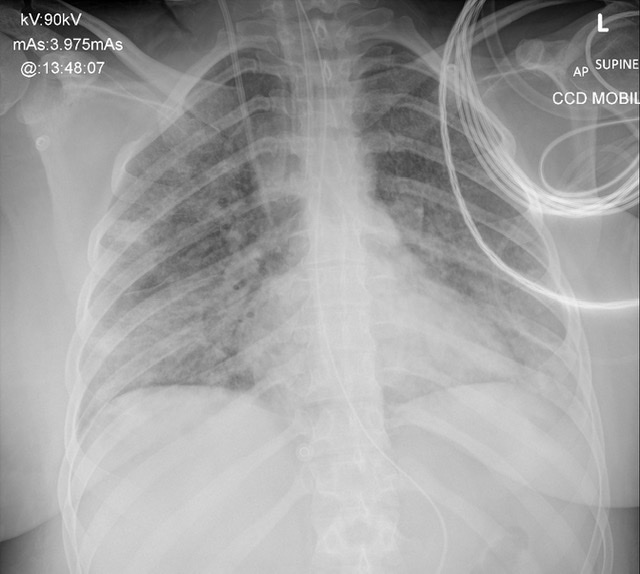}
            \textit{Real Day 7 Data}
        \end{minipage}
        \begin{minipage}{0.008\textwidth}
            \centering
            \scriptsize
            \includegraphics[width=1.0\textwidth]{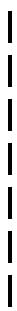}
        \end{minipage}
        \begin{minipage}{0.140\textwidth}
            \centering
            \scriptsize
            \includegraphics[width=\textwidth]{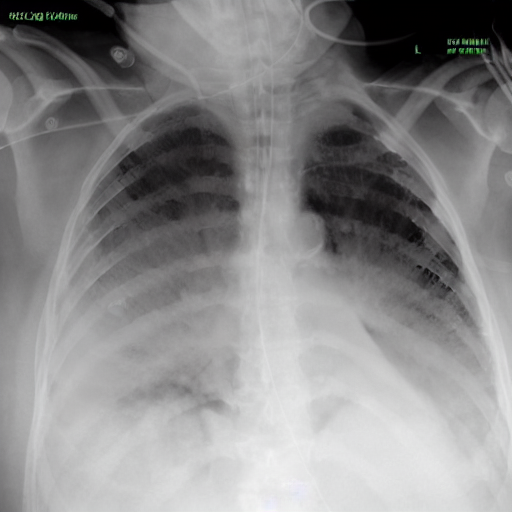}
            \textit{PIE Day 7 Data}
        \end{minipage}
    }
    \begin{minipage}{0.26\textwidth}
        \centering
        \scriptsize
        \includegraphics[width=\textwidth]{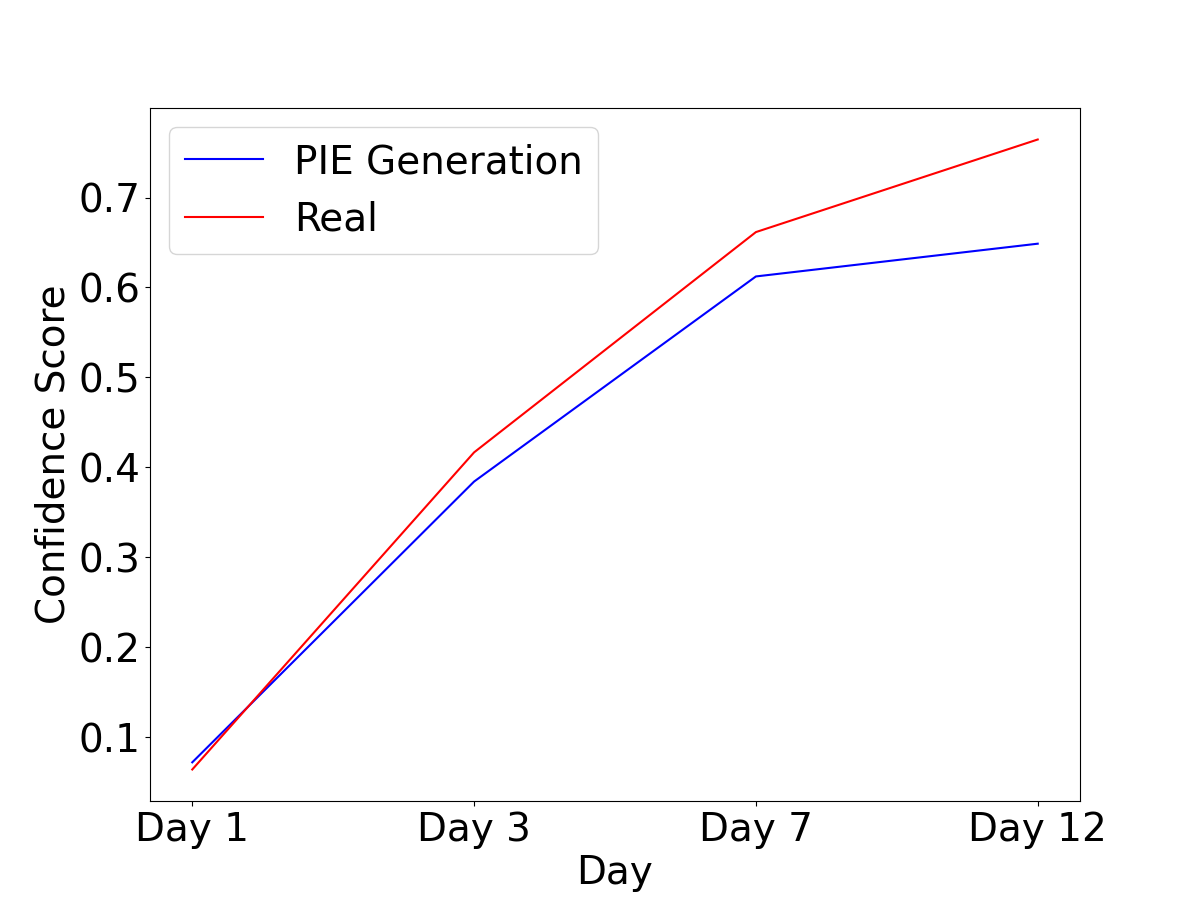}
        \textit{Real VS PIE Generation}
    \end{minipage}
    \end{minipage}
    \caption{Evaluating the confidence scores of PIE progression trajectories highlights the alignment with realistic progression. The mean absolute error between two trajectories is approximately $0.0658$.}
\label{fig:comparison_with_real}
\end{figure}

\begin{figure}[htbp]
    \centering
    \begin{minipage}[t]{0.95\textwidth}
    \centering
        
        \begin{minipage}{0.15\textwidth}
            \centering
            \scriptsize
            \includegraphics[width=\textwidth]{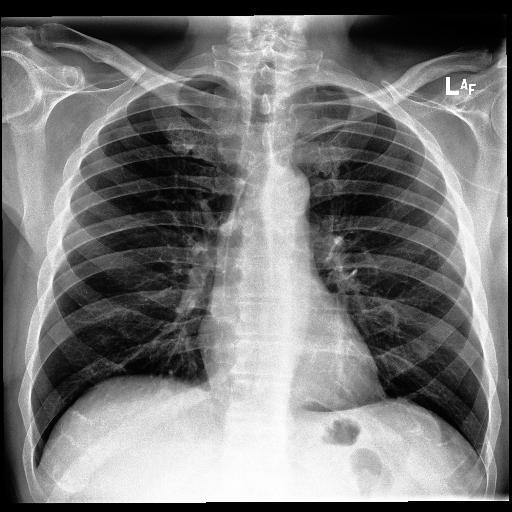}
            \textit{Input image}
        \end{minipage}
        \begin{minipage}{0.02\textwidth}
            \centering
            \scriptsize
            \includegraphics[width=1.0\textwidth]{figures/ar.png}
        \end{minipage}
        \begin{minipage}{0.15\textwidth}
            \centering
            \scriptsize
            \includegraphics[width=\textwidth]{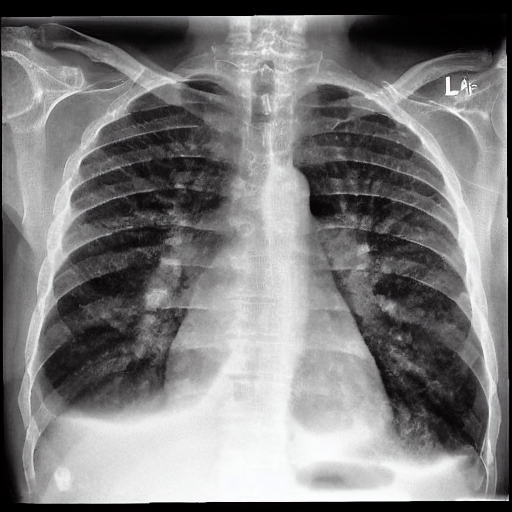}
            \textit{Step 4}
        \end{minipage}
        \begin{minipage}{0.02\textwidth}
            \centering
            \scriptsize
            \includegraphics[width=1.0\textwidth]{figures/ar.png}
        \end{minipage}
        \begin{minipage}{0.15\textwidth}
            \centering
            \scriptsize
            \includegraphics[width=\textwidth]{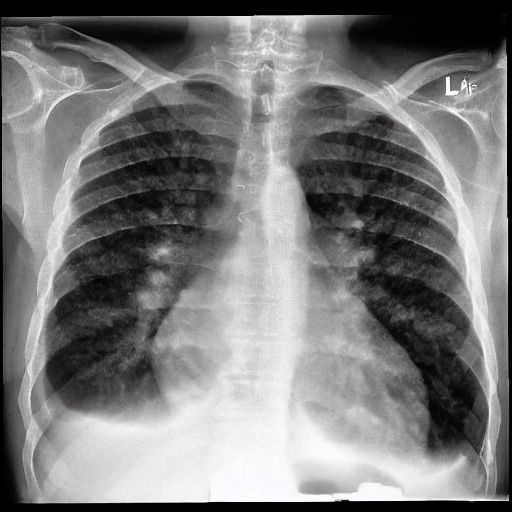}
            \textit{Step 10}
        \end{minipage}
        \begin{minipage}{0.25\textwidth}
            \centering
            \scriptsize
            \includegraphics[width=\textwidth]{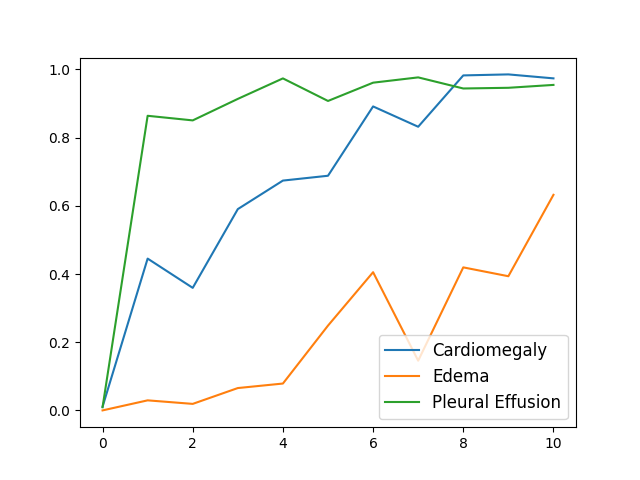}
            \textit{Confidence Score}
        \end{minipage}
    \end{minipage}
    \caption{PIE can successfully simulate co-occurring disease progression (Patent's clinical report shows high probability to be Cardiomegaly, Edema, Pleural Effusion at the same time).}
\label{fig:appendix:co-occurring}
\end{figure}

\textbf{Case study: co-occurring diseases.} PIE is capable of generating images for co-occurring diseases, although the performance slightly trails behind that of single disease generation. To evaluate this ability, we use 10 chest X-ray reports for co-occurring Cardiomegaly, Edema, and Pleural Effusion. 6 cases successfully obtained co-occurring diseases simulation sequence and agreed with experienced clinicians. Figure~\ref{fig:appendix:co-occurring} illustrates an example of disease progression simulation. After 10 steps, all diseases achieve a high confidence score, indicating successful simulation.

\subsection{User Study}
To further assess the quality of our generated images, we surveyed 35 physicians and radiologists with \emph{$14.4$} years of experience on average to answer a questionnaire on chest X-rays. 
The questionnaire includes disease classifications on the generated and real X-ray images and evaluations of the realism of generated disease progression sequences of  Cardiomegaly, Edema, and Pleural Effusion. More details of the questionnaire and the calculation of the statistics are presented in Supplementary \ref{appendix:extra_user_q}. 
The participating physicians have agreed with a probability of $\mathbf{76.2}\%$ that the simulated progressions on the targeted diseases fit their expectations.

\begin{wraptable}{R}{0.4\textwidth}
  \centering
  \caption{To quantitatively analyze the responses of experienced physicians, we consider each pathology class independent and calculate the precision, recall, and F1 score across all diseases and physicians.}
  \label{tab:user}
  \begin{tabular}{l|cc|c}
    \toprule
    \textbf{Data} & \textbf{Precision} & \textbf{Recall} & \textbf{F1} \\
    \midrule
    Real & 0.505 & 0.415 & 0.455 \\
    PIE & 0.468 & 0.662 & 0.549 \\
    \bottomrule
  \end{tabular}
\end{wraptable}

Table \ref{tab:user} provides an interesting insight into experienced physicians' performance in predicting the pathology on real and generated X-rays. Surprisingly, we find users' performance on generated X-rays is superior to their performance on real images, with substantially higher recall and F1. In addition, the statistical test suggests that the F1 scores of generated scans are significantly higher (p-value of $0.0038$) than the real scans.
One plausible explanation is  due to the nature of PIE, the result of running progressive image editing makes pathological features more evident. The aggregated results from the user study demonstrate our framework's ability to simulate disease progression to meet real-world standards.

%% file: tex_files/conclusion.tex
\section{Conclusion}
In conclusion, our proposed framework, Progressive Image Editing (PIE), holds great potential as a tool for medical research and clinical practice in simulating disease progression. By leveraging recent advancements in text-to-image generative models, PIE achieves high fidelity and personalized disease progression simulations. The theoretical analysis shows that the iterative refining process is equivalent to gradient descent with an exponentially decayed learning rate, and practical experiments on three medical imaging datasets demonstrate that PIE surpasses baseline methods, in several quantitative metrics. Furthermore, a user study conducted with veteran physicians confirms that the simulated disease progressions generated by PIE meet real-world standards. Despite current limitations due to the lack of large amounts of longitude data and detailed medical reports, our framework has vast potential in modeling disease trajectories over time, restoring missing data from previous records, predicting future treatment responses, and improving clinical education. Moving forward, we aim to incorporate more data with richer descriptions and different monitoring modalities, such as chemical biomarkers and physiological recordings, into fine-tuning generative models, enabling our framework to more precise control over disease simulation through text conditioning. 


%% file: tex_files/supplementary.tex
\clearpage

\appendix 

\tableofcontents

\section{Background}
\textbf{Denoising Diffusion Probabilistic Models} (DDPM) ~\citep{ho2020denoising} are a class of generative models that use a diffusion process to transform a simple initial distribution, such as a Gaussian, into the target data distribution. The model assumes that the data points are generated by iteratively applying a diffusion process to a set of latent variables $x_1, \ldots, x_T$ in a sample space $\chi$. At each time step $t$, Gaussian noise is added to the latent variables $x_t$, and the variables are then transformed back to the original space using a learned invertible transformation function. This process is repeated for a fixed number of steps to generate a final output. The latent variable models can be expressed in the following form,  
\begin{equation}
    p_\theta(x_0) = \int p_{\theta}(x_{0:T})dx_{1:T}, \ \ \ \ where\ \ \ p_{\theta}(x_{0:T})\prod_{t = 1}^{T}p_{\theta}^{(t)}(x_{t - 1} | x_t)
    \label{eq:background:1}
\end{equation}
Because of a special property of the forward process,
\begin{equation}
    q(x_t|x_0) = \int q(x_{1:t} | x_0) dx_{1:(t-1)} = \mathcal{N}(x_t; \sqrt{\alpha_t} x_0, (1 - \alpha_t)\cdot \mathbf{I})
    \label{eq:background:2}
\end{equation}
we can express $x_t$ as a linear combination of $x_0$ and a noise variable $\epsilon$, which is the key to enabling the image editing process.
\begin{equation}
    x_t = \sqrt{\alpha_{t}}\cdot x_{0} + \sqrt{1 - \alpha_{t}} \cdot \epsilon, \ \ \ \ where \ \ \ \ \epsilon \sim \mathcal{N}(\mathbf{0}, \mathbf{I})
    \label{eq:background:3}
\end{equation}

\textbf{Denoising Diffusion Implicit Models} (DDIM) ~\citep{song2020denoising} that uses a non-Markovian forward process to generate data. Unlike Denoising Diffusion Probabilistic Models (DDPM), DDIM does not require explicit modeling of the latent variables. Instead, the model generates samples by solving a non-linear differential equation, which defines a continuous-time evolution of the data distribution. We can express its forward process as follows,
\begin{equation}
    q_{\sigma} (x_{1:T} | x_0):= q_{\sigma}(x_T | x_0) \prod_{t = 2}^T q_{\sigma}(x_{t - 1}| x_t, x_0)
    \label{eq:background:4}
\end{equation}
where $q_{\sigma} (x_{t-1}| x_t, x_0) = \mathcal{N}(\sqrt{\alpha_T}x_0, (1 - \alpha_T), \mathbf{I})$ and for all $t > 1$
\begin{equation}
    q_{\sigma}(x_{t - 1} | x_t, x_0) = \mathcal{N} (\sqrt{\alpha_{t - 1} x_0} + \sqrt{1 - \alpha_{t - 1} - \sigma_t^2}\cdot \frac{x_t - \sqrt{\alpha_t}x_0}{\sqrt{1 - \alpha_t}}, \sigma_t^2 \mathbf{I})
    \label{eq:background:5}
\end{equation}
Setting $\sigma_t = 0$, it defines a generation process going from $x_{t}$ to $x_{t - 1}$ as follows
\begin{equation}
    x_{t - 1} = \sqrt{\alpha_{t - 1}} (\frac{x_t - \sqrt{1 - \alpha_t} \epsilon_{\theta}^{(t)}(x_t)}{\sqrt{\alpha_t}}) + \sqrt{1 - \alpha_{t - 1}}\cdot \epsilon^{(t)}_{\theta}(x_t)
    \label{eq:background:6}
\end{equation}
where the $\epsilon_{\theta}^{(t)}(x_t)$ is a model that attempts to predict $\epsilon_t \sim \mathcal{N}(\mathbf{0}, \mathbf{I})$ from $x_t$

\section{Theoretical Analysis} \label{appendix:proof:theorem1}
\subsection{Proof of Proposition 1}
In this proof, we follow the conventions and definitions in ~\cite{song2020denoising}
\begin{equation}
    x_{t-1} = \sqrt{\alpha_{t-1}}(\frac{x_{t} - \sqrt{1 - \alpha_t}\epsilon_{\theta}^{(t)}(x_t, y)}{\sqrt{\alpha_t}}) + \sqrt{1 - \alpha_{t - 1}}\cdot \epsilon_{\theta}^{(t)}(x_t, y)
    \label{eq:proof1:1}
\end{equation}

Now given a base image denoted as $\mathbf{x^{(0)}_{0}}$, we wish to perform diffusion-based editing recursively for $N$ times. The roll-back (to the $k$ th-steps, where $k \geq 1$) according to (\ref{eq:background:3}):

\begin{equation}
    x^{(n)}_{k} = \sqrt{\alpha_{k}}\cdot x^{(n - 1)}_{0} + \sqrt{1 - \alpha_{k}} \cdot \epsilon
    \label{eq:proof1:2}
\end{equation}
where $\epsilon \sim \mathcal{N}(0, \mathcal{I})$.
Plugging (\ref{eq:proof1:2}) into (\ref{eq:proof1:1}),
\begin{equation}
    \label{eq:proof1:3}
    \begin{split}
      x_{k - 1}^{(n)} & = \sqrt{\alpha_{k-1}}(\frac{x_{k}^{(n)} - \sqrt{1 - \alpha_k}\epsilon_{\theta}^{(k)}(x_k^{(n)}, y)}{\sqrt{\alpha_k}}) + \sqrt{1 - \alpha_{k - 1}}\cdot \epsilon_{\theta}^{(k)}(x_k^{(n)}, y)\\
        & = \sqrt{\alpha_{k-1}}\cdot x_{0}^{(n - 1)} + \sqrt{\frac{\alpha_{k - 1}(1 - \alpha_k)}{\alpha_k}}\cdot (\epsilon - \epsilon_{\theta}^{(k)}(x_k^{(n)}, y)) + \sqrt{1 - \alpha_{k - 1}}\cdot \epsilon_{\theta}^{(k)}(x_k^{(n)}, y)
    \end{split}
\end{equation}
Setting $k = 1$ to perform the revert diffusion process
\begin{equation}
\label{eq:proof1:4}
\begin{split}
    x_{0}^{(n)} & = \sqrt{\alpha_{0}}\cdot x_{0}^{(n - 1)} + \sqrt{\frac{\alpha_{0}(1 - \alpha_1)}{\alpha_1}}\cdot (\epsilon - \epsilon_{\theta}^{(1)}(x_1^{(n)}, y)) + \sqrt{1 - \alpha_{0}}\cdot \epsilon_{\theta}^{(1)}(x_1^{(n)}, y)
\end{split}    
\end{equation}

Unrolling this recursion in (\ref{eq:proof1:4}), we have
\begin{equation}
    \label{eq:proof1:5}
    \begin{gathered}
    \begin{split}
        x_{0}^{(N)} = (\sqrt{\alpha_0})^N \cdot x_0^{(0)} + \sqrt{\frac{\alpha_0(1 - \alpha_1)}{\alpha_1}}\cdot \sum_i^N (\sqrt{\alpha_0})^{i}\cdot \epsilon \\ + \frac{\sqrt{\alpha_1 - \alpha_0\alpha_1} - \sqrt{\alpha_0 - \alpha_0\alpha_1}}{\sqrt{\alpha_{1}}} \cdot \sum_i^N (\sqrt{\alpha_0})^{i}\cdot \epsilon^{(i)}_{\theta}(x_1^{{(i)}}, y)
    \end{split}
    \end{gathered}
\end{equation}

Typically, $\alpha_0$ is set to a number close to but less than 1, where in the case of stable diffusion $0.9999$, $\alpha_1$ is $0.9995$, assuming $50$ step schedule. 

The second term in (\ref{eq:proof1:5}) is a sampling from Gaussian distributions with geometrically decreasing variances.

\begin{equation}
    \label{eq:proof1:6}
    \begin{split}
    \lim_{N \rightarrow \infty}\sum_i^N (\sqrt{\alpha_0})^{i}\cdot \epsilon = 0
    \end{split}
\end{equation}
Given a large enough N,
\begin{equation}
    \label{eq:proof1:7}
    \begin{split}
        x_{0}^{(N)} \approx (\sqrt{\alpha_0})^N \cdot x_0^{(0)} + \frac{\sqrt{\alpha_1 - \alpha_0\alpha_1} - \sqrt{\alpha_0 - \alpha_0\alpha_1}}{\sqrt{\alpha_{1}}} \cdot \sum_i^N (\sqrt{\alpha_0})^{i}\cdot \epsilon^{(i)}_{\theta}(x_1^{{(i)}}, y)
    \end{split}
\end{equation}
 Proposition 1 in   ~\cite{song2020denoising} declares "optimal $\epsilon^{(i)}_{\theta}$ has an equivalent probability flow ODE corresponding to the “Variance-Exploding” SDE in \cite{song2020score}". Hence, $\epsilon^{(i)}_{\theta}(x_1^{{(i)}}, y) := \nabla_{x} \log p(x | y)$. This can be seen as gradient descent with a geometrically decaying learning rate with a factor of $\sqrt{\alpha_0}$, with "base learning rate" $\frac{\sqrt{\alpha_1 - \alpha_0\alpha_1} - \sqrt{\alpha_0 - \alpha_0\alpha_1}}{\sqrt{\alpha_{1}}}$.

Notice that in (\ref{eq:proof1:7}), there is a decaying factor on the initial image $x_0^{(0)}$, as N grows the original image will surely diminish. Therefore, some other empirical measures are required to preserve the structure of the image, such as segmentation masking,  edit strength scheduling and etc.

\subsection{Additional Theoretical Analysis}

\subsubsection{Proof of Proposition 2}

\textbf{Proof}\:
    Continuing on \ref{eq:proof1:7}, for any $n > 1$, we have
\begin{equation}
    \begin{split}
        \|x_{0}^{n} - x_{0}^{n - 1}\| =
        \|(\sqrt{\alpha_0})^{n}\cdot[(1 - \frac{1}{\sqrt{\alpha_0}})\cdot x_0^{(0)} - \lambda \cdot \epsilon_{\theta}^{(n)}(x_{1}^{(n)}, y)]\|\\
        \leq (\sqrt{\alpha_0})^{n}\cdot[\|(1 - \frac{1}{\sqrt{\alpha_0}})\cdot x_0^{(0)}\| + \lambda \cdot \| \epsilon_{\theta}^{(n)}(x_{1}^{(n)}, y)\|]\\
        \leq (\sqrt{\alpha_0})^{n} [(\frac{1}{\sqrt{\alpha_0}} - 1)\cdot C_1 + \lambda \cdot C_2]
    \end{split}
    \label{eq:step_diff}
\end{equation}
to guarantee $\|x_{0}^{n} - x_{0}^{n - 1}\|\leq \delta$, we just need to set,
\begin{equation}
    \begin{split}
        (\sqrt{\alpha_0})^{n} [(\frac{1}{\sqrt{\alpha_0}} - 1)\cdot C_1 + \lambda \cdot C_2] < \delta \\
        \frac{n}{2}\log(\alpha_0) + \log((\frac{1}{\sqrt{\alpha_0}} - 1)\cdot C_1 + \lambda \cdot C_2) < log(\delta)
    \end{split}
\end{equation}
hence, given that $\alpha_0 < 1$ and $\log(\alpha_0) < 0$
\begin{equation}
    \begin{split}
        n > \frac{2}{\log(\alpha_0)}\cdot (log(\delta) - \log((\frac{1}{\sqrt{\alpha_0}} - 1)\cdot C_1 + \lambda \cdot C_2))
    \end{split}
\end{equation}

Let $C = \log((\frac{1}{\sqrt{\alpha_0}} - 1)\cdot C_1 + \lambda \cdot C_2)$, we have
\begin{equation}
    n > \frac{2}{\log(\alpha_0)}\cdot (log(\delta) - C)
\end{equation}
From above, we can conclude that as n grows bigger the changes between steps would grow smaller. The difference between steps will get arbitrarily small.

\subsection{Proof of Proposition 3}

\textbf{Proof}\:
From \ref{eq:step_diff} and applying triangle inequality, we observe that the difference is a sum of a geometric sequence scaled by a constant factor,
\begin{equation}
\begin{split}
    \|x_0^{(N)} - x_0^{(0)}\|\leq \sum_{n = 1}^{N}\|x_0^{(n)} - x_0^{(n -1)}\|\\
    \leq \sum_{n = 1}^{N} (\sqrt{\alpha_0})^{n} [(\frac{1}{\sqrt{\alpha_0}} - 1)\cdot C_1 + \lambda \cdot C_2]\\
    = \frac{1 - (\sqrt{\alpha_0})^N}{1 - \sqrt{\alpha_0}} \cdot [(\frac{1}{\sqrt{\alpha_0}} - 1)\cdot C_1 + \lambda \cdot C_2]
\end{split}
\end{equation}
As $N$ goes to infinity,
\begin{equation}
\begin{split}
    \lim_{N\rightarrow \infty} \|x_0^{(N)} - x_0^{(0)}\| \leq \frac{1}{1 - \sqrt{\alpha_0}} \cdot [(\frac{1}{\sqrt{\alpha_0}} - 1)\cdot C_1 + \lambda \cdot C_2] = \kappa
\end{split}
\end{equation}

\section{Baselines} \label{appendix:baseline}

\begin{figure}[t]%
\centering
\begin{minipage}{\textwidth}
    \centering
    \begin{minipage}{\sizecci\textwidth}
        \centering
        \scriptsize
        \includegraphics[width=\sizeccj\linewidth]{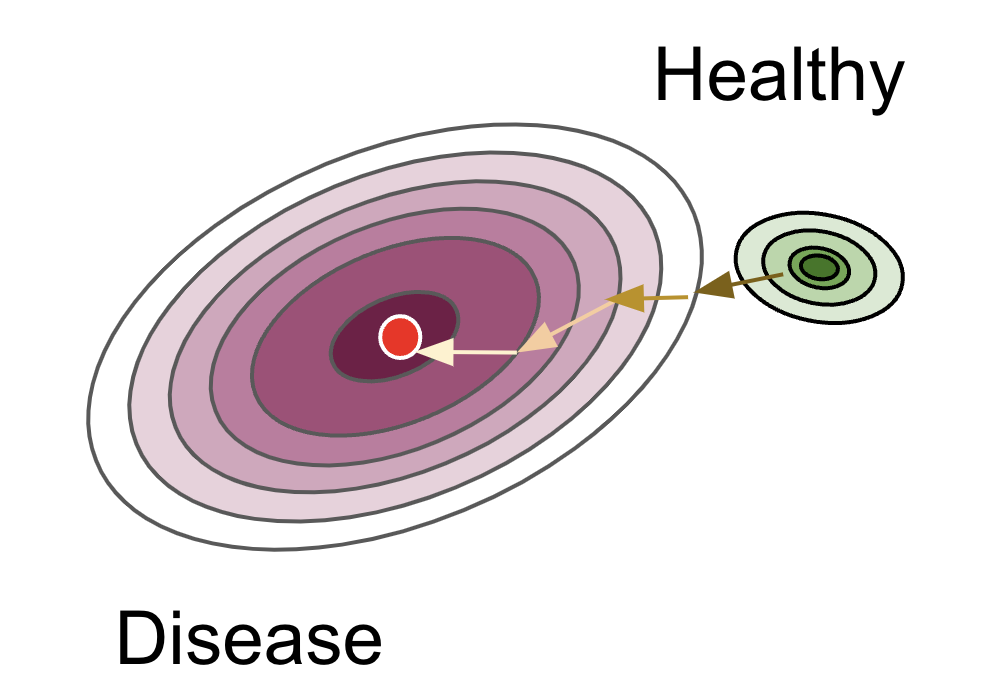}
        \textit{PIE}
    \end{minipage}
    \begin{minipage}{\sizecci\textwidth}
        \centering
        \scriptsize
        \includegraphics[width=\sizeccj\linewidth]{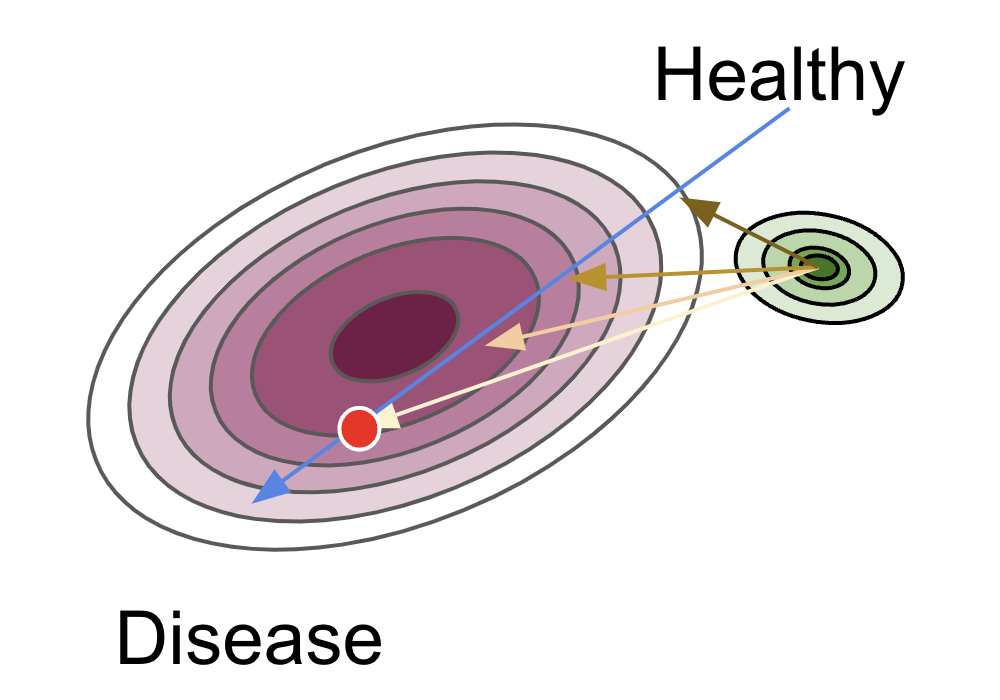}
        \textit{SD Video}
    \end{minipage}
    \begin{minipage}{\sizecci\textwidth}
        \centering
        \scriptsize
        \includegraphics[width=\sizeccj\linewidth]{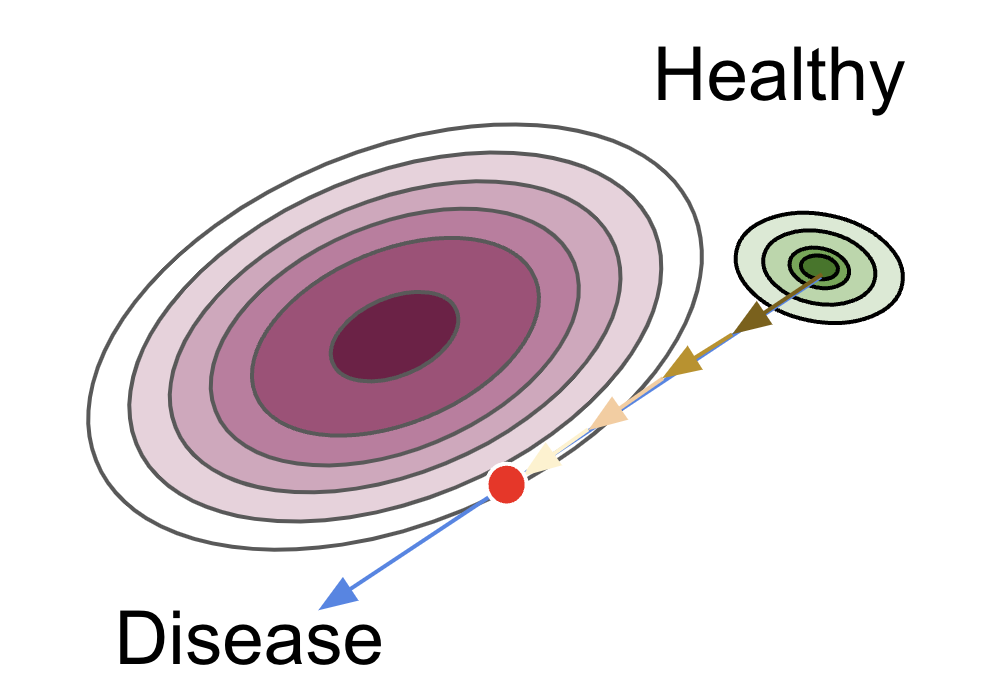}
        \textit{Extrapolation}
    \end{minipage} 
\end{minipage} 
\caption{Editing path of PIE, SD Video, and Extrapolation. }
\label{fig:appendix:baseline}
\end{figure}

Stable Diffusion Video (SD Video) and Style-Based Manifold Extrapolation (Extrapolation) are two leading techniques in the field of progressive video generation / medical image editing, each displaying promising results within specific domains. However, their applicability remains confined to these particular domains and poses a challenge in extending to the broader scope of different medical imaging data. To illustrate this, Figure~\ref{fig:appendix:baseline} provides a comparative visualization of single-step editing using these three techniques.

\textbf{Stable Diffusion Video Implementation} (SD Video)~\citep{nateraw2022} control the multi-step denoising process in the Stable Diffusion Videos pipeline. By smoothly and randomly traversing through the sampled latent space, SD Video demonstrates its capability to generate a series of images that progressively align with a given text prompt (see Figure~\ref{appendix:baseline}). 

As the state-of-the-art publicly available pipeline, SD Video can generate sequential imaging data by interpolating the latent space via multi-step Stable Diffusion. Though SD Video is useful for general domain~\citep{nateraw2022}, it is not controllable for medical prompts.

\textbf{Style-Based Manifold Extrapolation Implementation} (Extrapolation) ~\citep{han2022image}, involves iteratively modifying images by extrapolating between two latent manifolds. To determine the directions of latent extrapolation, the nearest neighbors algorithm is employed on distributions of known trajectories. However, in cases where progression data is not readily available, as in the study at hand, the directions are obtained by randomly sampling and computing the mean of each manifold.

The actual interpolation of Extrapolation for each step can be defined as: 

\begin{equation}
    \Delta = \frac{1}{m} \sum_{m}^{i=1} \frac{\Delta t^i}{\Delta T} [ G^{-1}(x_{(0)}^{(n+1)}) - G^{-1}(x_{(0)}^{(n)})  ]
    \label{eq:baseline}
\end{equation}

where $G^{-1}(x_{(0)}^{(n)})$ is the corresponding latent vector of the image at stage $n$.

\section{Experimental Reproducibility} \label{appendix:experiment}

In the following sections, we show the experimental setup used for finetuning the Stable Diffusion model with medical domain-specific medical data. We also outline the design of our disease progression simulation experiment across three datasets, as well as provide an evaluation of the time costs.

\subsection{Implementation Details}

Both \emph{PIE} and the baselines use publicly available Stable Diffusion checkpoints (CompVis/stable-diffusion-v1-4) that we further fine-tune on the training sets of each of the target datasets. Our code and checkpoints will be publicly available upon publication.

\textbf{Stable Diffusion Training.} 
To fine-tune the Stable Diffusion model, we center-crop and resize the input images to $512\times 512$ resolution. We utilize the AdamW optimizer~\citep{loshchilov2017decoupled} with a weight decay set at 0.01. Additionally, we employ a cosine learning rate scheduler ~\citep{loshchilov2016sgdr}, with the base learning rate set at $5\times 10^{-5}$. All models undergo fine-tuning for 20,000 steps on eight NVIDIA A100 GPUs, with each GPU handling a batch size of 8. 

\textbf{DeepAUC DenseNet121 Training.} 
We have previously outlined the concept of a classification confidence score. In order to pre-train the DeepAUC DenseNet121 model, we utilize the original code repository provided by the authors. The model training process involves the use of an exponential learning rate scheduler, with the base learning rate set to $1\times 10^{-4}$. Initially, the model is pre-trained on a multi-class task for each respective dataset. Subsequently, we employ the AUC loss proposed in the study by ~\citep{yuan2021large} to finetune the binary classification task, distinguishing between negative (healthy) and positive (disease) samples. The AUC loss finetuning process involves the use of an exponential learning rate scheduler, with the base learning rate set to $1\times 10^{-2}$. All models used for calculating classification confidence score are finetuned for 10 epochs on one NVIDIA A100 GPU, with a batch size of 128. For each class, the final classification accuracy on the validation set is 95.0 \% using the finetuned DenseNet121 model.

\textbf{PIE progression simulation.}
The PIE progression simulation is tested with one NVIDIA A100 GPU. In the following sections, we will show the hyper-parameter search experiment for PIE and explain the insight to adjust them.

\begin{figure}[htbp]
    \centering
    \begin{minipage}[t]{0.27\textwidth}
        \begin{minipage}{0.95\textwidth}
            \centering
            \scriptsize
            \includegraphics[width=\textwidth]{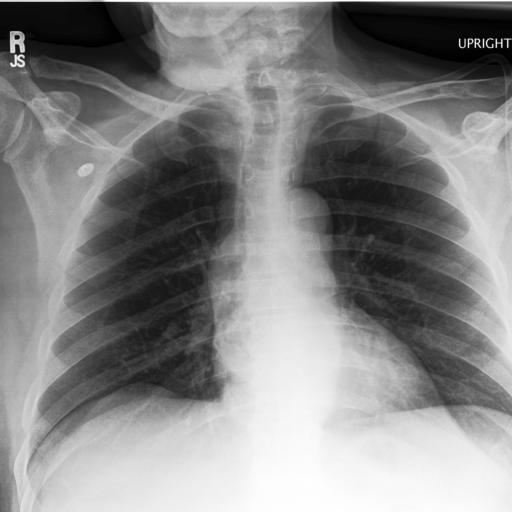}
        \end{minipage}
        \begin{minipage}{0.95\textwidth}
            \centering
            \scriptsize
            \includegraphics[width=\textwidth]{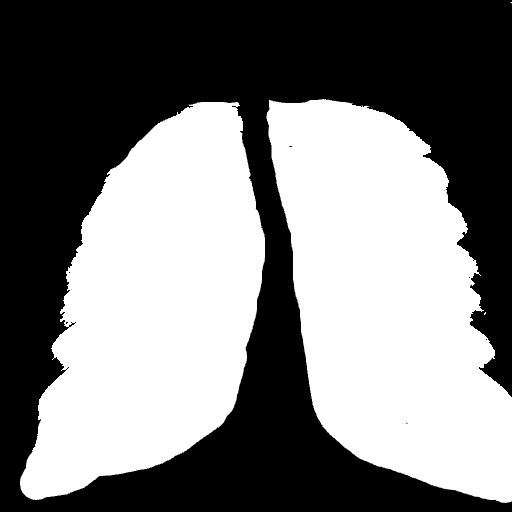}
        \end{minipage}
    \end{minipage}
    \begin{minipage}[t]{0.27\textwidth}
        \begin{minipage}{0.95\textwidth}
            \centering
            \scriptsize
            \includegraphics[width=\textwidth]{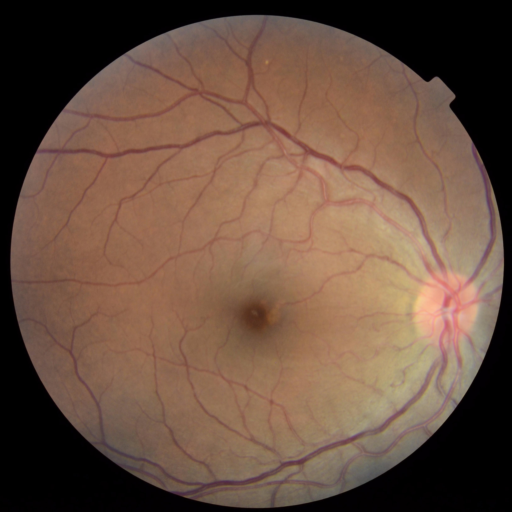}
        \end{minipage}
        \begin{minipage}{0.95\textwidth}
            \centering
            \scriptsize
            \includegraphics[width=\textwidth]{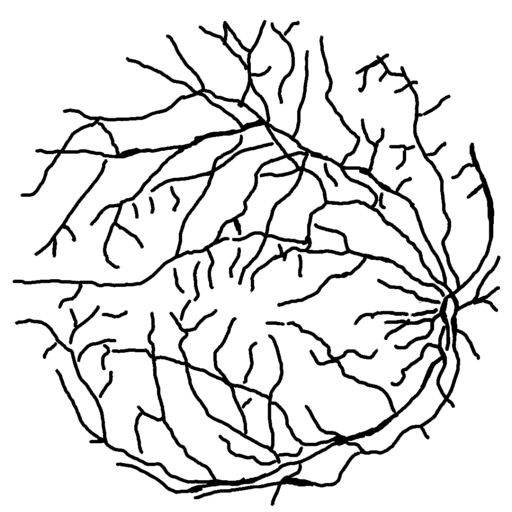}
        \end{minipage}
    \end{minipage}
    \begin{minipage}[t]{0.27\textwidth}
        \begin{minipage}{0.95\textwidth}
            \centering
            \scriptsize
            \includegraphics[width=\textwidth]{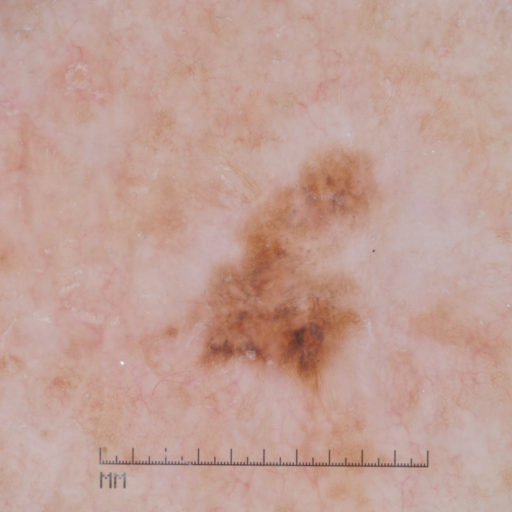}
        \end{minipage}
        \begin{minipage}{0.95\textwidth}
            \centering
            \scriptsize
            \includegraphics[width=\textwidth]{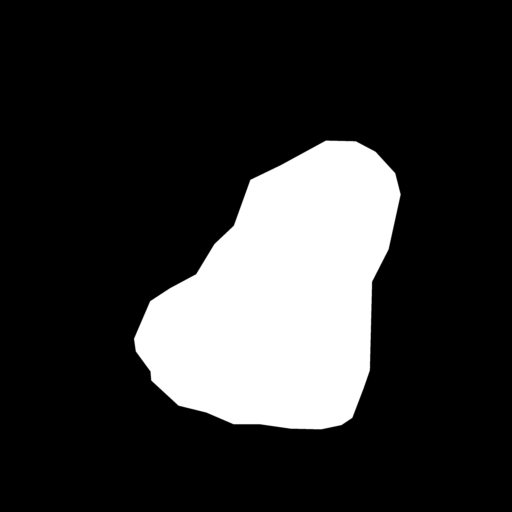}
        \end{minipage}
    \end{minipage}
\caption{ROI masks for three different domains. Note, the white part in the mask is the disease-related regions.}
\label{fig:appendix:show_mask}
\end{figure}

\subsection{ROI Mask Generation}

The ROI Masks used in experiments are generated by Segment Anything Model (SAM)~\cite{kirillov2023segment} and modified to smooth the edge. For different domains, since the region guide prior is different, the mask shape and size are also different. Figure~\ref{fig:appendix:show_mask} showcases examples of ROI masks utilized for simulation in the PIE and baseline models.

\subsection{Time Cost Analysis}

Given that both PIE and the baseline methods utilize the same Stable Diffusion backbone~\citep{rombach2022high}, a comparison of latency among each method is unnecessary. For simulating disease progression on an image of size $512 \times 512$, per step PIE requires approximately 0.078s to generate the subsequent stage when the strength parameter, $\gamma$, is set to 0.5, batch size is set to 1 and using one NVIDIA A100 (80GB).

\section{Ablation Study}\label{appendix:extra_ablation}

To investigate the influence of each hyper-parameter in PIE and analyse the progression visualization, we conduct several ablation studies for visualization, failure cases analysis and hyperparameter searching.

\begin{figure}[htbp]
    \begin{minipage}[t]{0.03\textwidth} 
        \rotatebox{90}{\!\!\!\!\!\! PIE}
    \end{minipage}
    \begin{minipage}[t]{0.95\textwidth}
        \begin{minipage}{0.19\textwidth}
            \centering
            \scriptsize
            \includegraphics[width=\textwidth]{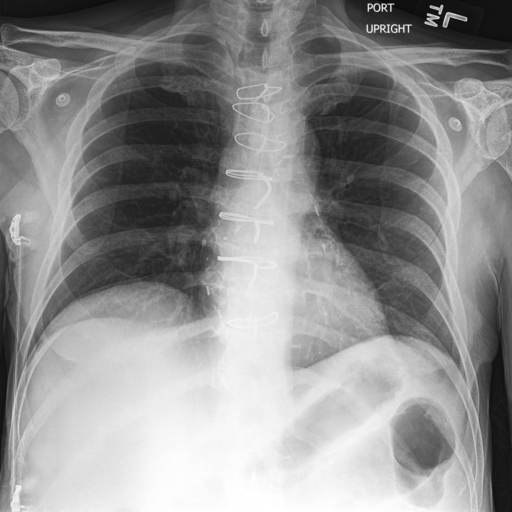}\\
        \end{minipage}
        \begin{minipage}{0.19\textwidth}
            \centering
            \scriptsize
            \includegraphics[width=\textwidth]{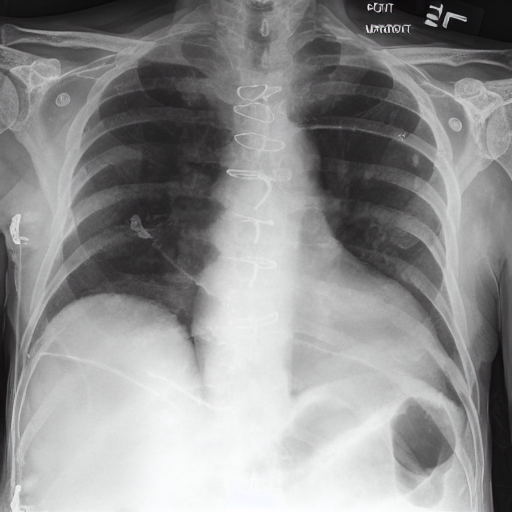}\\
        \end{minipage}
        \begin{minipage}{0.19\textwidth}
            \centering
            \scriptsize
            \includegraphics[width=\textwidth]{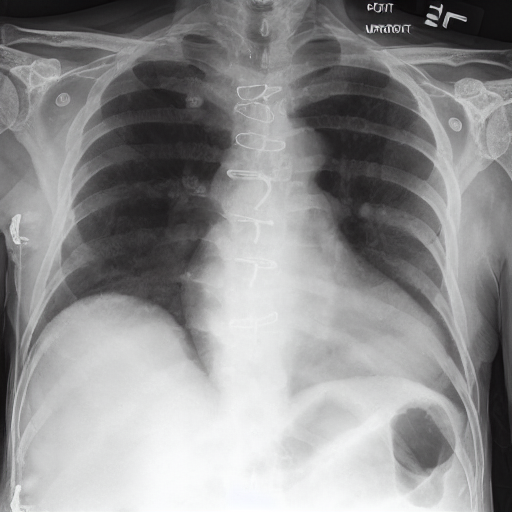}\\
        \end{minipage}
        \begin{minipage}{0.19\textwidth}
            \centering
            \scriptsize
            \includegraphics[width=\textwidth]{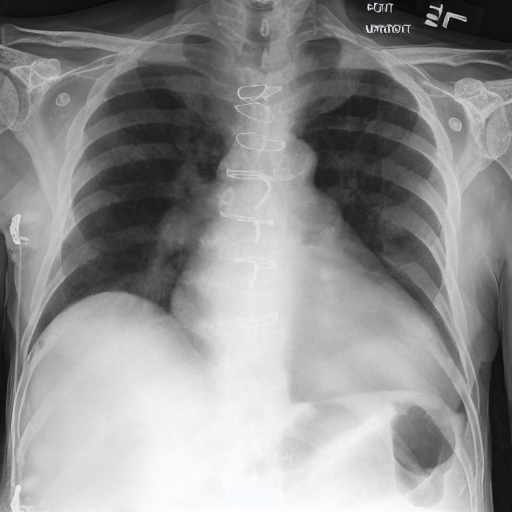}\\
        \end{minipage}
        \begin{minipage}{0.19\textwidth}
            \centering
            \scriptsize
            \includegraphics[width=\textwidth]{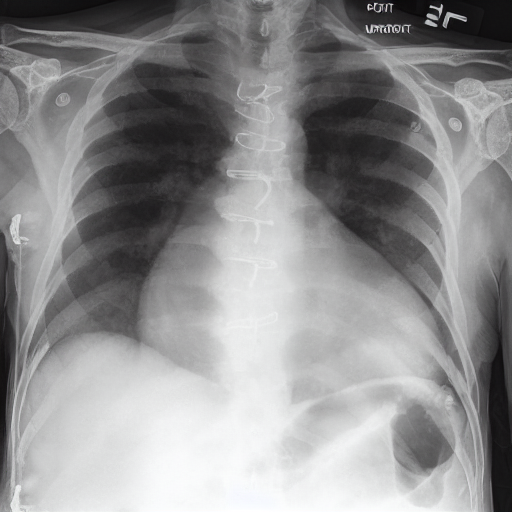}\\
        \end{minipage}
    \end{minipage}
    \\
    \begin{minipage}[t]{0.03\textwidth}
        \rotatebox{90}{\!\!\!\!\!\!\!\!\!\!\!\!\!\! SD Video}
    \end{minipage}
    \begin{minipage}[t]{0.95\textwidth}
        \begin{minipage}{0.19\textwidth}
            \centering
            \scriptsize
            \includegraphics[width=\textwidth]{figures/appendix/visualization/cardiomegaly/group1/0.png}\\
        \end{minipage}
        \begin{minipage}{0.19\textwidth}
            \centering
            \scriptsize
            \includegraphics[width=\textwidth]{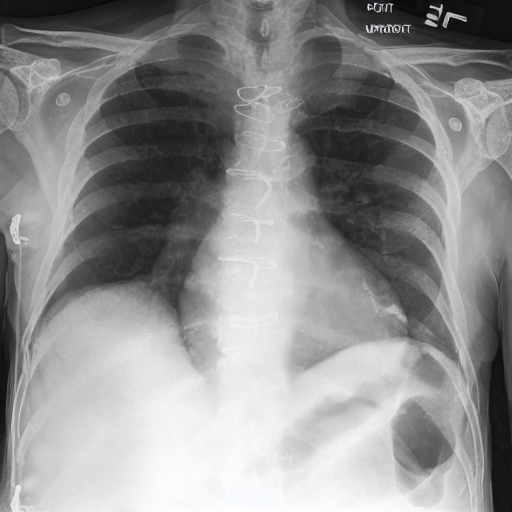}\\
        \end{minipage}
        \begin{minipage}{0.19\textwidth}
            \centering
            \scriptsize
            \includegraphics[width=\textwidth]{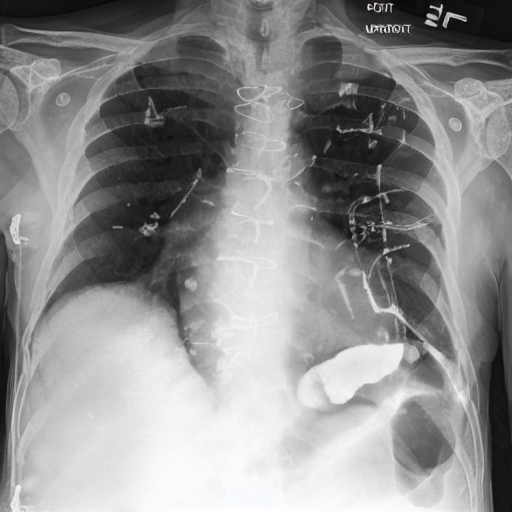}\\
        \end{minipage}
        \begin{minipage}{0.19\textwidth}
            \centering
            \scriptsize
            \includegraphics[width=\textwidth]{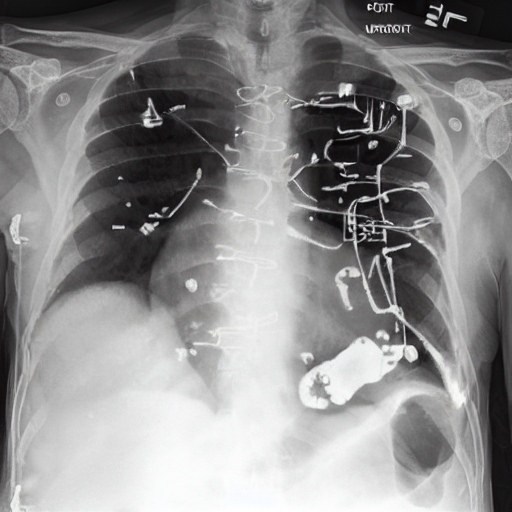}\\
        \end{minipage}
        \begin{minipage}{0.19\textwidth}
            \centering
            \scriptsize
            \includegraphics[width=\textwidth]{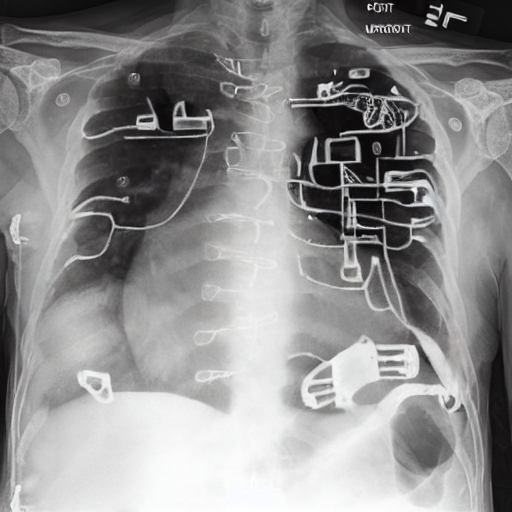}\\
        \end{minipage}
    \end{minipage}
    \\
    \begin{minipage}[t]{0.03\textwidth}
        \rotatebox{90}{\!\!\!\!\!\!\!\!\!\!\!\!\!\! Extrapolation}
    \end{minipage}
    \begin{minipage}[t]{0.95\textwidth}
        \begin{minipage}{0.19\textwidth}
            \centering
            \scriptsize
            \includegraphics[width=\textwidth]{figures/appendix/visualization/cardiomegaly/group1/0.png}\\
            \textit{Input Image}
        \end{minipage}
        \begin{minipage}{0.19\textwidth}
            \centering
            \scriptsize
            \includegraphics[width=\textwidth]{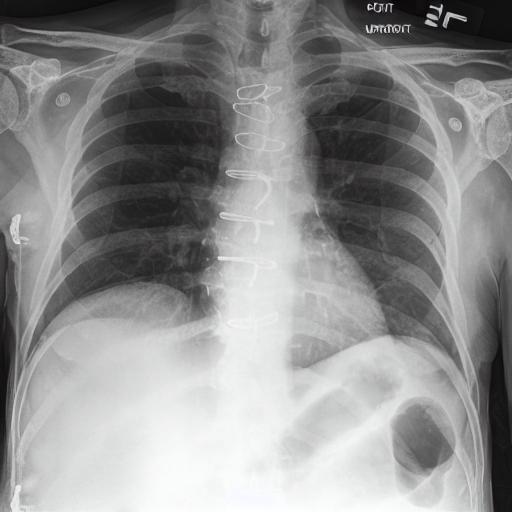}\\
            \textit{Step 1}
        \end{minipage}
        \begin{minipage}{0.19\textwidth}
            \centering
            \scriptsize
            \includegraphics[width=\textwidth]{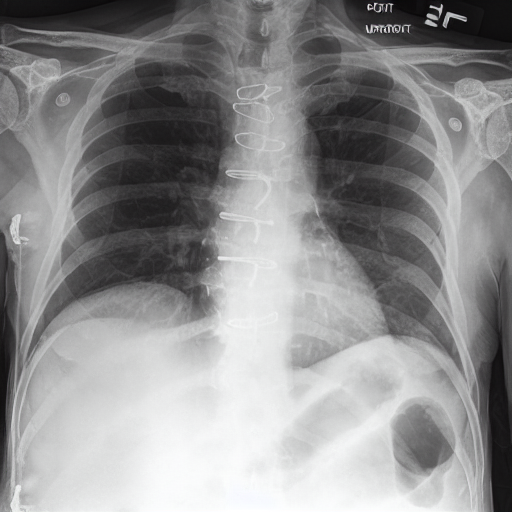}\\
            \textit{Step 2}
        \end{minipage}
        \begin{minipage}{0.19\textwidth}
            \centering
            \scriptsize
            \includegraphics[width=\textwidth]{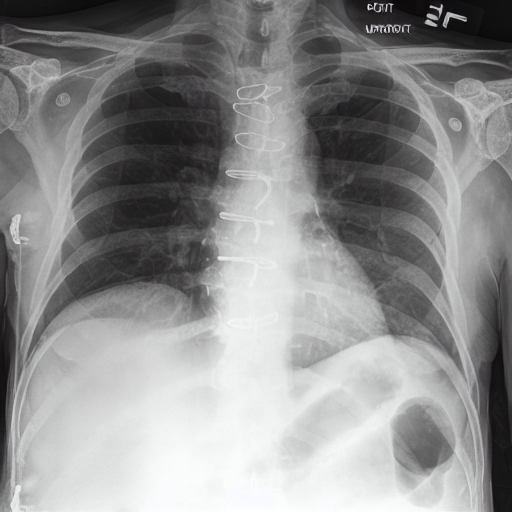}\\
            \textit{Step 4}
        \end{minipage}
        \begin{minipage}{0.19\textwidth}
            \centering
            \scriptsize
            \includegraphics[width=\textwidth]{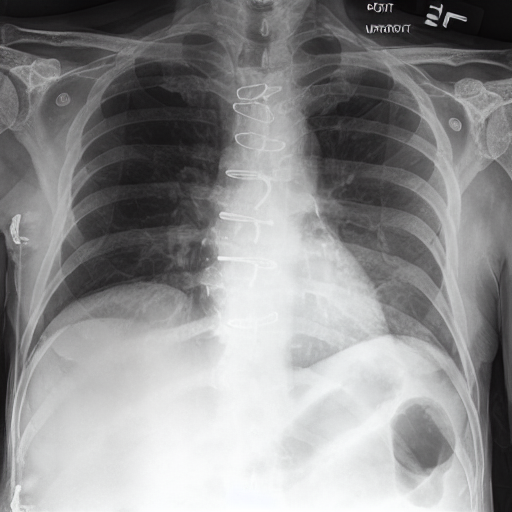}\\
            \textit{Step 10}
        \end{minipage}
    \end{minipage}
    \caption{Visualization of PIE, SD Video, Extrapolation to generate disease progression from Cardiomegaly clinical reports.}
\label{fig:progression_visualization_comp_cardiomegaly}
\end{figure}

\begin{figure}[htbp]
    \begin{minipage}[t]{0.03\textwidth}
        \rotatebox{90}{\!\!\!\!\!\! PIE}
    \end{minipage}
    \begin{minipage}[t]{0.95\textwidth}
        \begin{minipage}{0.19\textwidth}
            \centering
            \scriptsize
            \includegraphics[width=\textwidth]{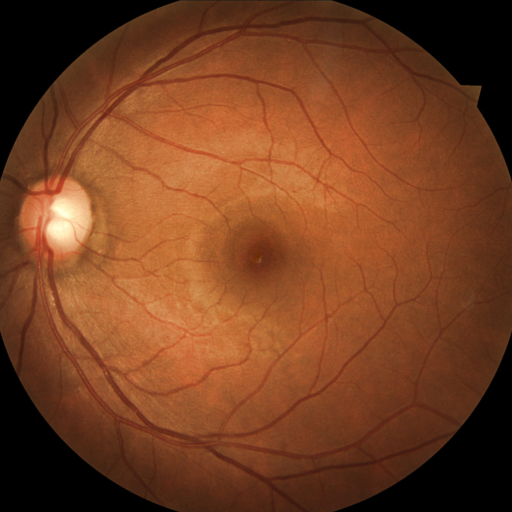}\\
        \end{minipage}
        \begin{minipage}{0.19\textwidth}
            \centering
            \scriptsize
            \includegraphics[width=\textwidth]{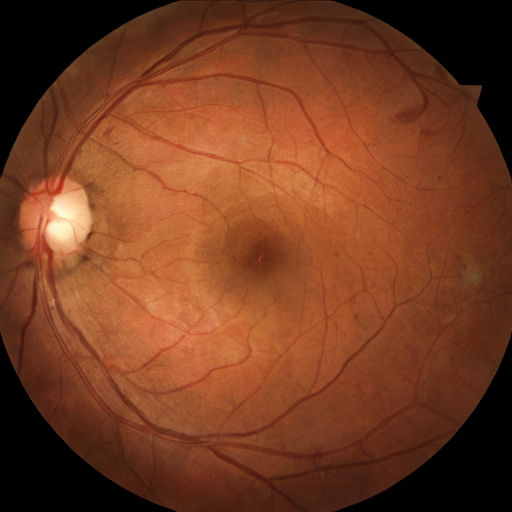}\\
        \end{minipage}
        \begin{minipage}{0.19\textwidth}
            \centering
            \scriptsize
            \includegraphics[width=\textwidth]{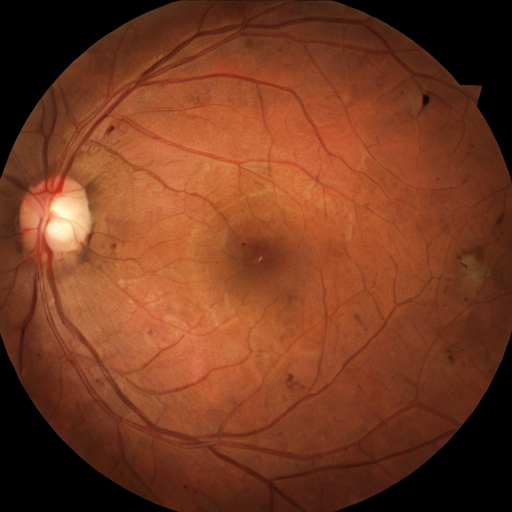}\\
        \end{minipage}
        \begin{minipage}{0.19\textwidth}
            \centering
            \scriptsize
            \includegraphics[width=\textwidth]{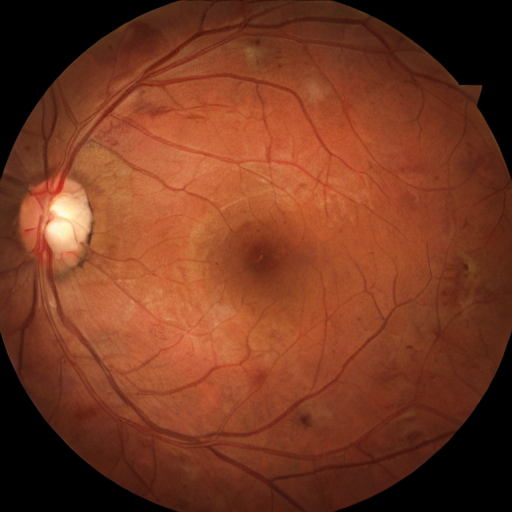}\\
        \end{minipage}
        \begin{minipage}{0.19\textwidth}
            \centering
            \scriptsize
            \includegraphics[width=\textwidth]{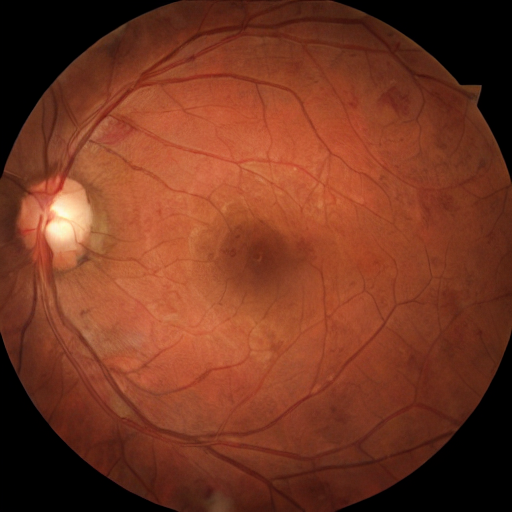}\\
        \end{minipage}
    \end{minipage}
    \\
    \begin{minipage}[t]{0.03\textwidth}
        \rotatebox{90}{\!\!\!\!\!\!\!\!\!\!\!\!\! SD Video}
    \end{minipage}
    \begin{minipage}[t]{0.95\textwidth}
        \begin{minipage}{0.19\textwidth}
            \centering
            \scriptsize
            \includegraphics[width=\textwidth]{figures/appendix/visualization/retina/group1/0.png}\\
        \end{minipage}
        \begin{minipage}{0.19\textwidth}
            \centering
            \scriptsize
            \includegraphics[width=\textwidth]{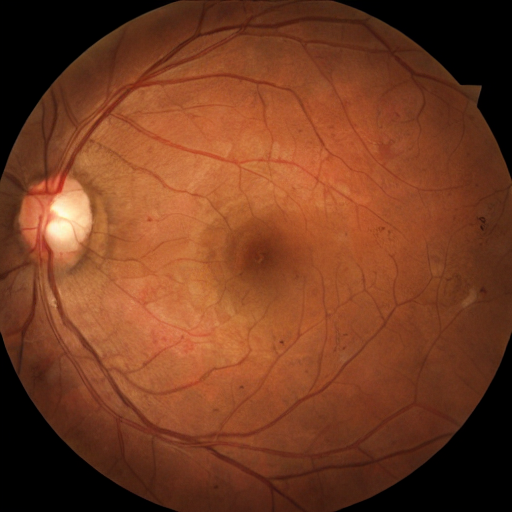}\\
        \end{minipage}
        \begin{minipage}{0.19\textwidth}
            \centering
            \scriptsize
            \includegraphics[width=\textwidth]{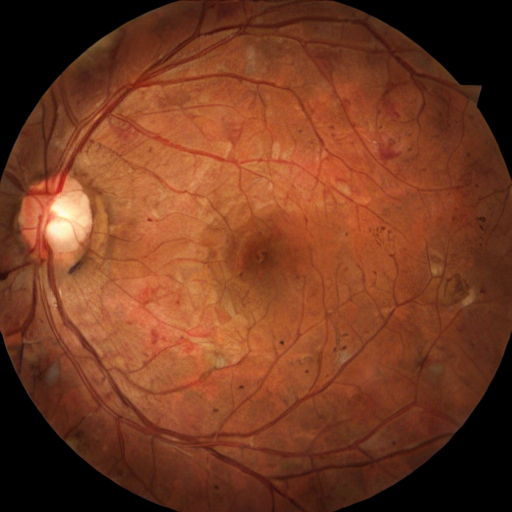}\\
        \end{minipage}
        \begin{minipage}{0.19\textwidth}
            \centering
            \scriptsize
            \includegraphics[width=\textwidth]{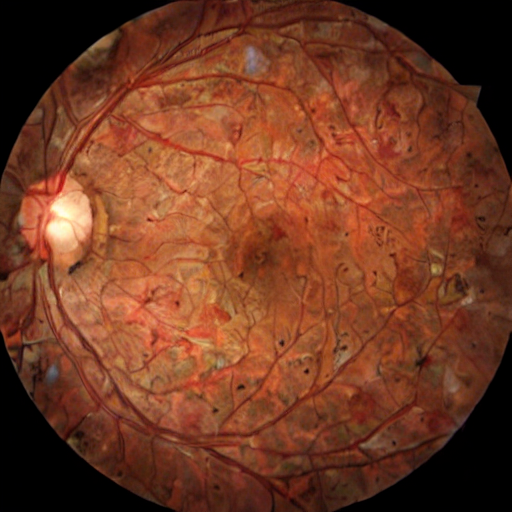}\\
        \end{minipage}
        \begin{minipage}{0.19\textwidth}
            \centering
            \scriptsize
            \includegraphics[width=\textwidth]{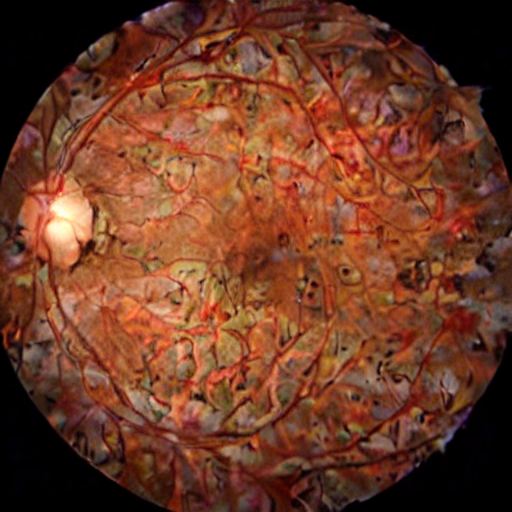}\\
        \end{minipage}
    \end{minipage}
    \\
    \begin{minipage}[t]{0.03\textwidth}
        \rotatebox{90}{\!\!\!\!\!\!\!\!\!\!\!\!\!\! Interpolation}
    \end{minipage}
    \begin{minipage}[t]{0.95\textwidth}
        \begin{minipage}{0.19\textwidth}
            \centering
            \scriptsize
            \includegraphics[width=\textwidth]{figures/appendix/visualization/retina/group1/0.png}\\
            \textit{Input Image}
        \end{minipage}
        \begin{minipage}{0.19\textwidth}
            \centering
            \scriptsize
            \includegraphics[width=\textwidth]{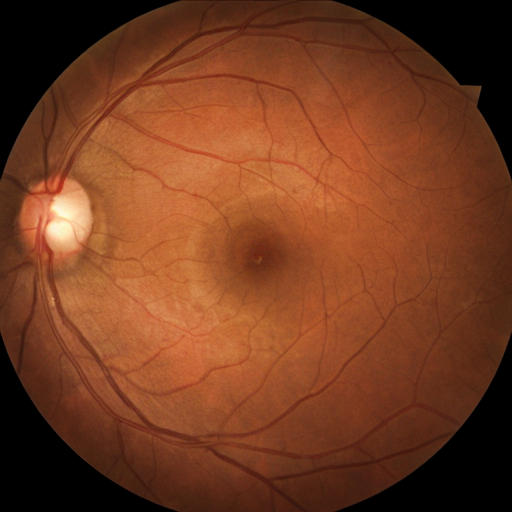}\\
            \textit{Step 1}
        \end{minipage}
        \begin{minipage}{0.19\textwidth}
            \centering
            \scriptsize
            \includegraphics[width=\textwidth]{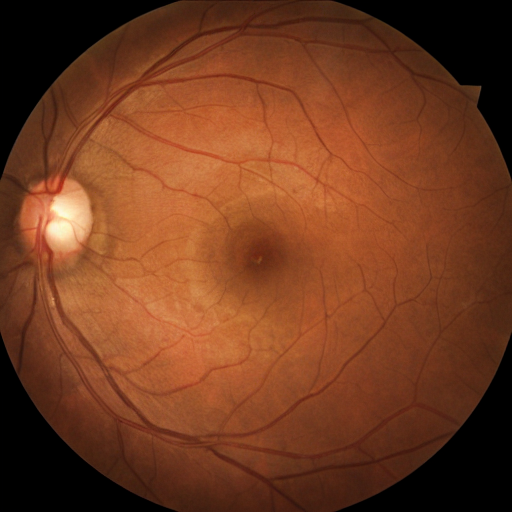}\\
            \textit{Step 2}
        \end{minipage}
        \begin{minipage}{0.19\textwidth}
            \centering
            \scriptsize
            \includegraphics[width=\textwidth]{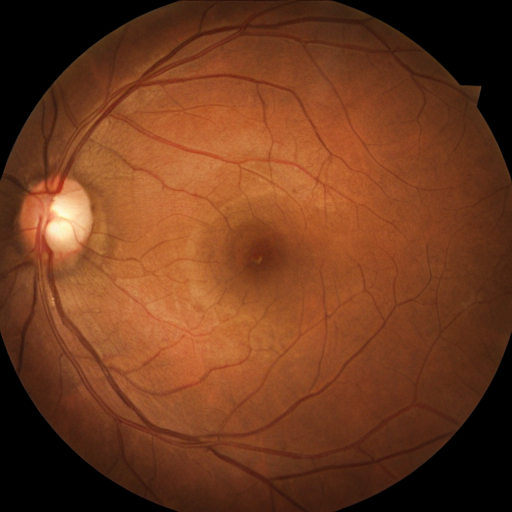}\\
            \textit{Step 4}
        \end{minipage}
        \begin{minipage}{0.19\textwidth}
            \centering
            \scriptsize
            \includegraphics[width=\textwidth]{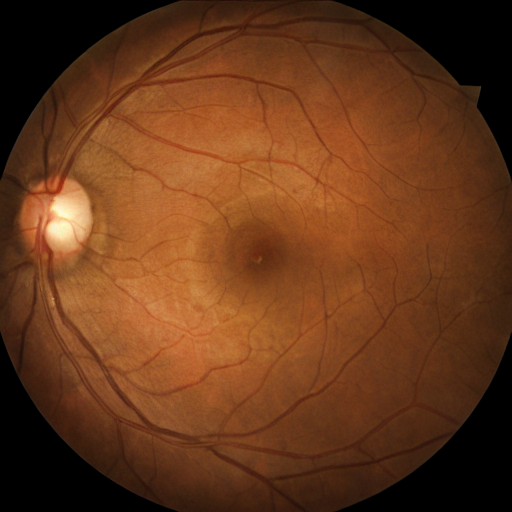}\\
            \textit{Step 10}
        \end{minipage}
    \end{minipage}
    \caption{Visualization of PIE, SD Video, Extrapolation to generate disease progression from Diabetic Retinopathy clinical reports.}
\label{fig:progression_visualization_comp_dr}
\end{figure}

\begin{figure}[htbp]
    \begin{minipage}[t]{0.03\textwidth}
        \rotatebox{90}{\!\!\!\!\!\! PIE}
    \end{minipage}
    \begin{minipage}[t]{0.95\textwidth}
        \begin{minipage}{0.19\textwidth}
            \centering
            \scriptsize
            \includegraphics[width=\textwidth]{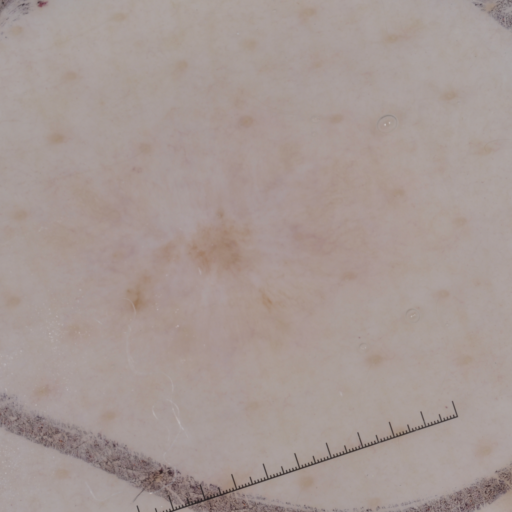}\\
        \end{minipage}
        \begin{minipage}{0.19\textwidth}
            \centering
            \scriptsize
            \includegraphics[width=\textwidth]{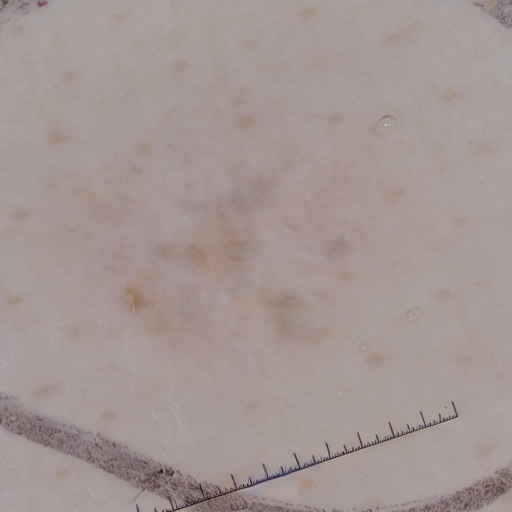}\\
        \end{minipage}
        \begin{minipage}{0.19\textwidth}
            \centering
            \scriptsize
            \includegraphics[width=\textwidth]{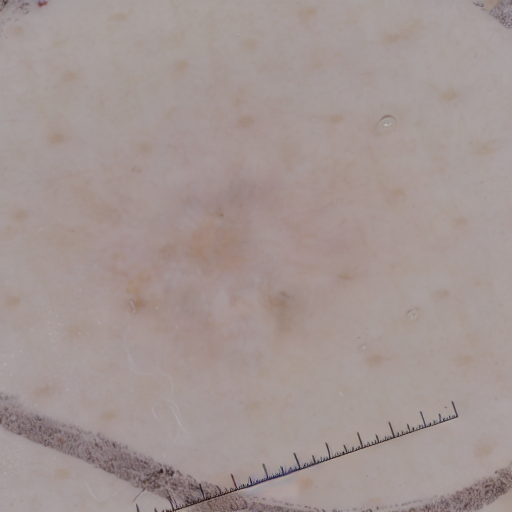}\\
        \end{minipage}
        \begin{minipage}{0.19\textwidth}
            \centering
            \scriptsize
            \includegraphics[width=\textwidth]{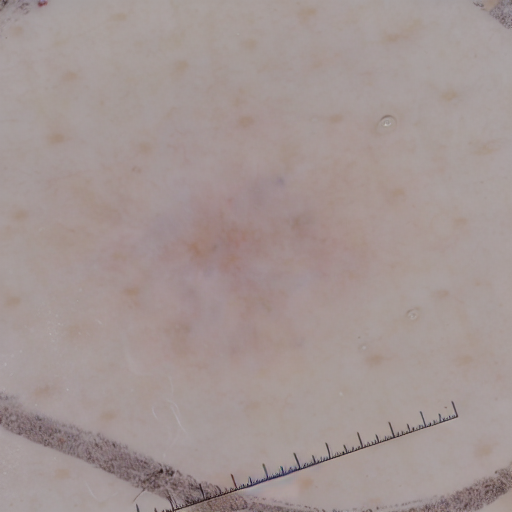}\\
        \end{minipage}
        \begin{minipage}{0.19\textwidth}
            \centering
            \scriptsize
            \includegraphics[width=\textwidth]{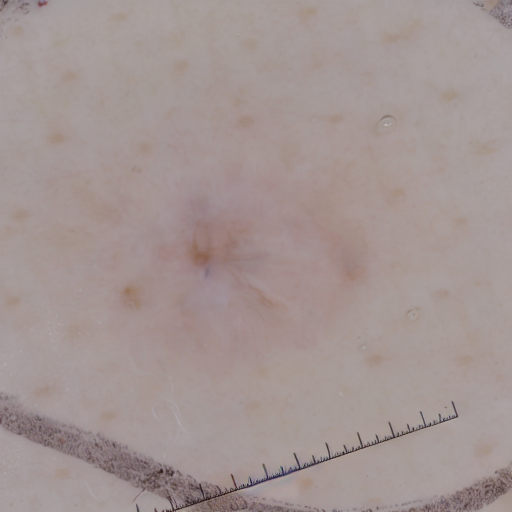}\\
        \end{minipage}
    \end{minipage}
    \\
    \begin{minipage}[t]{0.03\textwidth}
        \rotatebox{90}{\!\!\!\!\!\!\!\!\!\!\!\!\! SD Video}
    \end{minipage}
    \begin{minipage}[t]{0.95\textwidth}
        \begin{minipage}{0.19\textwidth}
            \centering
            \scriptsize
            \includegraphics[width=\textwidth]{figures/appendix/visualization/skin/group1/0.png}\\
        \end{minipage}
        \begin{minipage}{0.19\textwidth}
            \centering
            \scriptsize
            \includegraphics[width=\textwidth]{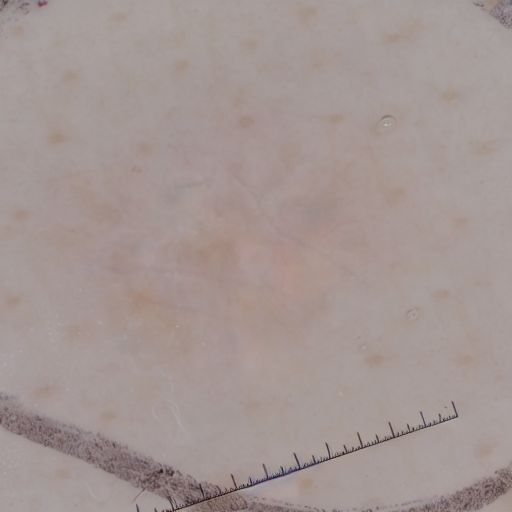}\\
        \end{minipage}
        \begin{minipage}{0.19\textwidth}
            \centering
            \scriptsize
            \includegraphics[width=\textwidth]{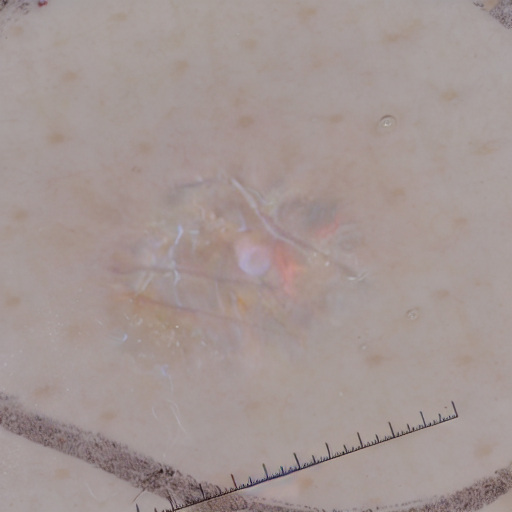}\\
        \end{minipage}
        \begin{minipage}{0.19\textwidth}
            \centering
            \scriptsize
            \includegraphics[width=\textwidth]{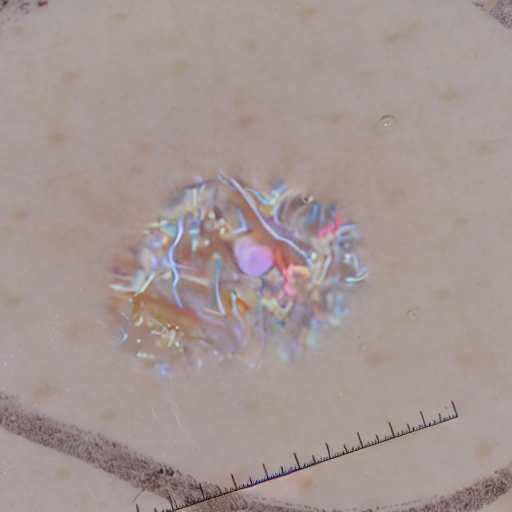}\\
        \end{minipage}
        \begin{minipage}{0.19\textwidth}
            \centering
            \scriptsize
            \includegraphics[width=\textwidth]{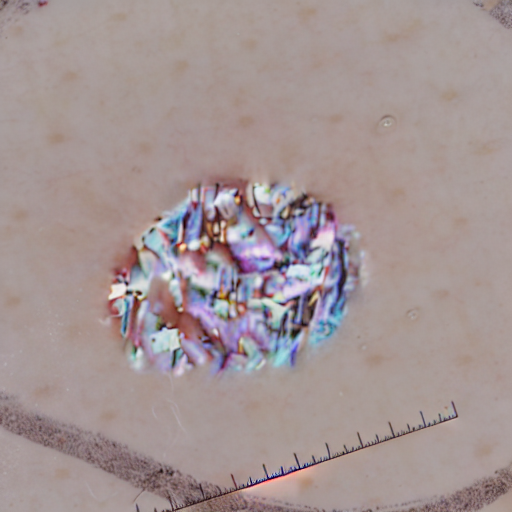}\\
        \end{minipage}
    \end{minipage}
    \\
    \begin{minipage}[t]{0.03\textwidth}
        \rotatebox{90}{\!\!\!\!\!\!\!\!\!\!\!\!\!\! Extrapolation}
    \end{minipage}
    \begin{minipage}[t]{0.95\textwidth}
        \begin{minipage}{0.19\textwidth}
            \centering
            \scriptsize
            \includegraphics[width=\textwidth]{figures/appendix/visualization/skin/group1/0.png}\\
            \textit{Input Image}
        \end{minipage}
        \begin{minipage}{0.19\textwidth}
            \centering
            \scriptsize
            \includegraphics[width=\textwidth]{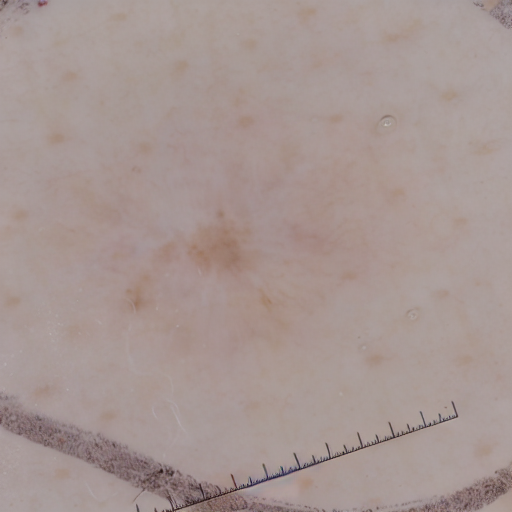}\\
            \textit{Step 1}
        \end{minipage}
        \begin{minipage}{0.19\textwidth}
            \centering
            \scriptsize
            \includegraphics[width=\textwidth]{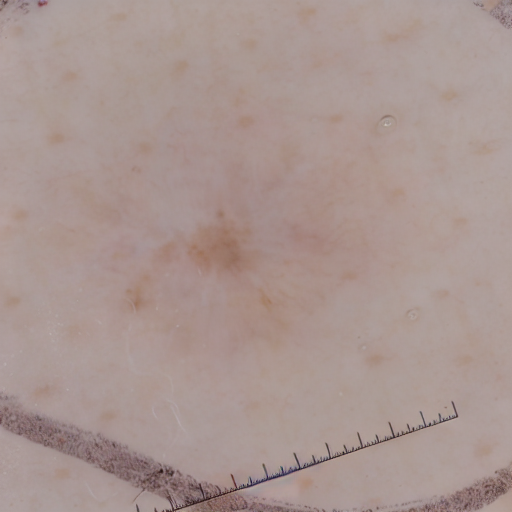}\\
            \textit{Step 2}
        \end{minipage}
        \begin{minipage}{0.19\textwidth}
            \centering
            \scriptsize
            \includegraphics[width=\textwidth]{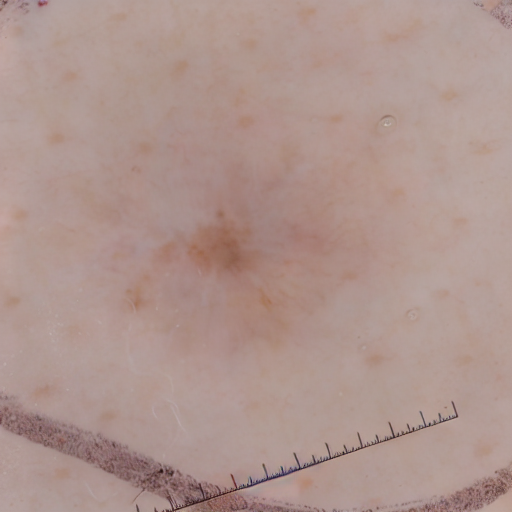}\\
            \textit{Step 4}
        \end{minipage}
        \begin{minipage}{0.19\textwidth}
            \centering
            \scriptsize
            \includegraphics[width=\textwidth]{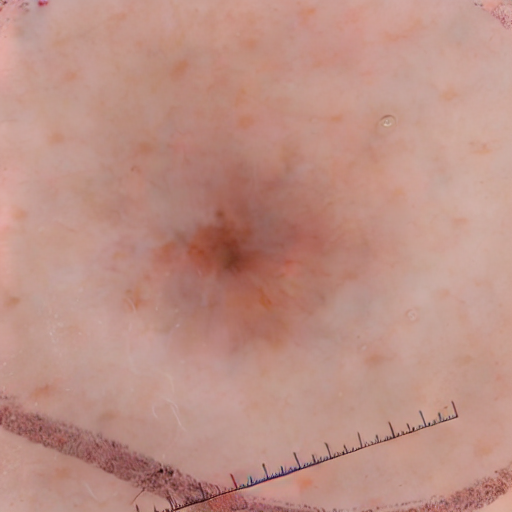}\\
            \textit{Step 10}
        \end{minipage}
    \end{minipage}
    \caption{Visualization of PIE, SD Video, Extrapolation to generate disease progression from Melanocytic Nevi clinical reports.}
\label{fig:progression_visualization_comp_skin}
\end{figure}

\subsection{Visualization for Three Medical Imaging Domains}

To provide an in-depth understanding of how PIE performs during different disease progression inference scenarios compared to baseline models, we present detailed visualizations demonstrating PIE's advantages. PIE consistently maintains the realism of the input image even after 10 steps of progression, excelling in most scenarios. Figure~\ref{fig:progression_visualization_comp_cardiomegaly} displays a comparison among three methods simulating Cardiomegaly progression. PIE outperforms both SD Video and Extrapolation by expanding the heart without introducing noise after 2 steps. Figure~\ref{fig:progression_visualization_comp_dr} displays a comparison among three methods simulating Diabetic Retinopathy progression. PIE outperforms both SD Video and Extrapolation by adding more bleeding (red) and small blind (white) regions without introducing noise after 2 steps. Figure~\ref{fig:progression_visualization_comp_skin} displays a comparison among three methods simulating Melanocytic Nevi progression. PIE outperforms the other two methods as it keeps the color and shape of the patient's skin but continually enhances the feature for Melanocytic Nevi.

SD Video, while it can interpolate within the prompt latent space, tends to generate noise. Although it can generate convincing video sequences in the general domain, it struggles to simulate authentic disease progression. On the other hand, Extrapolation, despite being effective for bone X-rays, faces challenges with more complex medical imaging domains like chest X-rays. Extrapolation's editing process is considerably slower than the other two methods, and it fails to be controlled effectively by clinical reports.

\subsection{Failure Case Analysis}

\begin{figure}[htbp]
    \begin{minipage}[t]{0.95\textwidth}
        \begin{minipage}{0.3\textwidth}
            \centering
            \scriptsize
            \includegraphics[width=\textwidth]{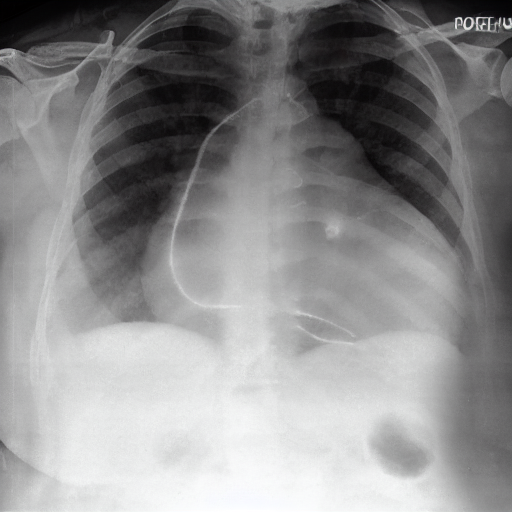}
            \textit{Step n-1}
        \end{minipage}
        \begin{minipage}{0.03\textwidth}
            \centering
            \scriptsize
            \includegraphics[width=1.0\textwidth]{figures/ar.png}
        \end{minipage}
        \begin{minipage}{0.3\textwidth}
            \centering
            \scriptsize
            \includegraphics[width=\textwidth]{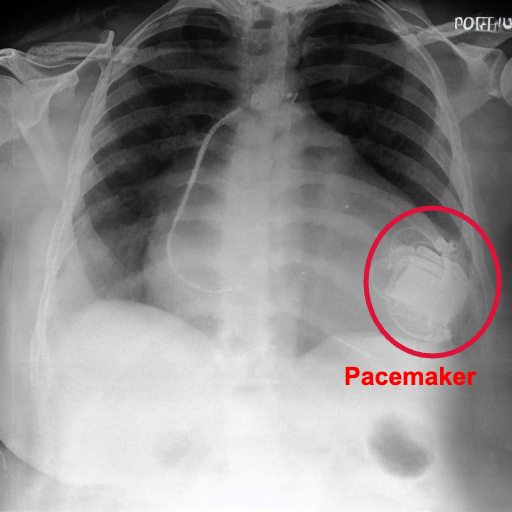}
            \textit{Step n}
        \end{minipage}
        \begin{minipage}{0.03\textwidth}
            \centering
            \scriptsize
            \includegraphics[width=1.0\textwidth]{figures/ar.png}
        \end{minipage}
        \begin{minipage}{0.3\textwidth}
            \centering
            \scriptsize
            \includegraphics[width=\textwidth]{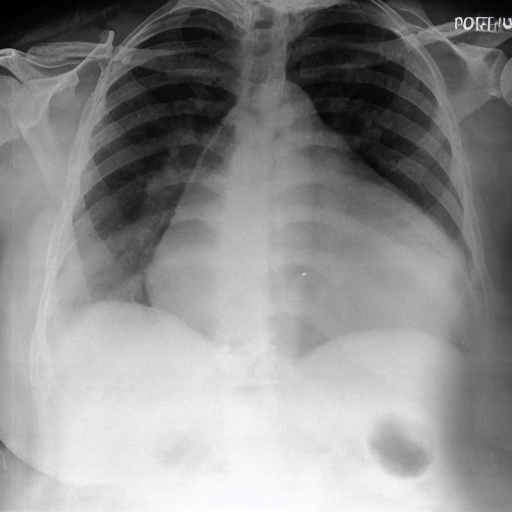}
            \textit{Step n+1}
        \end{minipage}
    \end{minipage}
    \caption{A failure case of the PIE model in preserving the features of a pacemaker during the simulation of Cardiomegaly disease progression. Pacemaker is usually used by patients with severe Cardiomegaly. At Step n-1, the X-ray displays an electronic line. At Step n, both the electronic line and the pacemaker are visible. However, by Step n+1, all the medical device features, including the pacemaker, have vanished from the simulation. It's important to note that the input clinical prompt did not contain any information regarding the pacemaker, making it difficult for the model to retain this crucial feature. This illustrates the challenges faced by models like PIE in dealing with significant but unmentioned clinical features in the input data. It underscores the need for incorporating comprehensive and detailed clinical data to ensure accurate and realistic disease progression simulations.}
\label{fig:appendix:failcase1}
\end{figure}

\begin{figure}[htbp]
    \begin{minipage}[t]{0.95\textwidth}
    \begin{minipage}{0.18\textwidth}
            \centering
            \scriptsize
            \includegraphics[width=\textwidth]{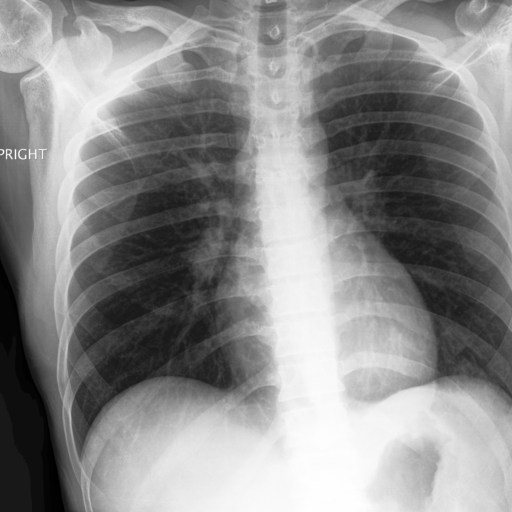}
        \end{minipage}
        \begin{minipage}{0.03\textwidth}
            \centering
            \scriptsize
            \includegraphics[width=1.0\textwidth]{figures/ar.png}
        \end{minipage}
        \begin{minipage}{0.18\textwidth}
            \centering
            \scriptsize
            \includegraphics[width=\textwidth]{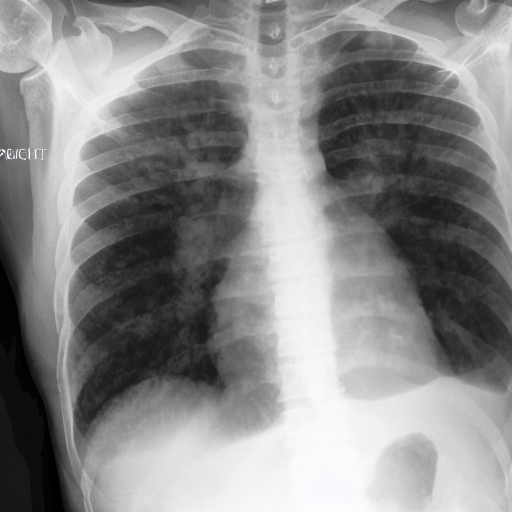}
        \end{minipage}
        \begin{minipage}{0.03\textwidth}
            \centering
            \scriptsize
            \includegraphics[width=1.0\textwidth]{figures/ar.png}
        \end{minipage}
        \begin{minipage}{0.18\textwidth}
            \centering
            \scriptsize
            \includegraphics[width=\textwidth]{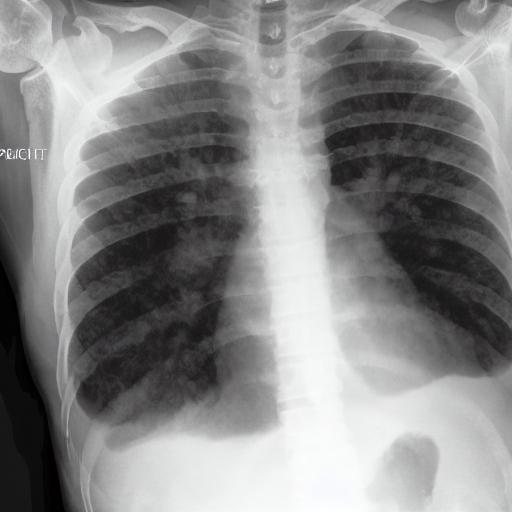}
        \end{minipage}
        \begin{minipage}{0.35\textwidth}
            \centering
            \scriptsize
            \includegraphics[width=\textwidth]{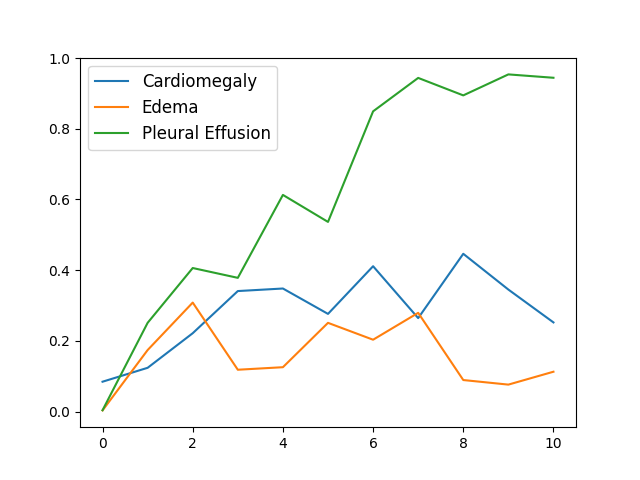}
        \end{minipage}
        
        \begin{minipage}{0.18\textwidth}
            \centering
            \scriptsize
            \includegraphics[width=\textwidth]{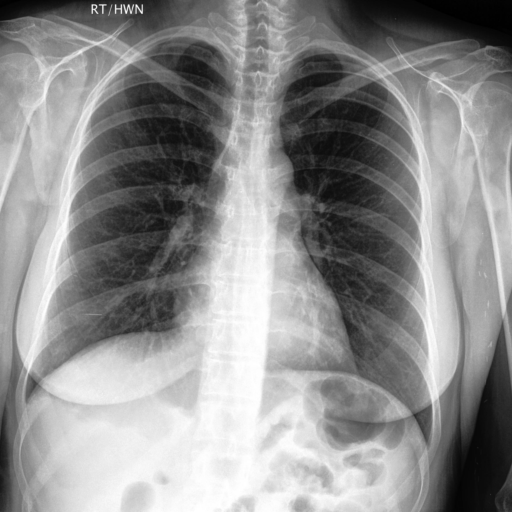}
            \textit{Input image}
        \end{minipage}
        \begin{minipage}{0.03\textwidth}
            \centering
            \scriptsize
            \includegraphics[width=1.0\textwidth]{figures/ar.png}
        \end{minipage}
        \begin{minipage}{0.18\textwidth}
            \centering
            \scriptsize
            \includegraphics[width=\textwidth]{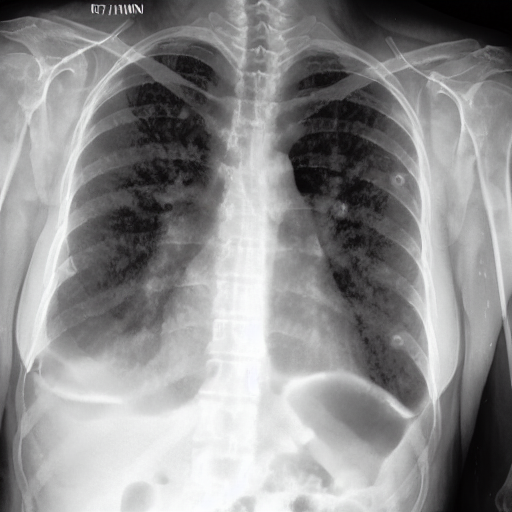}
            \textit{Step 4}
        \end{minipage}
        \begin{minipage}{0.03\textwidth}
            \centering
            \scriptsize
            \includegraphics[width=1.0\textwidth]{figures/ar.png}
        \end{minipage}
        \begin{minipage}{0.18\textwidth}
            \centering
            \scriptsize
            \includegraphics[width=\textwidth]{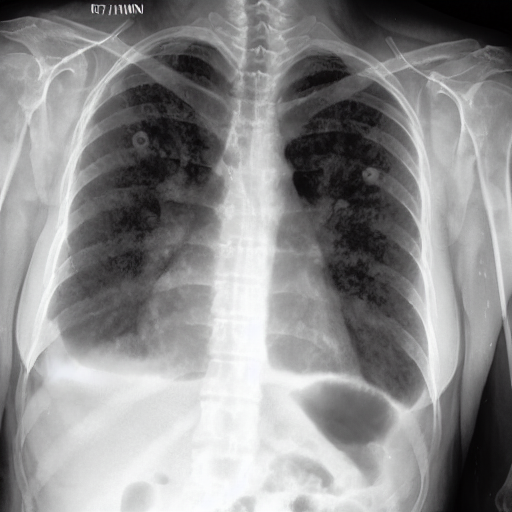}
            \textit{Step 10}
        \end{minipage}
        \begin{minipage}{0.35\textwidth}
            \centering
            \scriptsize
            \includegraphics[width=\textwidth]{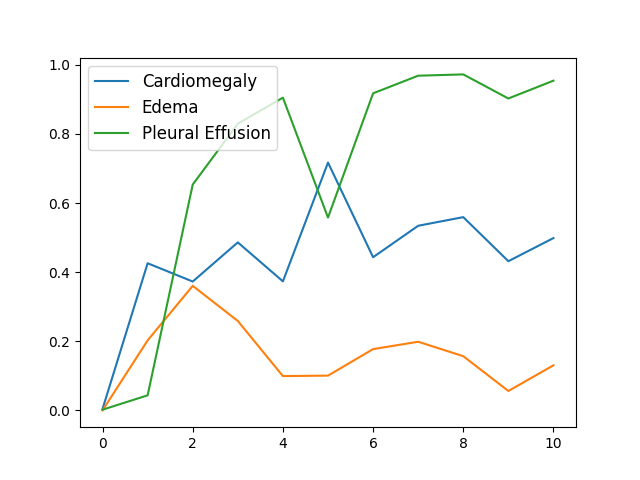}
            \textit{Confidence Score}
        \end{minipage}
    \end{minipage}
    \caption{Two failure cases of the PIE model in simulating co-occurring diseases progression for Cardiomegaly, Edema, and Pleural Effusion, only the features for Pleural Effusion are captured. These failure cases arise from the issue related to imbalanced label distribution in the training data. Specifically, the prevalence of Pleural Effusion is significantly higher than the other four classes, leading to an inherent bias in the model's simulations for co-occurring diseases. This imbalance emphasizes the need for a more diversified and balanced training dataset for more accurate simulation of co-occurring diseases.}
\label{fig:appendix:failcase2}
\end{figure}

Despite outperforming baseline models, PIE still faces limitations tied to data sensitivity issues. For instance, imbalances in the distribution of training data for Stable Diffusion can limit PIE's capability to edit rare diseases. In some cases, PIE might generate essential medical device features but fail to preserve them in subsequent stages of progression simulation, as observed with features like pacemakers. Figure~\ref{fig:appendix:failcase1} shows a good example of pacemaker disappearance during simulation. Besides, the co-occurring diseases simulation also has some failure cases (see Figure~\ref{fig:appendix:failcase2}). 

Fundamentally, these shortcomings could be addressed with a larger and label-equal distribution dataset. However, given that the volume of medical data is often smaller than in other domains, it is also important to explore fine-tuning PIE's diffusion backbone through few-shot learning under extremely imbalanced label distribution.

\begin{figure}[htbp]
    \begin{minipage}[t]{0.03\textwidth} 
        \rotatebox{90}{\!\!\!\!\!\!\!\!\!\!\!\!\!\!\!\!\!\! with ROI mask}
    \end{minipage}
    \begin{minipage}[t]{0.95\textwidth}
        \begin{minipage}{0.19\textwidth}
            \centering
            \scriptsize
            \includegraphics[width=\textwidth]{figures/appendix/visualization/cardiomegaly/group1/0.png}\\
        \end{minipage}
        \begin{minipage}{0.19\textwidth}
            \centering
            \scriptsize
            \includegraphics[width=\textwidth]{figures/appendix/visualization/cardiomegaly/group1/pie_1.png}\\
        \end{minipage}
        \begin{minipage}{0.19\textwidth}
            \centering
            \scriptsize
            \includegraphics[width=\textwidth]{figures/appendix/visualization/cardiomegaly/group1/pie_2.png}\\
        \end{minipage}
        \begin{minipage}{0.19\textwidth}
            \centering
            \scriptsize
            \includegraphics[width=\textwidth]{figures/appendix/visualization/cardiomegaly/group1/pie_4.png}\\
        \end{minipage}
        \begin{minipage}{0.19\textwidth}
            \centering
            \scriptsize
            \includegraphics[width=\textwidth]{figures/appendix/visualization/cardiomegaly/group1/pie_10.png}\\
        \end{minipage}
    \end{minipage}
    
    \begin{minipage}[t]{0.03\textwidth} 
        \rotatebox{90}{\!\!\!\!\!\!\!\!\!\!\!\!\!\!\!\!\!\! w/o ROI mask}
    \end{minipage}
    \begin{minipage}[t]{0.95\textwidth}
        \begin{minipage}{0.19\textwidth}
            \centering
            \scriptsize
            \includegraphics[width=\textwidth]{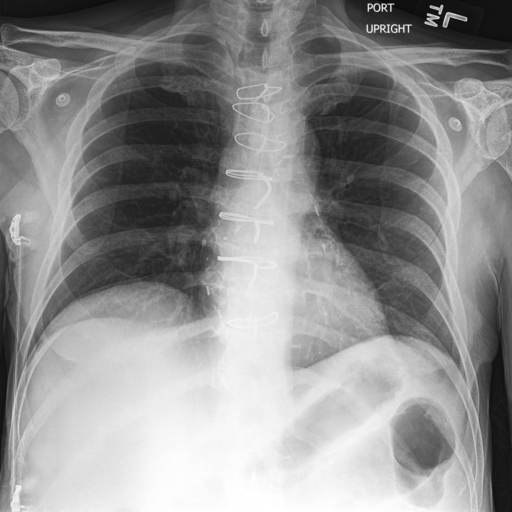}\\
        \end{minipage}
        \begin{minipage}{0.19\textwidth}
            \centering
            \scriptsize
            \includegraphics[width=\textwidth]{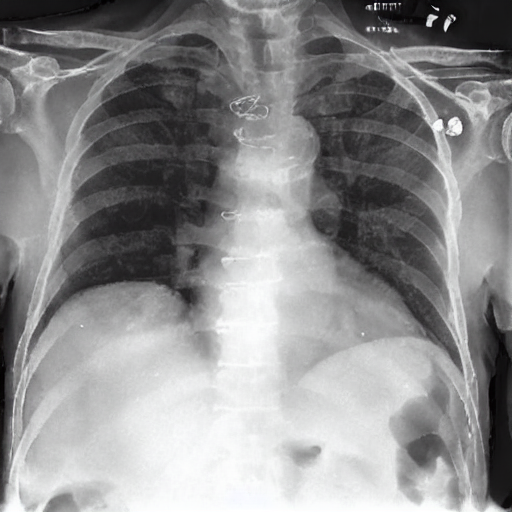}\\
        \end{minipage}
        \begin{minipage}{0.19\textwidth}
            \centering
            \scriptsize
            \includegraphics[width=\textwidth]{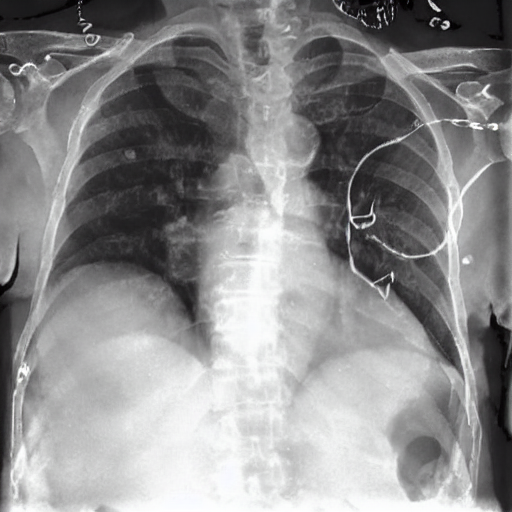}\\
        \end{minipage}
        \begin{minipage}{0.19\textwidth}
            \centering
            \scriptsize
            \includegraphics[width=\textwidth]{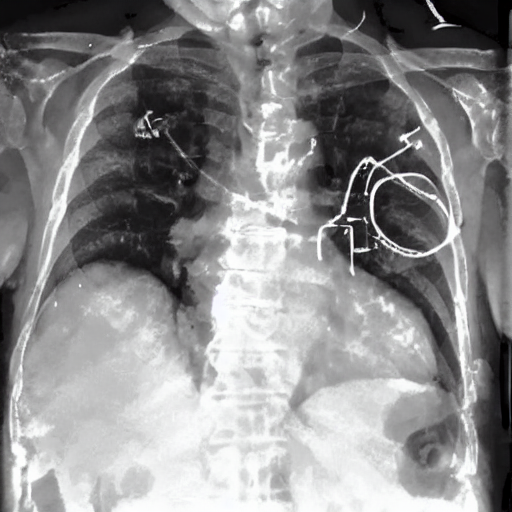}\\
        \end{minipage}
        \begin{minipage}{0.19\textwidth}
            \centering
            \scriptsize
            \includegraphics[width=\textwidth]{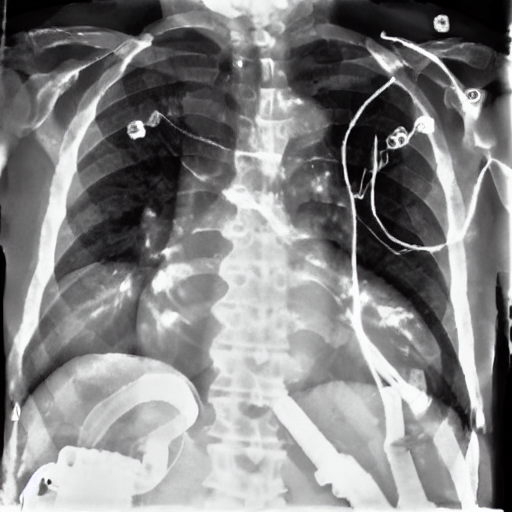}\\
        \end{minipage}
    \end{minipage}

    \begin{minipage}[t]{0.03\textwidth} 
        \rotatebox{90}{\!\!\!\!\!\!\!\!\!\!\!\!\!\!\!\!\!\! with ROI mask}
    \end{minipage}
    \begin{minipage}[t]{0.95\textwidth}
        \begin{minipage}{0.19\textwidth}
            \centering
            \scriptsize
            \includegraphics[width=\textwidth]{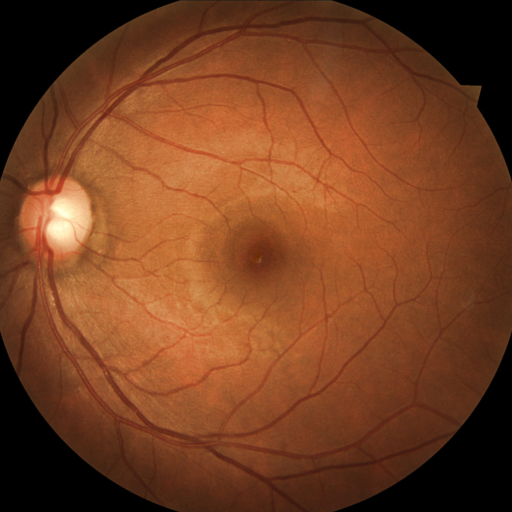}\\
        \end{minipage}
        \begin{minipage}{0.19\textwidth}
            \centering
            \scriptsize
            \includegraphics[width=\textwidth]{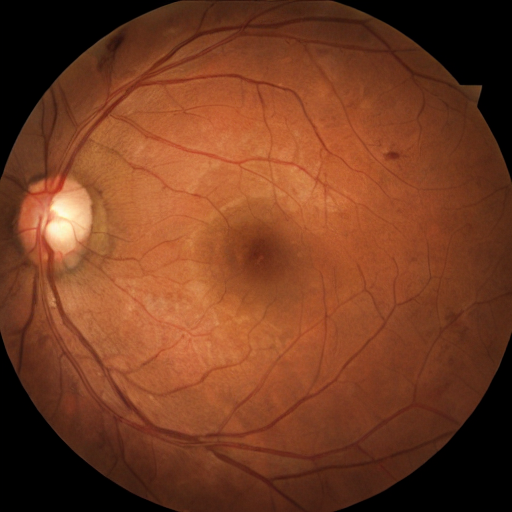}\\
        \end{minipage}
        \begin{minipage}{0.19\textwidth}
            \centering
            \scriptsize
            \includegraphics[width=\textwidth]{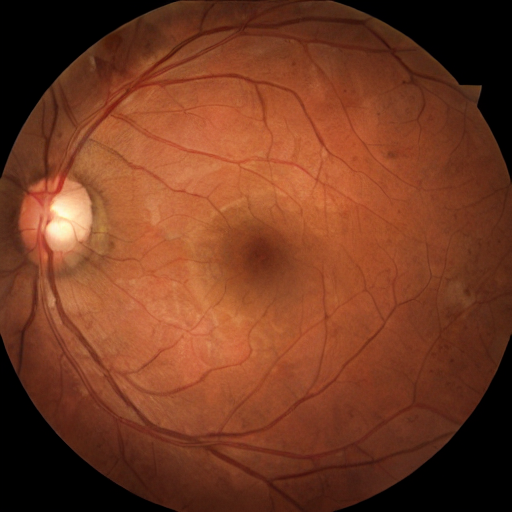}\\
        \end{minipage}
        \begin{minipage}{0.19\textwidth}
            \centering
            \scriptsize
            \includegraphics[width=\textwidth]{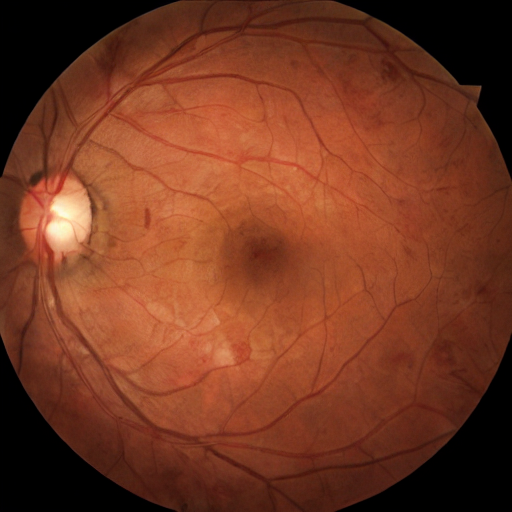}\\
        \end{minipage}
        \begin{minipage}{0.19\textwidth}
            \centering
            \scriptsize
            \includegraphics[width=\textwidth]{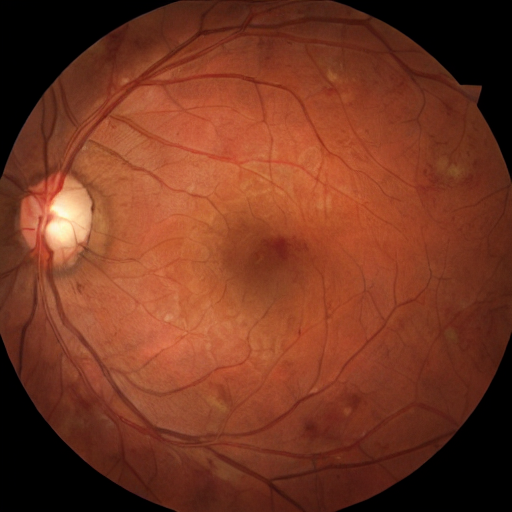}\\
        \end{minipage}
    \end{minipage}
    
    \begin{minipage}[t]{0.03\textwidth} 
        \rotatebox{90}{\!\!\!\!\!\!\!\!\!\!\!\!\!\!\!\!\!\! w/o ROI mask}
    \end{minipage}
    \begin{minipage}[t]{0.95\textwidth}
        \begin{minipage}{0.19\textwidth}
            \centering
            \scriptsize
            \includegraphics[width=\textwidth]{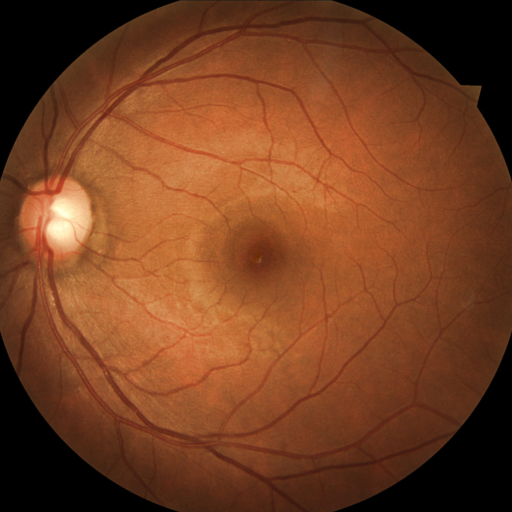}\\
        \end{minipage}
        \begin{minipage}{0.19\textwidth}
            \centering
            \scriptsize
            \includegraphics[width=\textwidth]{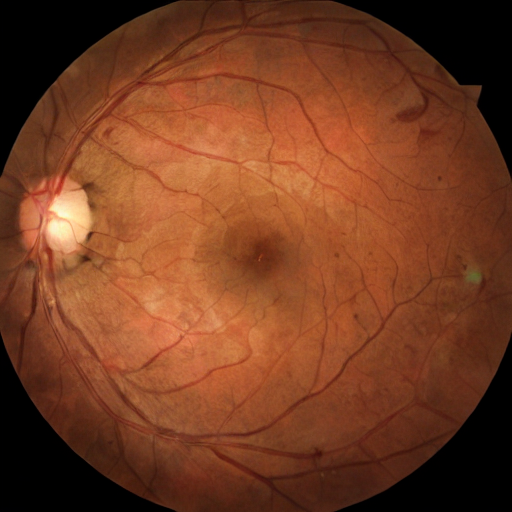}\\
        \end{minipage}
        \begin{minipage}{0.19\textwidth}
            \centering
            \scriptsize
            \includegraphics[width=\textwidth]{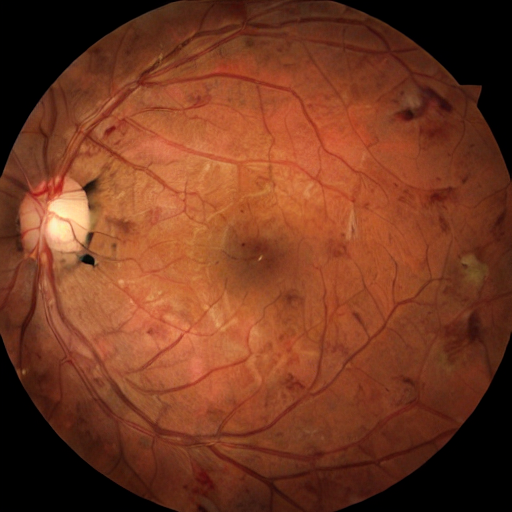}\\
        \end{minipage}
        \begin{minipage}{0.19\textwidth}
            \centering
            \scriptsize
            \includegraphics[width=\textwidth]{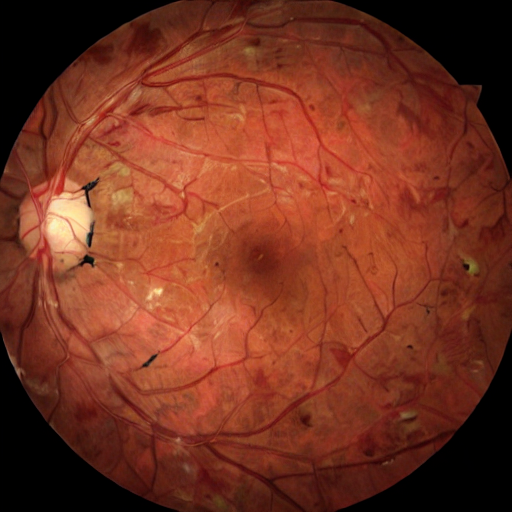}\\
        \end{minipage}
        \begin{minipage}{0.19\textwidth}
            \centering
            \scriptsize
            \includegraphics[width=\textwidth]{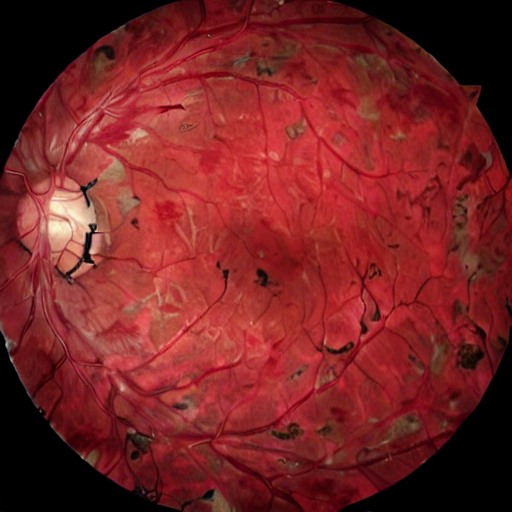}\\
        \end{minipage}
    \end{minipage}

    \begin{minipage}[t]{0.03\textwidth} 
        \rotatebox{90}{\!\!\!\!\!\!\!\!\!\!\!\!\!\!\!\!\!\! with ROI mask}
    \end{minipage}
    \begin{minipage}[t]{0.95\textwidth}
        \begin{minipage}{0.19\textwidth}
            \centering
            \scriptsize
            \includegraphics[width=\textwidth]{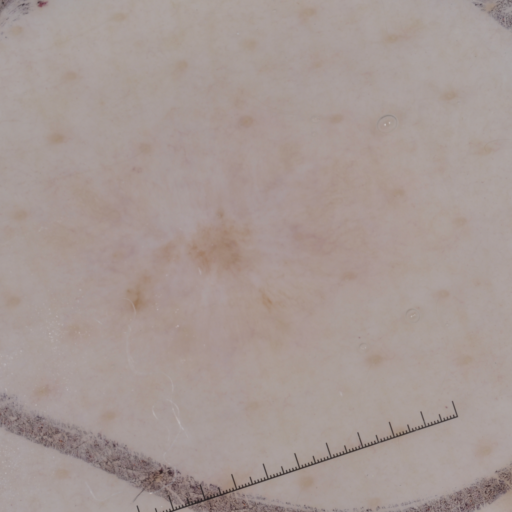}\\
        \end{minipage}
        \begin{minipage}{0.19\textwidth}
            \centering
            \scriptsize
            \includegraphics[width=\textwidth]{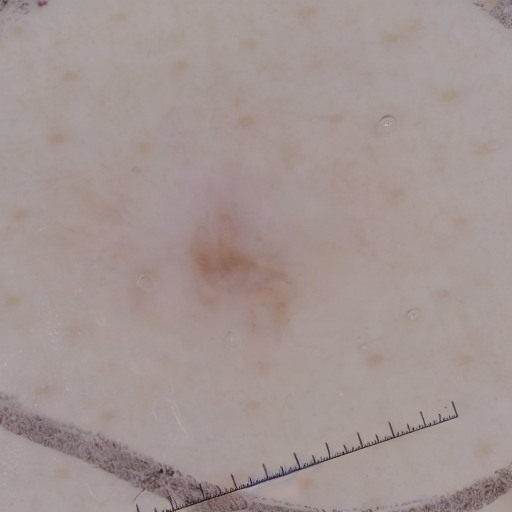}\\
        \end{minipage}
        \begin{minipage}{0.19\textwidth}
            \centering
            \scriptsize
            \includegraphics[width=\textwidth]{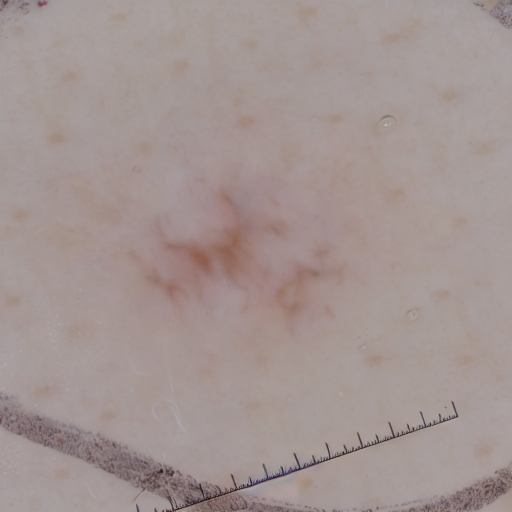}\\
        \end{minipage}
        \begin{minipage}{0.19\textwidth}
            \centering
            \scriptsize
            \includegraphics[width=\textwidth]{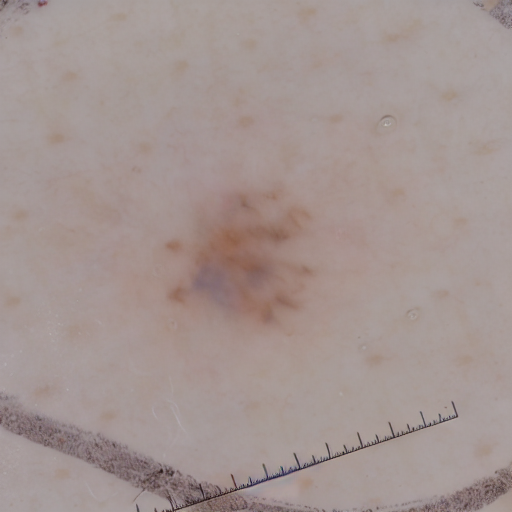}\\
        \end{minipage}
        \begin{minipage}{0.19\textwidth}
            \centering
            \scriptsize
            \includegraphics[width=\textwidth]{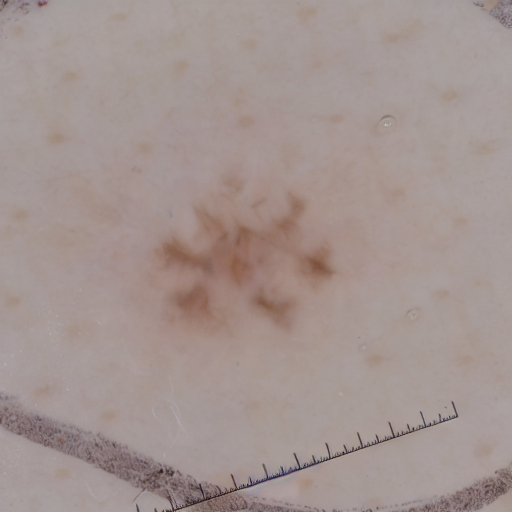}\\
        \end{minipage}
    \end{minipage}
    
    \begin{minipage}[t]{0.03\textwidth} 
        \rotatebox{90}{\!\!\!\!\!\!\!\!\!\!\!\!\!\!\!\!\!\! w/o ROI mask}
    \end{minipage}
    \begin{minipage}[t]{0.95\textwidth}
        \begin{minipage}{0.19\textwidth}
            \centering
            \scriptsize
            \includegraphics[width=\textwidth]{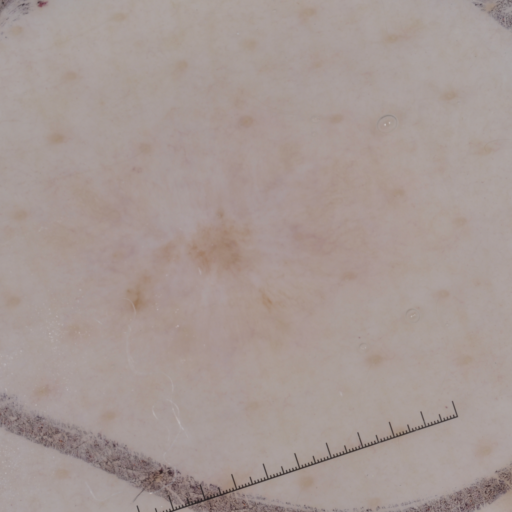}\\
        \end{minipage}
        \begin{minipage}{0.19\textwidth}
            \centering
            \scriptsize
            \includegraphics[width=\textwidth]{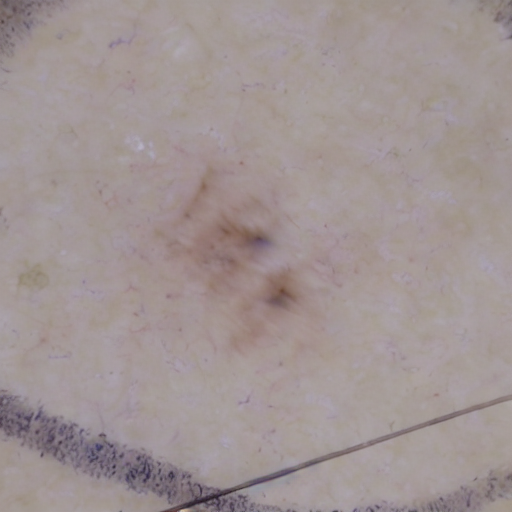}\\
        \end{minipage}
        \begin{minipage}{0.19\textwidth}
            \centering
            \scriptsize
            \includegraphics[width=\textwidth]{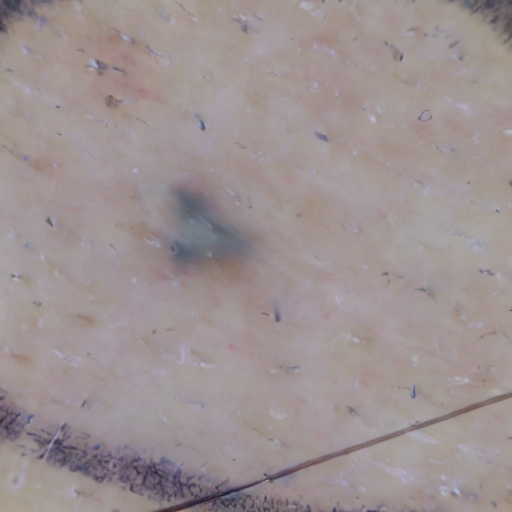}\\
        \end{minipage}
        \begin{minipage}{0.19\textwidth}
            \centering
            \scriptsize
            \includegraphics[width=\textwidth]{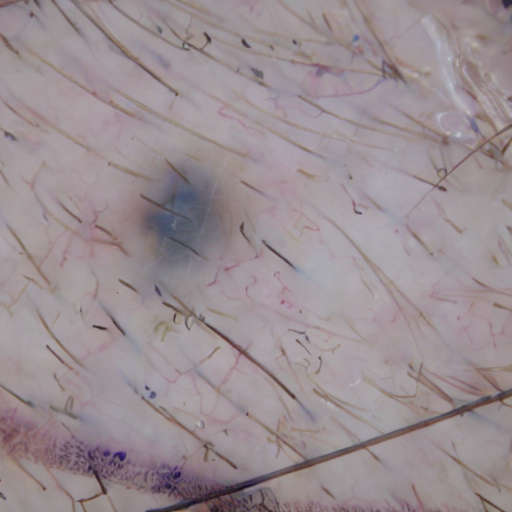}\\
        \end{minipage}
        \begin{minipage}{0.19\textwidth}
            \centering
            \scriptsize
            \includegraphics[width=\textwidth]{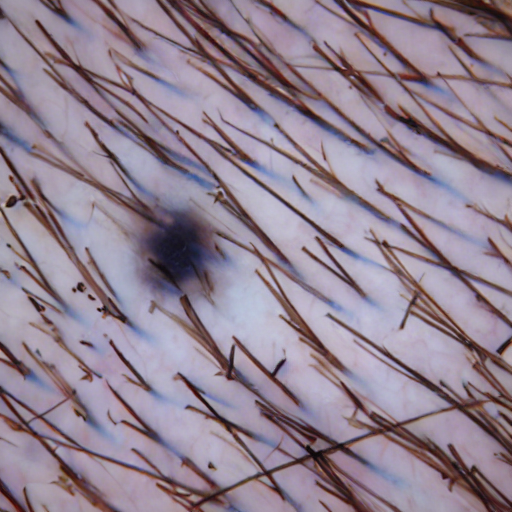}\\
        \end{minipage}
    \end{minipage}
\caption{Visualization comparison between ROI mask influence for three medical imaging domains.}
\label{fig:appendix:mask_nomask}
\end{figure}

\subsection{Hyperparameter Search \& Analysis}
While the Fréchet Inception Distance (FID) and Kernel Inception Distance (KID) metrics are widely utilized in the evaluation of generative models, they do not necessarily align with the perceptual quality of images, making them unsuitable for evaluating progression simulation tasks. Nevertheless, for the subsequent hyperparameter search and ablation study, we added the results of KID score, given its ability to provide insight into the diversity and distributional closeness of the generated images in relation to the real data.

To demonstrate the significance of the ROI mask as a key control factor in the PIE, an experiment was conducted to compare the performance of models using and not using ROI masks across three domains, with each model tested using 5 different random seeds. The evaluation focused on the classification confidence score and CLIP score. Results, as shown in Table~\ref{tab:ablation_mask}, revealed that while removing the ROI mask could sometimes increase the confidence score (observed for chest X-ray and skin imaging), it consistently led to a decrease in the CLIP score. Further visual analysis depicted in Figure~\ref{fig:appendix:mask_nomask} shows that the ROI mask crucially helps in preserving the basic shape of the medical imaging during the PIE process. Consequently, these findings suggest that the ROI mask, alongside clinical reports, serves as a critical medical prior for simulating disease progression. It helps the PIE to concentrate on disease-related regions while maintaining the realism of the input image.

We present a comprehensive examination of various hyperparameters, namely strength ($\gamma$ in Algorithm~\ref{algo:1}), step ($N$), $\beta_1$, and $\beta_2$, and their respective tradeoffs as demonstrated in tables \ref{tab:ablation_strength}, \ref{tab:ablation_multistep}, and \ref{tab:ablation_beta}. Notably, a discernible tradeoff exists between classification confidence score and CLIP-I/KID, wherein an increase in classification confidence score results in a decrease in CLIP-I. In the subsequent section, we offer an intricate analysis of each hyperparameter.

Table \ref{tab:ablation_strength} illustrates a positive correlation between the strength of PIE and classification confidence score, while revealing a negative relationship between strength and CLIP-I/KID. From an intuitive standpoint, as the strength of PIE increases, more features are directed to align with the pathologies within the original images. Consequently, the classifier exhibits greater confidence in accurately predicting the specific disease class, leading to a more significant deviation from the initial starting point. As a result, the classification confidence score value increases while the CLIP-I value decreases, reflecting the inverse relationship between the two metrics.

Table \ref{tab:ablation_multistep} presents a similar pattern initially. However, both classification confidence score and CLIP-I/KID reach a state of stability after a certain number of steps. This observation aligns with our theoretical analysis as outlined in Proposition 2. Moreover, it can be interpreted as the convergence of a Cauchy geometric sequence, wherein the discrepancy between successive steps gradually diminishes as the value of $N$ tends towards infinity.

Lastly, in Table \ref{tab:ablation_beta}, we explore the interplay between $\beta_1$ and $\beta_2$, which serve as parameters governing the rate of progression within and outside the region of interest (ROI) respectively. Our findings reveal that $\beta_2$, responsible for regulating the pace of progression within the ROI, exerts a more pronounced influence on classification confidence score, while $\beta_1$ exhibits a stronger impact on CLIP-I/KID. This outcome can be intuitively comprehended, considering that the ROI typically encompasses a smaller area of paramount importance for aligning with disease-specific features. Conversely, the areas outside the ROI exert a significantly greater influence on the realism captured by CLIP-I/KID.

In summary, our study elucidates the inherent trade-offs and offers valuable practical insights, thereby furnishing meaningful guidance for effectively utilizing PIE in practice.
\begin{table}[!h]
\centering
\caption{Mask, w/o mask guidance comparisons.}
\label{tab:ablation_mask}
\begin{tabular}{lcccccc}
\toprule
\multirow{2}{*}{\textbf{Method}} & \multicolumn{2}{c}{\textbf{Chest X-ray}} & \multicolumn{2}{c}{\textbf{Retinopathy}} & \multicolumn{2}{c}{\textbf{Skin Lesion Image}} \\ \cmidrule(lr){2-3} \cmidrule(lr){4-5} \cmidrule(lr){6-7}
& \textbf{Conf $(\uparrow)$} & \textbf{CLIP-I  $(\uparrow)$} & \textbf{Conf $(\uparrow)$} & \textbf{CLIP-I $(\uparrow)$} & \textbf{Conf $(\uparrow)$} & \textbf{CLIP-I  $(\uparrow)$} \\ 
\midrule
w/o mask & 0.729 & 0.93 & 0.163 & 0.96 & 0.666 & 0.85 \\ 
with mask & 0.690 & 0.96 & 0.807 & 0.99 & 0.453 & 0.95 \\ 
\bottomrule
\end{tabular}
\end{table}

\begin{table}[!h]
\centering
\caption{Strength $\gamma$ selection for $N=10$.}
\label{tab:ablation_strength}
\begin{tabular}{lccc}
\toprule
\textbf{Strength} & \textbf{Conf $(\uparrow)$} & \textbf{CLIP-I $(\uparrow)$} & \textbf{KID $(\downarrow)$} \\
\midrule
0.05 & 0.038 & 0.966 & 0.0685 \\ 
0.1 & 0.120 & 0.969 & 0.0638 \\ 
0.2 & 0.273 & 0.969 & 0.0885 \\ 
0.3 & 0.455 & 0.967 & 0.1033 \\ 
0.4 & 0.746 & 0.965 & 0.1142 \\ 
0.5 & 0.977 & 0.962 & 0.1389 \\ 
0.6 & 0.995 & 0.956 & 0.1549 \\ 
0.7 & 0.998 & 0.955 & 0.1533 \\ 
0.8 & 0.999 & 0.951 & 0.1629 \\ 
\bottomrule
\end{tabular}
\end{table}

\begin{table}[!h]
\centering
\caption{Simulation steps $N$ selection with $\gamma=0.5$.}
\label{tab:ablation_multistep}
\begin{tabular}{lccc}
\toprule
\textbf{Step ($N$)} & \textbf{Conf $(\uparrow)$} & \textbf{CLIP-I $(\uparrow)$} & \textbf{KID $(\downarrow)$} \\
\midrule
1 & 0.491 & 0.965 & 0.094 \\ 
2 & 0.731 & 0.964 & 0.098 \\ 
5 & 0.881 & 0.963 & 0.121 \\ 
10 & 0.978 & 0.962 & 0.142 \\ 
20 & 0.989 & 0.961 & 0.111 \\ 
50 & 0.975 & 0.962 & 0.130 \\ 
100 & 0.959 & 0.962 & 0.115 \\ 
\bottomrule
\end{tabular}
\end{table}

\begin{table}[!h]
\centering
\caption{Beta selection.}
\label{tab:ablation_beta}
\begin{tabular}{lcccc}
\toprule
\textbf{$\beta_{1}$} & \textbf{$\beta_{2}$} & \textbf{Conf $(\uparrow)$} & \textbf{CLIP-I $(\uparrow)$} & \textbf{KID $(\downarrow)$} \\
\midrule
0.01 & 1.0 & 0.954 & 0.946 & 0.133 \\ 
0.01 & 0.75 & 0.977 & 0.948 & 0.140 \\ 
0.01 & 0.5 & 0.554 & 0.965 & 0.090 \\ 
\midrule
0.1 & 1.0 & 0.960 & 0.965 & 0.126 \\ 
0.1 & 0.75 & 0.976 & 0.962 & 0.140 \\ 
0.1 & 0.5 & 0.554 & 0.962 & 0.089 \\ 
\midrule
0.2 & 1.0 & 0.963 & 0.947 & 0.134 \\ 
0.2 & 0.75 & 0.977 & 0.964 & 0.137 \\ 
0.2 & 0.5 & 0.556 & 0.962 & 0.089 \\ 
\bottomrule
\end{tabular}
\end{table}

\section{User Study}\label{appendix:extra_user}


\begin{figure}[h]
  \centering
  \begin{minipage}[b]{0.45\textwidth}
    \includegraphics[width=1.0\textwidth]{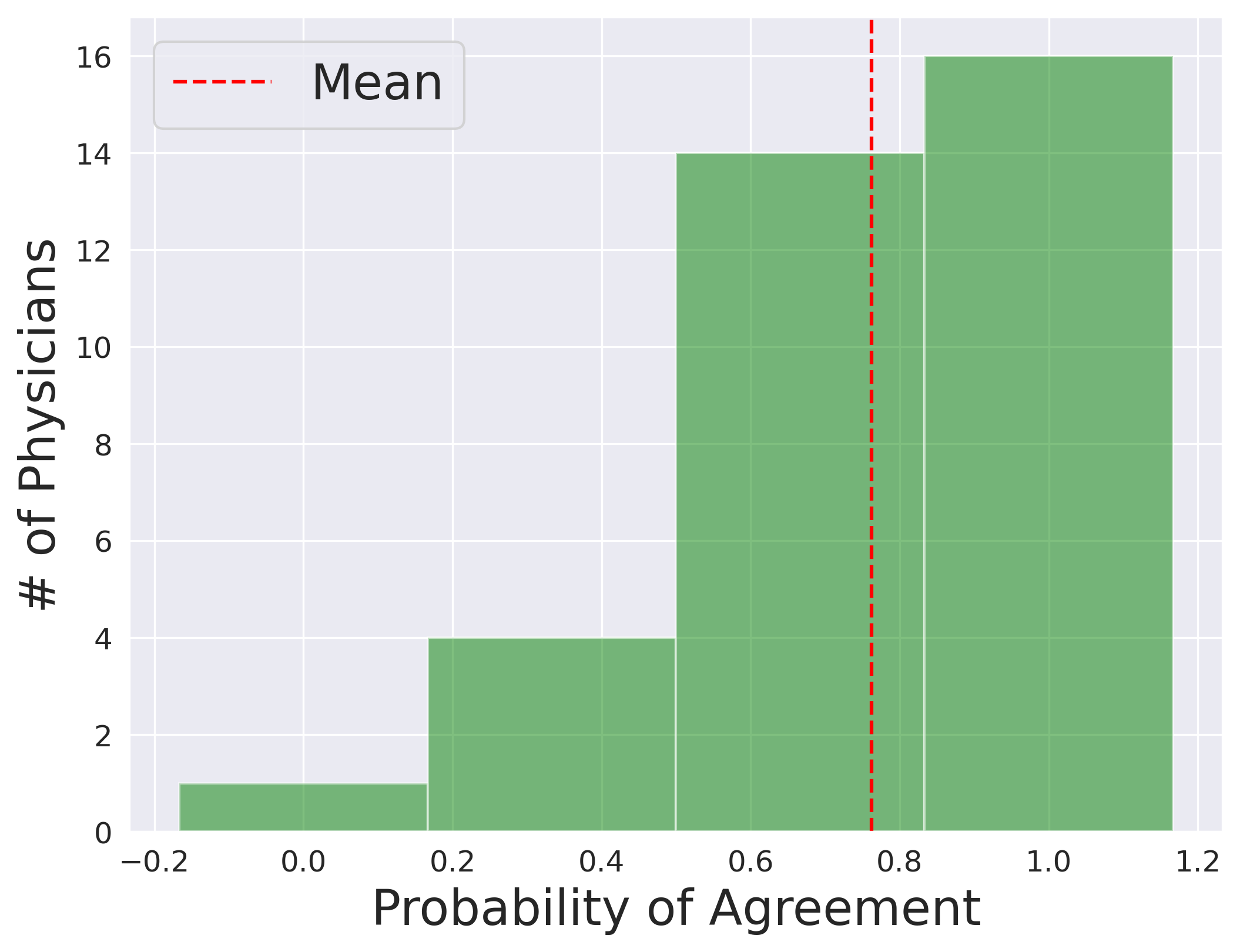}
    \caption{Distribution of probability of agreement to the generated progressions among 35 veteran physicians.}
    \label{fig:wrap_user}
  \end{minipage}
  \begin{minipage}[b]{0.08\textwidth}
    \textit{}
  \end{minipage}
  \begin{minipage}[b]{0.45\textwidth}
    \includegraphics[width=1.0\textwidth]{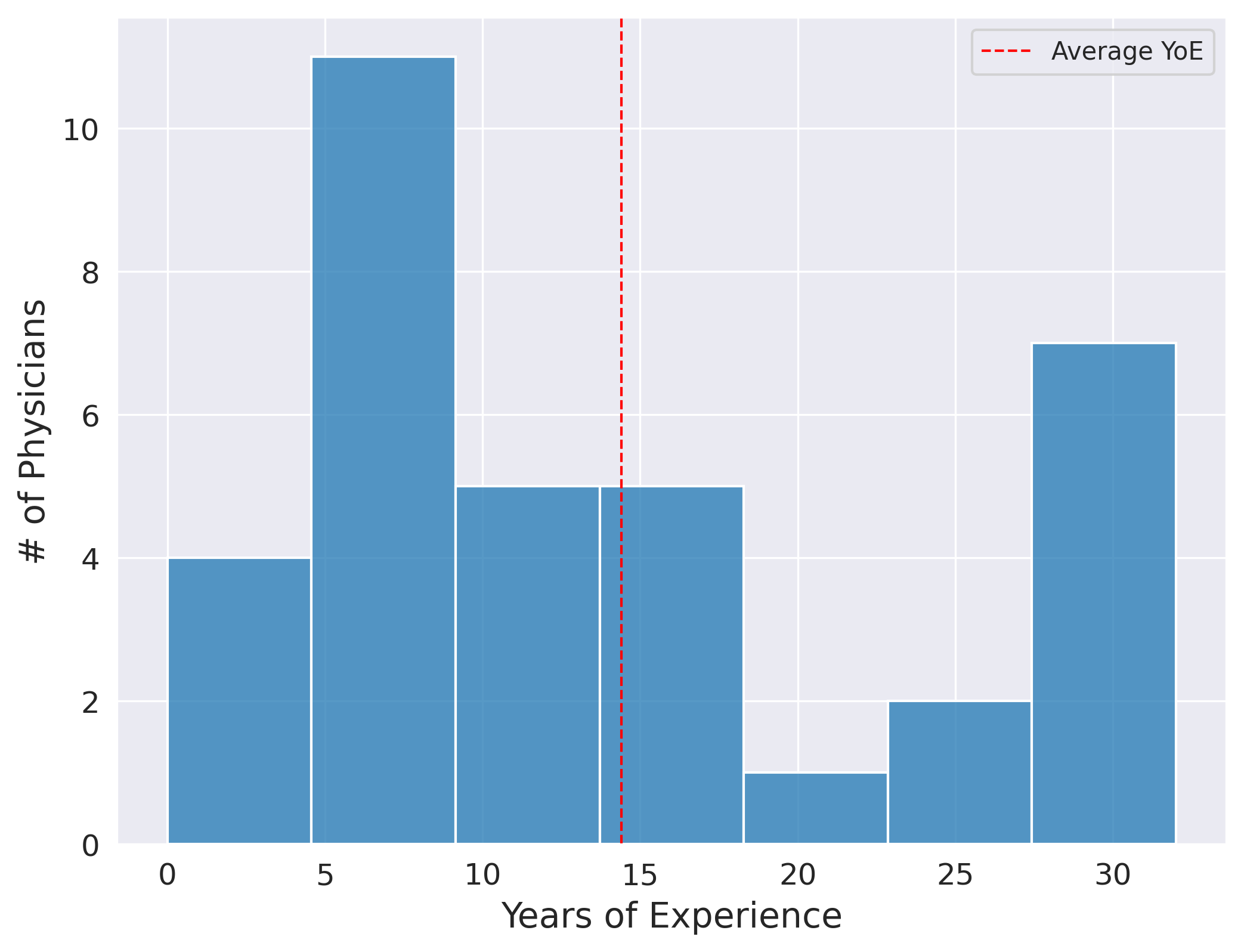}
    \caption{The distribution of the years of experience from the group of physicians who participated in the user study. The average number of years of experience is $14.4$ years. }
    \label{fig:yoe}
  \end{minipage}
\end{figure}

\subsection{Questionnaire}
\label{appendix:extra_user_q}

The questionnaire in the survey is approved by Affiliated clinical research institute. The questionnaire includes 2 parts. Part one consists of $20$ multiple choices of single image classifications, $10$ single-step generations, and $10$ real X-ray images sampled from the training set. Part two consists of $3$ generated disease progressions consisting of Cardiomegaly, Edema as well Pleural Effusion. Each progression runs 10 steps. For each single image classification, we ask "Please determine the pathologies of the following patient" and let the user pick from 6 options \{No findings, Cardiomegaly, Consolidation, Edema, Effusion, Atelectasis\} with possible co-occurrence, while for each of the $10$-step progressions, we ask "Does the below disease progression fit your expectation?" and let the user input a binary answer of yes or no.

\begin{figure}[h]
  \centering
  \begin{minipage}[b]{0.45\textwidth}
    \includegraphics[width=1.0\textwidth]{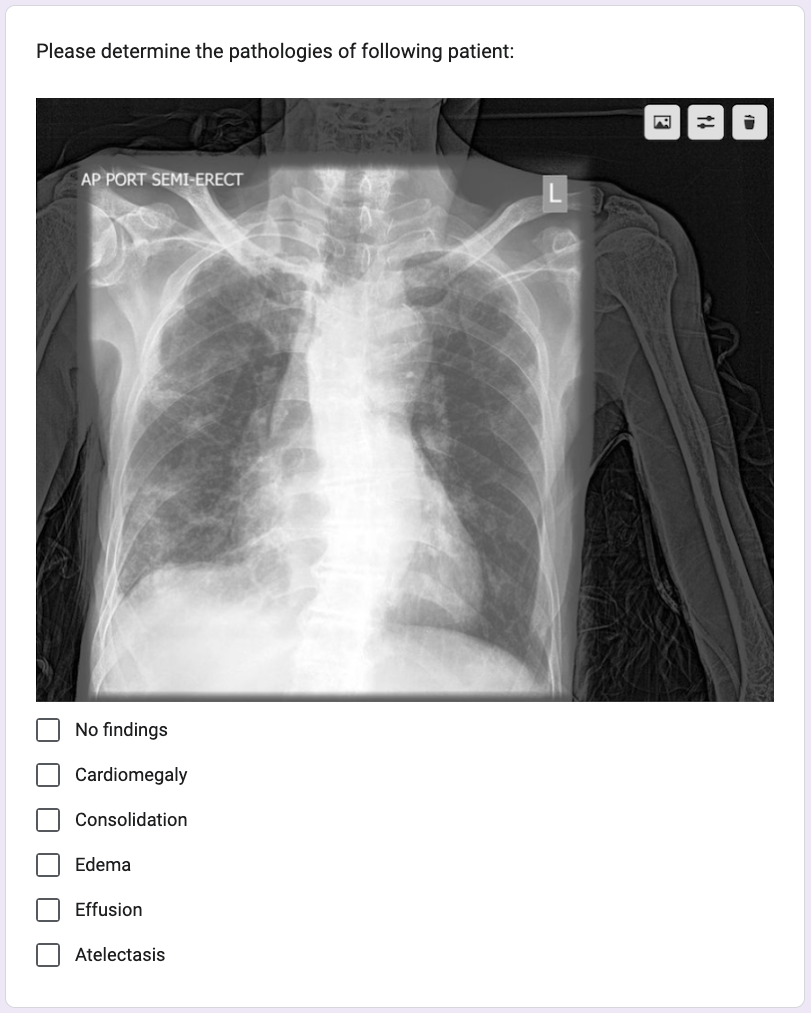}
    \caption{Example of User Study I: we ask the physicians to pick from the 6 options and it's possible to pick more than one option.}
    \label{fig:user_study_example_1}
  \end{minipage}
  \begin{minipage}[b]{0.08\textwidth}
    \textit{}
  \end{minipage}
  \begin{minipage}[b]{0.45\textwidth}
    \includegraphics[width=1.0\textwidth]{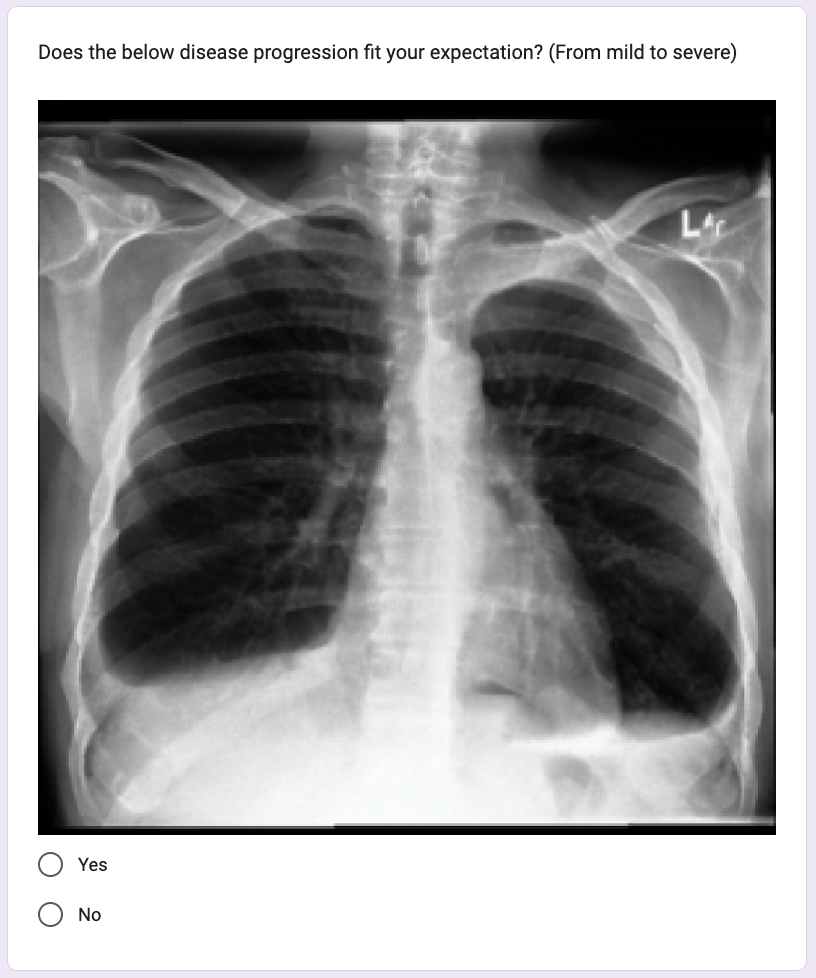}
    \caption{Example of User Study II: we ask the physicians to decide if the generated progression of the disease is credible or not.}
    \label{fig:user_study_example_2}
  \end{minipage}
\end{figure}

Below we include the full instructions we gave for our user study both in English and Chinese. Here we only include the English version of questions. The examples are shown in Figure~\ref{fig:user_study_example_1} and \ref{fig:user_study_example_2}. 
\begin{enumerate}
    \item Please read the instructions and inspect the images carefully before answering.
    \item Please provide your years of experience
    \item For the first 20 questions, please determine the pathologies from the X-ray images (you can choose more than one answer). For the last 3 questions, please answer if the disease progression shown fits your expectations.
\end{enumerate}


\subsection{Statistics}
\label{appendix:extra_user_stats}

The distribution of the years of experience from the group of physicians who participated in the user study. The average number of years of experience is $14.4$ years. Over half of the group have more than 10 years of experience. This data attests that our surveyees are highly professional and experienced. 

To show the significance of our findings, we also performed the paired t-test on the F1 scores of real and generated scans over the 35 users. Our finding is significant with a p-value of $0.0038$.

To quantitatively analyze their responses, we treat each class of pathologies as an independent class and compute precision, recall, and F1 over all the pathologies and physicians.

\section{Discussion and Ethics Statement}\label{appendix:discussion}

The proposed framework is subject to several limitations, with one of the primary constraints being the limited scope of Stable Diffusion. Due to the model's pre-training on general domain data, the absence of detailed medical reports poses a significant challenge to the model's ability to accurately and reliably edit medical images based on precise text conditioning. Moreover, the framework's overall performance may be influenced by the quality and quantity of available data, which can limit the model's accuracy and generalizability. Furthermore, the absence of surgical or drug intervention data further restricts the framework's ability to simulate medical interventions. Moving forward, it would be beneficial to explore ways of integrating more detailed descriptions of medical data in the fine-tuning process of Stable Diffusion to improve the framework's performance and precision in disease simulation through text conditioning. Additional details on the framework's limitations, including an analysis of failure cases, can be found in Supplementary~\ref{appendix:extra_ablation}.

Progressive Image Editing (PIE) holds promise as a technology for simulating disease progression, but it also raises concerns regarding potential negative social impacts. One crucial concern revolves around the ethical use of medical imaging data, which may give rise to privacy and security issues. To address this, healthcare providers must take measures to safeguard patient privacy and data security when utilizing PIE. An effective mitigation strategy involves employing anonymized or de-identified medical imaging data, while also adhering to ethical guidelines and regulations like those outlined by HIPAA (Health Insurance Portability and Accountability Act).

Another concern relates to the accuracy of the simulations generated by the framework, as errors could lead to misdiagnosis or incorrect treatment decisions. To alleviate this concern, rigorous testing and evaluation of the technology must be conducted before its implementation in a clinical setting. Enhancing the accuracy of simulations can be achieved by incorporating additional data sources, such as patient history, clinical notes, and laboratory test results. Additionally, healthcare providers should receive adequate training on effectively utilizing the technology and interpreting its results.

Discrimination against certain groups, based on factors such as race, gender, or age, poses yet another potential concern. Healthcare providers must ensure the fair and unbiased use of the technology. This can be accomplished by integrating diversity and inclusion considerations into the technology's development and training processes, as well as regular monitoring and auditing its usage to identify any signs of bias or discrimination.

The cost and accessibility of the technology present further concerns, potentially restricting its availability to specific groups or geographic regions. To tackle this issue, healthcare providers should strive to make the technology accessible and affordable to all patients, regardless of socioeconomic status or geographic location. This can be achieved through the creation of cost-effective models, partnerships with healthcare providers, and government funding initiatives.

Lastly, there is a risk of excessive reliance on technology, leading to a diminished reliance on clinical judgment and expertise. To mitigate this concern, healthcare providers must be trained to view the technology as a tool that complements their clinical judgment and expertise, rather than relying solely on it for diagnostic or treatment decisions. The technology should be used in conjunction with other data sources and clinical expertise to ensure a comprehensive understanding of disease progression.

In conclusion, while the use of PIE comes with potential negative social impacts, there are viable mitigations that can address these concerns. Healthcare providers must be aware of the ethical implications associated with this technology and take appropriate measures to ensure its safe and responsible utilization.